\documentclass{aa}
\usepackage{graphicx}
\usepackage{txfonts}
\usepackage{amssymb}
\usepackage[figuresright]{rotating}
\usepackage[switch]{lineno}  

\begin{document}

  \title{Tracing the potential planet-forming regions around \\
    seven pre-main-sequence stars\thanks{Based on observations made with
      Telescopes of the European Organisation for Astronomical Research in the
      Southern Hemisphere (ESO) at the Paranal Observatory, Chile, under the programs
      074.C-0342(A), 075.C-0064(A,B), 075.C-0413(A,B), and 076.C-0356(A).}} 

  \author{A. A. Schegerer\inst{1,2}, S. Wolf\inst{3,2}, C. A. Hummel\inst{4},
    S. P. Quanz\inst{2}, A. Richichi\inst{4}} 
  
  \offprints{A. A. Schegerer, \email{schegerer@mpia-hd.mpg.de}}
  
  \institute{
    Helmholtz Zentrum M\"unchen, German Research Center for
    Environmental Health, Ingolst\"adter Landstra{\ss}e 1, 85758 Neuherberg,
    Germany \and 
    Max Planck Institute for Astronomy (MPIA), 
    K\"onigstuhl 17, 69117 Heidelberg, Germany \and
    University of Kiel, Institute
    of Theoretical Physics and Astrophysics, Leibnizstra{\ss}e 15, 24098
    Kiel, Germany \and European Organisation for Astronomical Research in the
      Southern Hemisphere (ESO),
    Karl-Schwarzschild-Stra{\ss}e 2, 85748 Garching, Germany}
  
  \date{Received 11 August 2008 / Accepted 7 April 2009 }
  
  \abstract
  {}
  {We investigate the nature of the innermost regions
    with radii of several AUs of seven 
    circumstellar disks around pre-main-sequence stars, T\,Tauri stars in particular. 
    Our object sample contains disks apparently at various stages of their evolution.
    Both single stars and spatially resolved binaries are
    considered. In particular, we search for inner disk gaps as proposed
    for several young stellar objects (YSOs). When 
    analyzing the underlying dust  
    population in the atmosphere of circumstellar disks, the shape of the
    $10\,\mathrm{\mu m}$ feature 
    should additionally be investigated. } 
  {We performed interferometric observations in N band ($8-13\,\mathrm{\mu m}$) 
    with the Mid-Infrared Interferometric Instrument (MIDI) at the Very Large
    Telescope Interferometer (VLTI) using baseline lengths of between
    $54\,\mathrm{m}$ and $127\,\mathrm{m}$.
    The data analysis is based on radiative-transfer simulations using the
    Monte Carlo code MC3D by modeling 
    simultaneously the spectral energy distribution (SED), N band
    spectra, and interferometric visibilities. Correlated
    and uncorrelated N band spectra are compared to investigate the
    radial distribution of the dust composition of the disk atmosphere.}
  {Spatially resolved mid-infrared (MIR) emission was detected
    in all objects. For four objects (DR\,Tau, RU\, Lup, S\,CrA\,N, and
      S\,CrA\,S), the observed N band visibilities and corresponding SEDs 
      could be simultaneously simulated using a parameterized active
      disk-model. For the more evolved objects of our sample, HD\,72106 and HBC\,639, a
    purely passive disk-model provides the closest fit. The visibilities
    inferred for the source RU\,Lup allow the presence of an inner disk
    gap. For the YSO GW\,Ori, one of two visibility measurements could not be 
      simulated by our modeling approach.   
    All uncorrelated spectra reveal the $10\,\mathrm{\mu m}$ silicate emission
    feature. 
    In contrast to this, some correlated spectra of the observations of the
    more evolved objects do not show this feature, indicating a lack of small
    silicates in the inner 
    versus the outer regions of these disks. We conclude from this observational
    result that more evolved dust grains can be found in the more central disk
    regions. }
  {}  
  \keywords{Infrared: stars -- Accretion disks -- Planetary systems: protoplanetary disks
    -- Astrochemistry -- 
    Instrumentation: interferometer -- Radiative transfer } 
  
  \authorrunning{Schegerer et al.}
  \titlerunning{Tracing the planet-forming regions around YSO} 
  \maketitle
 
  \section{Introduction}\label{section:introduction}
  
  In this study we present the results of highly spatially resolved
  observations of seven pre-main-sequence stars acquired with the
  instrument MIDI at the VLTI. 
  Our study is focused on YSOs, T\,Tauri stars in particular, because this
  class of object is considered to represent the  
  progenitors of solar-type stars. Studying the evolution of the disks around
  these stars with MIDI 
  will provide us with an insight into the innermost and most dynamical
  regions where the earliest stages of the formation of
  a planetary system can be found.

  In terms of spatial resolution, sensitivity, and spectral coverage, MIDI is the only
  instrument suitable to the simultaneous investigation of the temperature and density
  distributions in the innermost disk regions, accretion effects, and
  dust properties, i.\/e., in determining the evolutionary status of
  circumstellar 
  disks. Even the presence of a (stellar) companion and its location can be
  investigated using MIDI. The analysis of (interferometric)
  observations of more than one YSO at different evolutionary stages provide the
  opportunity of verifying and constraining disk evolution and planet-formation
  models (e.g., Leinert et al.~\cite{leinert}).

  MIDI, operating in the atmospheric N band, is sensitive to warm
  ($\gtrapprox$$300\,\mathrm{K}$) matter, dust in particular, 
  located in the inner few AUs of young circumstellar disks. 
  It is generally assumed that the formation 
  of planetary systems starts in these inner disk regions (Klahr~\cite{klahr};
  W\"unsch et al.~\cite{wuensch}; Nagasawa et al.~\cite{nagasawa}).
  As a consequence of the planet-formation process, the disk structure and the
  composition 
  of the dust population is assumed to evolve 
  (Lissauer~\cite{lissauer}; Gail~\cite{gail}; Wolf~\cite{wolf-habil}). 
  The dust composition and density distribution of old
  T\,Tauri objects with an age of $>$$10\,\mathrm{Myrs}$ are assumed to have been
  strongly modified. In these systems, the accretion onto the star is strongly
  reduced or has even stopped (Hartmann et al.~\cite{hartmann}). Furthermore,
  the  former circumstellar envelope of Class I objects as well as the
  innermost disk regions (Calvet et 
  al.~\cite{calvetII}) may have disappeared in these
  objects. In contrast,  
  in YSOs with an age of $\lse$$1\,\mathrm{Myr}$, the accretion process 
  usually plays a dominant role. 

  Binarity is a common phenomenon in 
  YSOs. Extensive studies in the near-infrared (NIR) wavelength range,
  where a spatial resolution of between {$0.005$\arcsec} and {$10$\arcsec} was
  reached, showed that between $23$\% and $55$\% of YSOs are binaries or even multiple
  systems (Simon et al.~\cite{simon}; K\"ohler \& Leinert~\cite{koehler};
  D\^uchene et al.~\cite{ducheneII}; Ratzka et 
  al.~\cite{ratzkaII}). In a theoretical study, Artymowicz \&
  Lubow~(\cite{artymowicz}) highlighted modifications to the structure of
  circumstellar and circumbinary disks resulting from a close
  companion. Furthermore, the nature of the infrared companions of T\,Tauri stars
  remains a matter of debate in terms of the structure and orientation of
  their circumstellar disks as well as their evolutionary stage (Koresko et
  al.~\cite{koreskoII}; McCabe et al.~\cite{cabe}).

  Simultaneous modeling of the spectral energy distribution (SED) -- averaged
  over the entire disk --  
  and of interferometric measurements -- sensitive to the inner, warm
  regions -- is the main goal of this study. 
  Modeling the SED alone does not allow us to constrain unambiguously the disk
  density and temperature structure (e.g., Thamm et al.~\cite{thamm}). 
  Spatially resolved observations, such as interferometric observations
  with MIDI are needed to reduce the ambiguity caused by considering solely the
  SED. The
  modeling approach should be as simple as possible to reduce the 
  mathematical underdetermination of the model. A key issue of this study is
  to determine whether
  all objects in our heterogeneous sample can be modeled with one
  single approach or if modifications and extensions of the model must also 
  be considered. Finally,
  it has not yet been shown whether any quantities of the objects, such as the
  stellar luminosity, stellar temperature, and mass   
  accretion rate derived in previous, spatially unresolved
  observations can also be confirmed with interferometric, 
  small-scale (range of AU) observations. 
  
  Any deviations in the outcome of a homogeneous disk model from the (spectral and
  spatial) measurements
  point to (small-scale) perturbations of the disk 
  structure. A non-representative selection of potential perturbations are an
  inner-disk gap (Calvet et al.~\cite{calvetII};
  Mathieu et al.~\cite{mathieuIII}), vortices (Klahr~\cite{klahr}), a
  puffed-up inner-rim wall (Dullemond et al.~\cite{dullemond}), and a close
  stellar companion (Mathieu et al.~\cite{mathieu}; Artymowicz \&
  Lubow~\cite{artymowicz}). Non-homologous evolution
  of the disk produces these perturbations (Wood et al.~\cite{wood}; McCabe et
  al.~\cite{cabe}).    
  
  MIDI provides both correlated, i.\/e., coherent spectra sensitive to the inner
  part, and uncorrelated spectra sensitive to regions that can be
  non-interferometrically resolved. For the T\,Tauri objects TW\,Hya (Ratzka
  et al.~\cite{ratzkaIII}) and RY\,Tau (Schegerer et al.~\cite{schegerer}), an
  increase in the re\-lative crystalline mass contribution to the
  $10\,\mathrm{\mu m}$ silicate feature could be determined for an increasing
  spatial resolution obtained with MIDI. This result was even found for
  the active object FU\,Ori (Quanz et al.~\cite{quanz}). This
  finding was explained by the 
  hypothesis that silicate dust evolves more rapidly in inner disk regions, i.\/e.,
  at small radii $r$, than in outer regions of the disk (e.\/g., van Boekel et 
  al.~\cite{boekel}) assuming similar initial dust compositions. It is not
  known, so far, if this increase occurs
  only for intermediate-mass YSOs such as RY\,Tau or if such a
  correlation exists even for fainter ($\sim$$1\,\mathrm{L_{\odot}}$),
  younger ($\sim$$1\,\mathrm{Myrs}$ old), and less active T\,Tauri
  objects. Radial mixing could reverse the increase in evolved dust
  towards the more inner regions (Gail~\cite{gail}). In this study, we 
  compare silicate features that effectively arise at different disk
  radii within an object. By focusing on one single object, the results
  of this investigation do not depend on disk properties such as the
  evolutionary stage, initial composition, and disk inclination but only 
  on the origin of the feature. 

  This paper is structured in the following way. In
  Sect.~\ref{section:observation}, we present the object sample and briefly
  describe our observations and data reduction. In
  Sect.~\ref{section:modeling}, our modeling approach is outlined. The derived
  individual disk parameters are presented in Sect.~\ref{section:models}. An
  analysis of the $10\,\mathrm{\mu m}$ silicate feature is given in
  Sect.~\ref{section:dust composition}, followed by a summary in
  Sect.~\ref{section:conclusion}. The Appendix lists some further properties
  of the objects in our sample.  
  
  \section{Interferometric observations and data reduction}\label{section:observation}

  \begin{table*}
    \centering
    \begin{minipage}{0.92\textwidth} 
      \caption{Object properties derived in previous measurements or in this
        study. Different models of an object are provided in roman
        digits. Here, the 
        coordinates (RA and DEC in J2000.0), distance ($d$), visual extinction
        ($A_\mathrm{V}$), spectral type (SpTyp), stellar mass ($M_\star$),
        effective stellar temperature ($T_\star$), stellar luminosity
        ($L_\star$), and age are listed. Footnotes are references and point
        to the previous measurements. The footnotes ``det.'' and ``ass.''
        represent parameter values that were determined and assumed in this study,
        respectively.} 
      \begin{tabular}{lllllllllc}\hline\hline
        Object (no.) &  $RA$ [h\ m\ s] & $DEC$ [$^{\circ}$\ \arcmin\ \arcsec] &
        $d$ [pc] & $A_\mathrm{V}$ [mag] & SpTyp & $M_\star$ [M$_\mathrm{\odot}$]
        & $T_\star$ [K] & $L_\star$ [L$_\mathrm{\odot}$] & $Age$ [Myr] \\  \hline
        DR\,Tau (I) & & & & & & & & $1.7_\mathrm{(det.)}$ & \\[1.0ex] 
        DR\,Tau (II) & $04\ 47\ 06.2_\mathrm{(2)}$ & $+16\ 58\ 43_\mathrm{
          (2)}$ & $140_\mathrm{(ass.)}$ & $1.6_\mathrm{(3)}$ & K$7_\mathrm{(3,4)}$ &
        $0.80_\mathrm{(4)}$ & $4000_\mathrm{(4)}$ & $0.9_\mathrm{(3)}$ &
        $3_\mathrm{(5)}$ \\[1.0ex] 
        DR\,Tau (III) & & & & & & & & $0.9_\mathrm{(3)}$ & \\[1.0ex] 
        GW\,Ori (I) & \raisebox{-1.5ex}{$05\ 29\ 08.4_\mathrm{(1)}$} &
        \raisebox{-1.5ex}{$+11\ 52\ 13_\mathrm{(1)}$} &
        \raisebox{-1.5ex}{$440_\mathrm{(6)}$} &
        \raisebox{-1.5ex}{$1.3_\mathrm{(6)}$} &
        \raisebox{-1.5ex}{G$0_\mathrm{(6)}$} &
        \raisebox{-1.5ex}{$3.7_\mathrm{(6)}$} &   
        \raisebox{-1.5ex}{$6000_\mathrm{(6)}$} & $40_\mathrm{(det.)}$ &
        \raisebox{-1.5ex}{$1_\mathrm{(6)}$}\\[1.0ex] 
        GW\,Ori (II) & & & & & & & & $62_\mathrm{(6)}$ & \\[1.0ex]
        HD\,72106\,B & $08\ 29\ 34.9_\mathrm{(1)}$ & $-38\ 36\ 21_\mathrm{
          (1)}$ & $290_\mathrm{ (9)}$ & $0.0_\mathrm{ (7)}$ & A$0_\mathrm{ (9)}$ & 
        $1.8_\mathrm{ (8)}$ & $9500_\mathrm{ (7)}$ & $28_\mathrm{ (det.)}$ &
        $10_\mathrm{ (8)}$ \\[1.0ex]
        RU\,Lup & $15\ 56\ 42.3_\mathrm{ (1)}$ & $-37\ 49\ 16_\mathrm{
          (1)}$ &
        $127_\mathrm{ (11)}$ & $0.50_\mathrm{ (11)}$ & K$8_\mathrm{ (11)}$ &
        $0.80_\mathrm{ (11)}$ & $4000_\mathrm{ (11)}$ & $1.3_\mathrm{ (11)}$ &  
        $1_\mathrm{ (11)}$ \\[1.0ex]
        HBC\,639 & $16\ 26\ 23.4_\mathrm{ (2)}$ & $-24\ 20\ 60_\mathrm{ (2)}$
        & $170_\mathrm{ (11)}$ & $5.7_\mathrm{ (11)}$ & K$0_\mathrm{ (11)}$ &
        $2.0_\mathrm{ (11)}$ & $4800_\mathrm{ (det.)}$ &  
        $8.5_\mathrm{ (11)}$ & $2_\mathrm{ (11)}$ \\[1.0ex]
        S\,CrA\,N (I,II) & \raisebox{-1.5ex}{$19\ 01\ 08.6_\mathrm{ (2)}$} &
        \raisebox{-1.5ex}{$-36\ 57\ 20_\mathrm{ (2)}$} &
        \raisebox{-1.5ex}{$130_\mathrm{ (12,13)}$} & $1.5_\mathrm{ (12)}$ &
        K$3_\mathrm{ (12)}$ & $1.5_\mathrm{ (12)}$ & $4400_\mathrm{ (12)}$ &
        $2.3_\mathrm{ (12)}$ & $3_\mathrm{ (12)}$ \\[1.0ex]
        S\,CrA\,S & & & & $1.0_\mathrm{ (12)}$ & M$0_\mathrm{ (12)}$
        & $0.60_\mathrm{ (12)}$ & $3800_\mathrm{ (12)}$ & $1.0_\mathrm{ (det.)}$
        & $1_\mathrm{ (12)}$ \\  
        \hline  
      \end{tabular} 
      \label{table:properties-midisurvey} 
      {\newline \scriptsize {References:} { 1}:
        Perryman~(\cite{perryman}); { 2}:  
        2\,MASS catalogue~(Cutri et al.~\cite{cutri}); { 3}: Muzerolle
        et al.~(\cite{muzerolle});  
        { 4}:~Mohanty et al.~(\cite{mohanty}); { 5}:
        Greaves~(\cite{greaves}); { 6}: Calvet et  
        al.~(\cite{calvet});  { 7}: Sch\"utz et al.~(\cite{schuetz});
        { 8}: Wade et  
        al.~(\cite{wade}); { 9}: Vieira et al.~(\cite{vieira}); {
          10}: Folsom~(\cite{folsom}); { 11}: 
        Gras-Vel\'azquez \& Ray~(\cite{gras-velazquez}); { 12}:~Prato
        et al.~(\cite{prato}); { 13}:  
        Johns-Krull et al.~(\cite{johns-krullII})}
    \end{minipage}
  \end{table*} 
  
  Most of our objects are well-known T\,Tauri stars. 
  They belong to various star-forming regions such as the Taurus-Auriga (e.\/g., DR\,Tau) and $\rho$ 
  Ophiuchi (e.\/g., HBC\,639) region. Selected object parameters are compiled
  in  
  Table~\ref{table:properties-midisurvey} (see Appendix~\ref{appendix} for more details).
  Apart from one, all objects are between $0.1$ and $3\,$million years old.  
  The exception is HD\,72106\,B with an age of $10\,$million years. 
  Highly spatially resolved measurements in the NIR and mid-infrared (MIR) wavelength range have
  shown that the objects DR\,Tau and  
  RU\,Lup do not have any stellar companion (e.g., Ghez et
  al.~\cite{ghez}). In contrast, HD\,72106 and S\,CrA at least are well-known binary systems. Considering the
  stellar mass, GW\,Ori with $M_{\star} = 3.7\,\mathrm{M_\mathrm{\odot}}$
  would be the only Herbig Ae/Be object ($M_\star 
  > 3\,\mathrm{M_{\odot}}$) in our sample, while in terms of spectral type only
  HD\,72106\,B would be a Herbig Ae/Be object (Vieira et
  al.~\cite{vieira}). 
  
  The observing time with MIDI, the projected baseline length ($B$),
  and the corresponding position angle  
  ($PA$) of the interferometer during observations are listed in 
  Table~\ref{table:observation-midisurvey}. 
  A detailed description of an observing sequence with MIDI is given in
  Leinert et al.~(\cite{leinert}).
  The calibration was completed using data for all calibrator stars that
  were observed during one night with the 
  same observational mode as the scientific target and
  show clear visibility signals. The error bars in the visibility curves, which are
  presented below, represent the standard deviation when using the ensemble of
  calibrator measurements. Since the sample contains various objects with
  known close-by infrared companions, we checked whether these companions affect the
  visibility measurements. This examination was necessary because of the
  relatively wide field-of-view of MIDI, of about $1\,\mathrm{''}$. The
  companions do not affect the measurements. 
  We used the MIA package (MIDI Interactive Analysis Software) for data
  reduction (Leinert et al.~\cite{leinert};
  Ratzka~\cite{ratzka}). The results were confirmed by using the independent
  EWS software. Both reduction software 
  packages are publicly available.\footnote{software packages are available at {\tt
      http://www.mpia-hd.mpg.de/MIDISOFT/} \\ 
    and \\ 
    {\tt http://www.strw.leidenuniv.nl/$^\sim$koehler/MIA+EWS-Manual/}} 
  Several target/calibrator pairs were observed at different airmasses
  (Table~\ref{table:observation-midisurvey}). Therefore, a correction to the
  airmass was performed by multiplying the uncorrelated spectra
  with the inverse of the cosine of the zenith angle. The visibility spectra
  do not need to be corrected because the correlated flux is
  normalized by the uncorrelated flux.

  \begin{table*}[!tb]
    \centering  
    \begin{minipage}{0.92\textwidth}  
    \caption{\label{table:observation-midisurvey} 
      Overview of our observations with MIDI. The date and the observing
      time ($UT$) as well as telescope pairs ($TP$), and both   
      the projected (effective) length ({\@}$B${\@}) and angle of the baseline
      ({\@}$PA${\@}, measured from North to East) during the 
      observations are presented for targets and corresponding calibrator
      stars. The given airmass ($AM$) is an average value for the 
      observation. In the following columns, we add the diameter of the
      calibrator star and the corresponding references. The diameters were
      derived from fitting the SED of the objects. In the last column, we
      also provide comments concerning the observation and data reduction.} 
    \begin{tabular}{lcccrrrrrl}
      \hline 
      \hline 
      Date & $UT$ & Object & $TP$ & $B$ [m] & $PA$ [$^\mathrm{\circ}$] 
      & $AM$ & Diameter [mas] & Ref. & Comments \\\hline
      Jan. $01^\mathrm{st}\,2005$ & $2:30 - 2:59$ & DR\,Tau & UT3-UT4 & $61$ & $106$ &
      $1.3$ & & & $\ast$, $\ast \ast \ast\ \ast$ \\ 
      Jan. $01^\mathrm{st}\,2005$ & $3:15 - 3:17$ & HD\,31421 & UT3-UT4 & $60$ & $106$ &
      $1.3$ & $2.74\pm0.03$ & 1 & \\
      Jan. $31^\mathrm{st}\,2005$ & $3:56 - 3:58$ & HD\,31421 & UT3-UT4 & $56$ & $104$ &
      $1.3$ & $2.74\pm0.03$ & 1 & \\ \hline

      Mar. $01^\mathrm{st}\,2005$ & $23:57 - 0:14$ & HD\,31421 & UT3-UT4 & $56$ & $105$ &
      $1.3$ & $2.74\pm0.03$ & 1 & \\
      Mar. $01^\mathrm{st}\,2005$ & $0:21 - 0:35$ & GW\,Ori & UT3-UT4 & $56$ & $105$ &
      $1.3$ & & & \smallskip\\

      Mar. $03^\mathrm{rd}\,2005$ & $0:02 - 0:13$ & HD\,31421 & UT2-UT4 & $87$ & $79$ &
      $1.3$ & $2.74\pm0.03$ & 1 & \\
      Mar. $03^\mathrm{rd}\,2005$ & $0:28 - 0:45$ & GW\,Ori & UT2-UT4 & $88$ & $79$ &
      $1.3$ & & & $\ast$ \\\hline
      
      Dec. $31^\mathrm{st}\,2005$ & $4:19 - 4:39$ & HD\,72106 & UT1-UT4 & $127$ & $41$
      & $1.1$ & & & $\ast \ast$\\
      Dec. $31^\mathrm{st}\,2005$ & $4:53 - 4:56$ & HD\,69142 & UT1-UT4 & $130$ & $48$ &
      $1.1$ & $2.18\pm0.18$ & 2 & \smallskip\\

      Mar. $11^\mathrm{th}\,2005$ & $0:29 - 1:01$ & HD\,72106 & UT2-UT4 & $88$ & $71$ &
      $1.1$ & & & $\ast \ast$\\
      Mar. $11^\mathrm{th}\,2005$ & $1:08 - 1:27$ & HD\,69142 & UT2-UT4 & $89$ & $78$ &
      $1.0$ & $2.18\pm0.18$ & 2 &  \\\hline

      Aug. $26^\mathrm{st}\,2005$ & $0:52 - 0:56$ & HD\,152885 & UT2-UT4 & $84$ & $96$ &
      $1.1$ & $2.88\pm0.09$ & 2 & \\
      Aug. $26^\mathrm{st}\,2005$ & $1:55 - 2:25$ & RU\,Lup & UT2-UT4 & $64$ & $122$ &
      $1.5$ & & & $\ast$ \\
      Aug. $26^\mathrm{st}\,2005$ & $2:38 - 2:40$ & HD\,152885 & UT2-UT4 & $69$ & $116$ &
      $1.4$ & $2.88\pm0.09$ & 2 & \\
      Aug. $26^\mathrm{st}\,2005$ & $3:02 - 3:04$ & HD\,178345 & UT2-UT4 & $84$ & $96$ &
      $1.1$ & $2.42\pm0.03$ & 2 & \\
      Aug. $26^\mathrm{st}\,2005$ & $4:48 - 4:50$ & HD\,178345 & UT2-UT4 & $69$ & $116$ &
      $1.4$ & $2.42\pm0.03$ & 2 & \smallskip \\

      May  $15^\mathrm{th}\,2006$ & $2:29 - 2:43$ & RU\,Lup & UT2-UT3 & $46$ & $17$ &
      $1.2$ & & & \\
      May  $15^\mathrm{th}\,2006$ & $4:51 - 5:03$ & HD\,152820 & UT2-UT3 & $46$ & $13$ &
      $1.3$ & $2.63\pm0.15$ & 2 & \smallskip\\

      May  $25^\mathrm{th}\,2005$ & $3:20 - 3:28$ & HD\,152820 & UT3-UT4 & $55$ & $94$ &
      $1.1$ & $2.63\pm0.15$ & 2 & \\
      May  $25^\mathrm{st}\,2005$ & $3:43 - 3:50$ & RU\,Lup & UT3-UT4 & $61$ & $103$ &
      $1.0$ & & & \smallskip\\
      May  $25^\mathrm{st}\,2005$ & $7:32 - 7:40$ & HD\,152820 & UT3-UT4 & $58$ & $130$ &
      $1.1$ & $2.63\pm0.15$ & 2 & \\
      May  $25^\mathrm{th}\,2005$ & $7:55 - 8:02$ & RU\,Lup & UT3-UT4 & $55$ & $149$ &
      $1.5$ & & & \\\hline

      Apr. $18^\mathrm{th}\,2005$ & $4:11 - 4:18$ & HD\,142804 & UT2-UT4 & $71$ & $66$
      & $1.3$ & $2.80\pm0.08$ & 2 & \\
      Apr. $18^\mathrm{th}\,2005$ & $4:39 - 4:58$ & HBC\,635 & UT2-UT4 & $75$ & $61$ &
      $1.2$ & & & \smallskip \\
      Apr. $19^\mathrm{th}\,2005$ & $3:19 - 4:33$ & HD\,142804 & UT2-UT4 & $60$ & $58$ &
      $1.5$ & $2.80\pm0.08$ &  2 & \\
      Apr. $19^\mathrm{th}\,2005$ & $3:44 - 4:11$ & HBC\,635 & UT2-UT4 & $66$ & $52$ &
      $1.4$ & & & $\ast$ \smallskip \\
      Aug. $25^\mathrm{th}\,2005$ & $23:19 - 23:21$ & HD\,142804 & UT2-UT4 & $60$ & $116$ &
      $1.0$ & $2.80\pm0.08$ & 2 & \\
      Aug. $25^\mathrm{th}\,2005$ & $23:45 - 23:46$ & HD\,142804 & UT2-UT4 & $58$ & $118$ &
      $1.0$ & $2.80\pm0.08$ & 2 & \\
      Aug. $26^\mathrm{th}\,2005$ & $0:08 - 0:27$ & HBC\,635 & UT2-UT4 & $59$ & $121$ &
      $1.1$ & & & $\ast$ \\ \hline

      May  $31^\mathrm{st}\,2005$ & $10:07 - 10:38$ & S\,CrA\,N & UT2-UT4 & $70$ &
      $109$ & $1.4$ & & & $\ast \ast \ast$ \\
      May  $31^\mathrm{st}\,2005$ & $10:48 - 10:49$ & HD\,178345 & UT2-UT4 & $66$
      & $120$ & $1.5$ & $2.42\pm0.03$ & 1 & \smallskip\\
      Jun. $28^\mathrm{th}\,2005$ & $7:57 - 7:58$ & HD\,178345 & UT3-UT4 & $59$ & $137$ &
      $1.2$ & $2.42\pm0.03$ & 1 & \\
      Jun. $28^\mathrm{th}\,2005$ & $8:17 - 8:39$ & S\,CrA\,N & UT3-UT4 & $56$ & $145$
      & $1.4$ & & & \\\hline

      Aug. $26^\mathrm{th}\,2005$ & $0:52 - 0:57$ & HD\,152885 & UT2-UT4 & $84$ & $96$
      & $1.1$ & $2.88\pm0.09$ & 2 & \\
      Aug. $26^\mathrm{th}\,2005$ & $2:38 - 2:40$ & HD\,152885 & UT2-UT4 & $69$ & $116$
      & $1.4$ & $2.88\pm0.09$ & 2 & \\
      Aug. $26^\mathrm{th}\,2005$ & $3:02 - 3:04$ & HD\,178345 & UT2-UT4 & $84$ & $96$
      & $1.1$ & $2.42\pm0.03$ & 1 & \\
      Aug. $26^\mathrm{th}\,2005$ & $4:04 - 4:35$ & S\,CrA\,S & UT2-UT4 & $72$ & $110$
      & $1.3$ & & & $\ast$ \\
      Aug. $26^\mathrm{th}\,2005$ & $4:48 - 4:50$ & HD\,178345 & UT2-UT4 & $69$ & $116$
      & $1.4$ & $2.42\pm0.03$ & 1 & \\\hline
      \hline 
    \end{tabular}
    {\newline \scriptsize $\ast$ bad weather conditions (not perfect beam overlap,
      noisy data) \hfill{} \newline 
      $\ast \ast$ data reduced using EWS package, only \hfill{} \newline
      $\ast \ast \ast$ results from MIA and EWS package are not consistent at short
      wavelength (deviation: $13$\%) \hfill{} \newline  
      $\ast \ast \ast\ \ast$ results from MIA and EWS package are not
      consistent for the whole band (devation: $14$\%)\hfill{}\newline 
      {References:} 1: Cohen et al.~(\cite{cohenV}); 2: MIDI
      consortium~(in prep.)}
  \end{minipage}
\end{table*} 
  
  \section{Modeling approach}\label{section:modeling}
  In the data analysis, we used the radiative-transfer code MC3D, which is
  based on a Monte Carlo method (Wolf et al.~\cite{wolfI}) and was
  more recently developed to its present form (Schegerer et al.~\cite{schegerer}). 
  A predefined, parameterized model of a passive disk is the basic ingredient
  of our modeling approach. 
  We use the disk model of Shakura \& Sunyaev~(\cite{shakura}), where the density distribution
  is given as:
  \begin{eqnarray}
    \label{eq:shakura}
    \hfill{}
    \rho(r,z)=\rho_{0} \left( \frac{R_{\star}}{r} \right)^{\alpha} 
    \exp \left[ - \frac{1}{2} \left( \frac{z}{h(r)} \right)^{2} \right].
    \hfill{}
  \end{eqnarray} 
  The quantities $r$ and $z$ are the radius measured from the disk center and the
  vertical distance from the disk midplane, respectively. The quantity
  $R_\mathrm{\star}$ is the stellar radius. 
  The function $h(r)$ represents the scale-height:\footnote
  {
    The scale-height is defined as the vertical distance from the midplane,
    where the density has  
    decreased by a factor $e \approx 2.718$ (Euler's constant).
  } 
  \begin{eqnarray}
    \label{eq:scaleheight}
    \hfill{}
    h(r)=h_\mathrm{100} \left( \frac{r}{100\,\mathrm{AU}} \right)^{\beta}
    \hfill{}
  \end{eqnarray} 
  with $h_\mathrm{100} = h(r=100\,\mathrm{AU})$. 
  The quantity $\rho_{0}$ allows us to scale the total disk mass $M_\mathrm{disk}$. 
  We assume a gas-to-dust mass ratio of 100:1. In our approach the exponents
  $\alpha$ and $\beta$ satisfy the following relation:
  \begin{eqnarray}
    \label{eq:alpha-beta}
    \hfill{}
    \alpha=3(\beta-\frac{1}{2}).
    \hfill{}
  \end{eqnarray}   
  The latter relation results from the coupling of surface density
  and temperature of the disk (Shakura \& Sunyaev~\cite{shakura}). We  note that
  the approach established by Shakura \& Sunyaev is correct only for
  geometrically thin 
  disks, i.\/e., where $z \ll r$. However, since the density exponentially decreases
  with increasing $z$ and the scale-height
  $h(r)$ is at least one order of magnitude smaller than the corresponding radius $r$,
  this approximation should be justified for the predominant mass fraction. The
  error in the density of regions $z \gtrapprox r$ is of the 
  order $z^5/r^7$ for Taylor's series. The approach of Shakura \&
  Sunyaev was already successfully used in modeling HH\,30\,IRS
  (Burrows et al.~\cite{burrows}; Wood et al.~\cite{woodII}; Cotera et
  al.~\cite{cotera}) and the Butterfly Star (Wolf et al.~\cite{wolfII}). In
  contrast to the more basic, but time-consuming, approach of hydrostatical
  equilibrium (Schegerer et al.~\cite{schegerer}), the predefined,
  parametrized disk model of Shakura \& Sunyaev~(\cite{shakura}) avoids the
  difficulties in handling regions of high optical depth (e.\/g., Sonnhalter
  et al.~\cite{sonnhalter}). However, even in predefined density distributions,
  the resulting temperature distribution must converge
  independently of the number of photon packages emitted.
  
  For all objects besides HD\,72106\,B and HBC\,639, accretion effects were
  also considered. The potential energy of a particle on its way
  towards the star is partly released in the disk midplane assuming the
  canonical modeling approach of an geometrically thin, active disk as
  formerly established by Lynden-Bell \& Pringle~(\cite{lynden}), and
  Pringle~(\cite{pringle}). Considering the "magnetically mediated" modeling
  approach (e.\/g., Uchida \& Shibata~\cite{uchida}; Bertout et
  al.~\cite{bertout}; Calvet \& Gullbring~\cite{calvet+gullbring}), the
  accretion disk is truncated by the stellar magnetic field at a radius
  $R_\mathrm{bnd}$. Thereupon, more than half of the potential energy of the
  accreting particles is released in a boundary region above the stellar
  surface. This boundary region is heated to a temperature of
  $T_\mathrm{bnd}$ by the accreting material. The effects of the this
  approach were described and applied by Schegerer et
  al.~(\cite{schegerer}) in modeling the circumstellar disk around
  RY\,Tau. Calvet \& Gullbring~(\cite{calvet+gullbring}) theoretically derived 
  reasonable ranges for the latter both accretion values,
  i.\/e., $5700\,\mathrm{K} < T_\mathrm{bnd} < 8800\,\mathrm{K}$ and
  $2\,R_{\star} < R_\mathrm{bnd} < 5\,R_{\star}$. We note
  that the effects of any variations in the boundary temperature
  $T_\mathrm{bnd}$ and the magnetic truncation radius $R_\mathrm{bnd}$ on the SED
  and MIR visibilities  can be neglected in the above-mentioned ranges
  (D'Alessio et al.~\cite{dalessio}). In our modeling approach, we always used
  standard values for the boundary temperature and truncation radius,
  i.\/e., $T_\mathrm{bnd}=8000\,\mathrm{K}$ and
  $R_\mathrm{bnd}=5\,R_{\star}$. These values were applied by
  Akeson et al.~(\cite{akeson}) and Schegerer et al.~(\cite{schegerer}) for 
  other T\,Tauri stars. The accretion rate $\dot{M}$, which determines the
  total accretion luminosity $L_\mathrm{acc}$, is not an independent model
  parameter. But its value is constrained by the results of previous
  measurements. Correspondingly, the stellar mass $M_\mathrm{\star}$, the
  effective stellar temperature $T_\mathrm{\star}$, and the stellar luminosity
  $L_\mathrm{\star}$, which were derived in previous studies were starting
  parameters in the modeling. In some cases, it was necessary to deviate from
  these results to be able to simulate the measurements considered
  in this study. These deviations are discussed below.
 
  In our modeling approach, we assumed a dust mixture of "astronomical
  silicate" and graphite with relative abundances of $62.5$\% and $37.5$\%,
  respectively (Draine \&
  Malhotra~\cite{draine}). 
  We used a two-layer disk model, i.\/e., the disk interior, where the optical 
  depth in N band fulfills the condition $\tau_\mathrm{N}>1$,
  containing dust grains with a maximum particle size 
  of $a_\mathrm{max}=1\,\mathrm{mm}$. A maximum particle size of
  $a_\mathrm{max}=1\,\mathrm{mm}$ was  
  found for several T\,Tauri stars with millimeter observations using
  the Very Large Array  
  (Rodmann et al.~\cite{rodmann}). Here, the optical depth $\tau_{N}$ is
  measured for constant radii, from the disk  
  atmosphere vertically to the disk midplane. The 
  surface regions consist of interstellar, non-evolved dust with
  $a_\mathrm{max}=0.25\,\mathrm{\mu m}$  
  as found in the interstellar medium (Mathis et al.~\cite{mrn}). In both the 
  disk interior and the disk surface layer a minimum grain size of 
  $a_\mathrm{min}=0.005\,\mathrm{\mu m}$ and a 
  grain size distribution of $a^{-3.5}$ for $a \in [a_\mathrm{min},
  a_\mathrm{max}]$ was adopted
  (MRN grain-size distribution; Mathis et al.~\cite{mrn}). 

  After the temperature distribution was determined assuming a
  density given by Eq.~\ref{eq:shakura}, the SED, the projected
  image of the star, and its circumstellar environment, assuming an
  inclination angle $\vartheta$, were calculated. In the modeled images from
  which the visibilities result, we generally used a spatial resolution of
  $<$$0.2\,\mathrm{AU}$ in the MIR range, which is a factor $\sim$$10$ higher
  than the spatial resolution reached in our MIDI observations. For the images
  from which we derived the NIR visibilities, we used a resolution of 
  $<$$0.03\,\mathrm{AU}$, i.\/e., a factor of $10$ higher than the
  resolution reached by any existing NIR interferometer.

  Independent model parameters of our approach are the outer radius
  $R_\mathrm{out}$, the scale-height $h_\mathrm{100}$, the exponent $\beta$,
  and the inclination angle $\vartheta$ of the circumstellar disk.\footnote{An
    inclination angle of {$0$\degr} corresponds to a face-on disk.} Assuming a
  certain dust mixture in the models, the disk mass is fixed predominantly by
  the flux in the millimeter wavelength range.
  Considering a stellar temperature $T_\star$, the quadratic
  distance law, a mean dust sublimation temperature of $1500\,\mathrm{K}$, and
  the specific absorption coefficients $\kappa$ of the adopted dust
  set, the temperature of single dust grains at each radius can be
  determined. However, our approach is only valid for optically thin
  media. For optically thick media, 
  the grain temperature at the inner-disk edge is even higher because
  back radiation from outer dust grains produces an additional heating of
  the grains at the inner edge.
  The resulting sublimation radius $R_\mathrm{sub}$ is the initial value of
  $R_\mathrm{in}$ in our modeling approach. 
  
  Initially, the simulation grid of independent parameters is scanned coarsely
  during modeling. Simultaneously, we consider parameter values that were
  derived in preceding studies. By means of manual modifications of the model
  parameters, we search for the model that reproduces most successfully the
  measured SED and
  MIR visibilities, simultaneously. The SED and the MIR visibilities that we
  try to model are complex functions of many different disk and dust
  parameters.  However, the effects of any modification of the parameters on
  modeled SED and visibilities can be estimated. A manual
  modification is necessary because a disk model with a specific SED and
  images at different wavelengths\footnote{used for the calculation of the
    modeled visibility} takes about an hour to compute. 
  In our search for the best-fit model the uniqueness of our final models
  cannot therefore be
  proven. However, we verified whether the modeling results can
  be improved by varying the model parameters using the following step
  widths: \medskip\\ 
  \begin{tabular}{llr}
    $\Delta L_\mathrm{\star} = 1\,\mathrm{L_\mathrm{\odot}}$, & $\Delta M_\mathrm{disk} =
    0.5\,\mathrm{M_\mathrm{disk}}$, & \\
    $\Delta R_\mathrm{out} = 10\,\mathrm{AU}$, & $\Delta R_\mathrm{in} =
    0.05\,\mathrm{AU}$, & \\
    $\Delta \beta = 0.1$, and  & $\Delta h_\mathrm{100} =
    1\,\mathrm{AU}$. & \\
  \end{tabular}\medskip\\
  Considering the error bars of the measurements, the modeling results could
  not be improved using finer step widths. The step widths can be considered as
  the precision to which the local 
  minimum in the $\chi^2$-surface can be determined. The determination of the
  modeling errors would require a fit of the $\chi^2$-surface in the
  simulation grid to the independent parameters around the determined local
  minimum. 

  \begin{table*}
    \centering
    \begin{minipage}{0.82\textwidth}  
    \caption{Parameters of the models that simulate the measured SED
      and MIR visibilities best. Footnotes of the mass
      accretion-rate point to 
      measurements in previous studies. Values that are marked with (det.) are
      determined in this study.}
    \begin{tabular}{llrrrrrrr}\hline\hline
      object (model-no.) & $M_\mathrm{disk}$ [M$_\mathrm{\odot}$] &
      $R_\mathrm{out}$ [AU] & $R_\mathrm{in}$ [AU] & $\beta$ &
      $h_\mathrm{100}$ [AU] & $\vartheta$ [$^\circ$] & $\dot{M}$
      [M$_\mathrm{\odot}$/a] \\ \hline  
      DR\,Tau (I) & & & $0.10$ & & & & $2.0 \cdot 10^{-8}_\mathrm{(det.)}$\\[1.0ex] 
      DR\,Tau (II) & $0.1$ & $90$ & $0.10$ & $0.75$ & $15$ & $20$ & $8.0 \cdot
      10^{-8}_\mathrm{(1)}$ \\[1.0ex]
      DR\,Tau (III) & & & $0.05$ & & & & $8.0 \cdot 10^{-8}_\mathrm{(det.)}$\\[1.0ex]
      GW\,Ori (I,II) & $1.0$ & $360$ & $0.35$ & $1.10$ & $22$ & $10$ & $2.5
      \cdot 10^{-7}_\mathrm{(2)}$\\[1.0ex] 
      HD\,72106\,B & $0.005$ & $40$ & $0.50$ & $1.30$ & $8.0$ & $60$ &
      {N \hfill o \hfill A \hfill c \hfill c \hfill r \hfill e \hfill t \hfill
        i \hfill o \hfill n} & \\[1.0ex] 
      RU\,Lup & $0.1$ & $100$ & $0.10$ & $0.90$ & $20$ & $$$28$ & $1.0 \cdot
      10^{-8}_\mathrm{(3)}$ \\[1.0ex] 
      HBC\,639 & $0.1$ & $120$ & $0.10$ & $1.00$ & $10$ & $65$ & {N \hfill o
        \hfill A \hfill c \hfill c \hfill r \hfill e \hfill t \hfill i \hfill
        o \hfill n} & \\[1.0ex] 
      S\,CrA\,N (I) & \raisebox{-1.5ex}{$0.03$} & \raisebox{-1.5ex}{$120$} &
      \raisebox{-1.5ex}{$0.05$} & $1.10$ & $9$ & \raisebox{-1.5ex}{$10$}&
      \raisebox{-1.5ex}{$4.0 \cdot 10^{-8}_\mathrm{(det.)}$} & \\[1.0ex] 
      S\,CrA\,N (II) & & & & $0.90$ & $12$ & $45$ & & \\[1.0ex]
      S\,CrA\,S & $0.03$ & $100$ & $0.10$ & $1.05$ & $8$ & $30$ & $4.0
      \cdot 10^{-8}_\mathrm{(det.)}$ & \\[1.0ex]
      \hline
    \end{tabular}
        {\newline \scriptsize {References: } 1: Akeson et
          al.~(\cite{akeson}); 2: Calvet  
          et al.~(\cite{calvetII}); 3: Herczeg et al.~(\cite{herczeg})}
        \label{table:properties-midisurveyII}
      \end{minipage}
  \end{table*}
  \section{Results}\label{section:models}  
  All scientific targets could be well resolved, i.\/e., the
  visibilities are lower than unity.
  Table~\ref{table:properties-midisurveyII} presents the parameters 
  resulting from our effort to model the circumstellar disks of seven
  YSOs. In the following, we present and discuss our modeling results  
  for each object in our sample, in detail. Previous observations, including photometric
  measurements, are mentioned in   
  Appendix~\ref{appendix}.
  
  \subsection{DR\,Tau}\label{section:drtau}
\begin{figure*}[!tb] 
    \center
    \resizebox{0.48\textwidth}{!}{\includegraphics{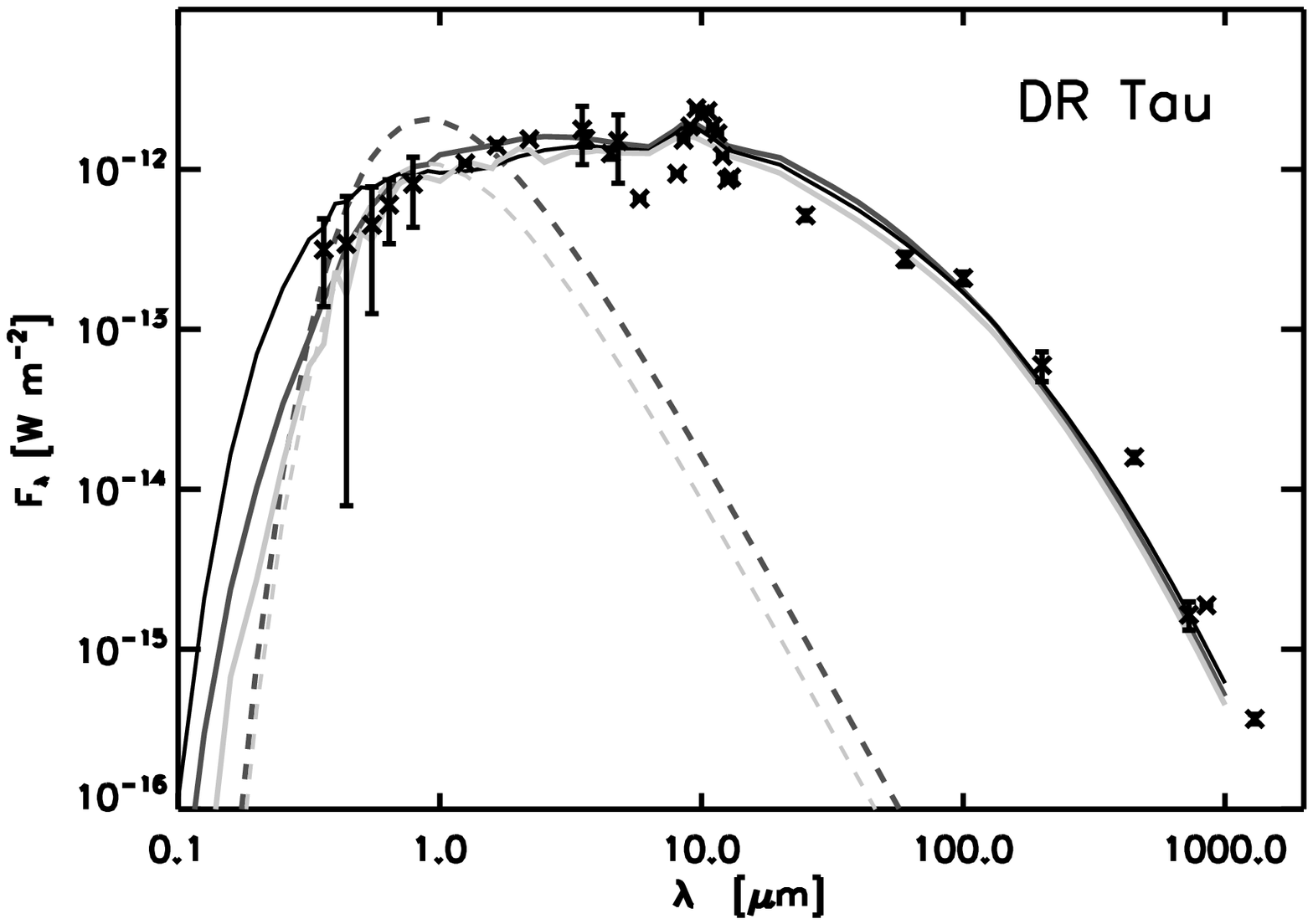}} 
    \resizebox{0.48\textwidth}{!}{\includegraphics{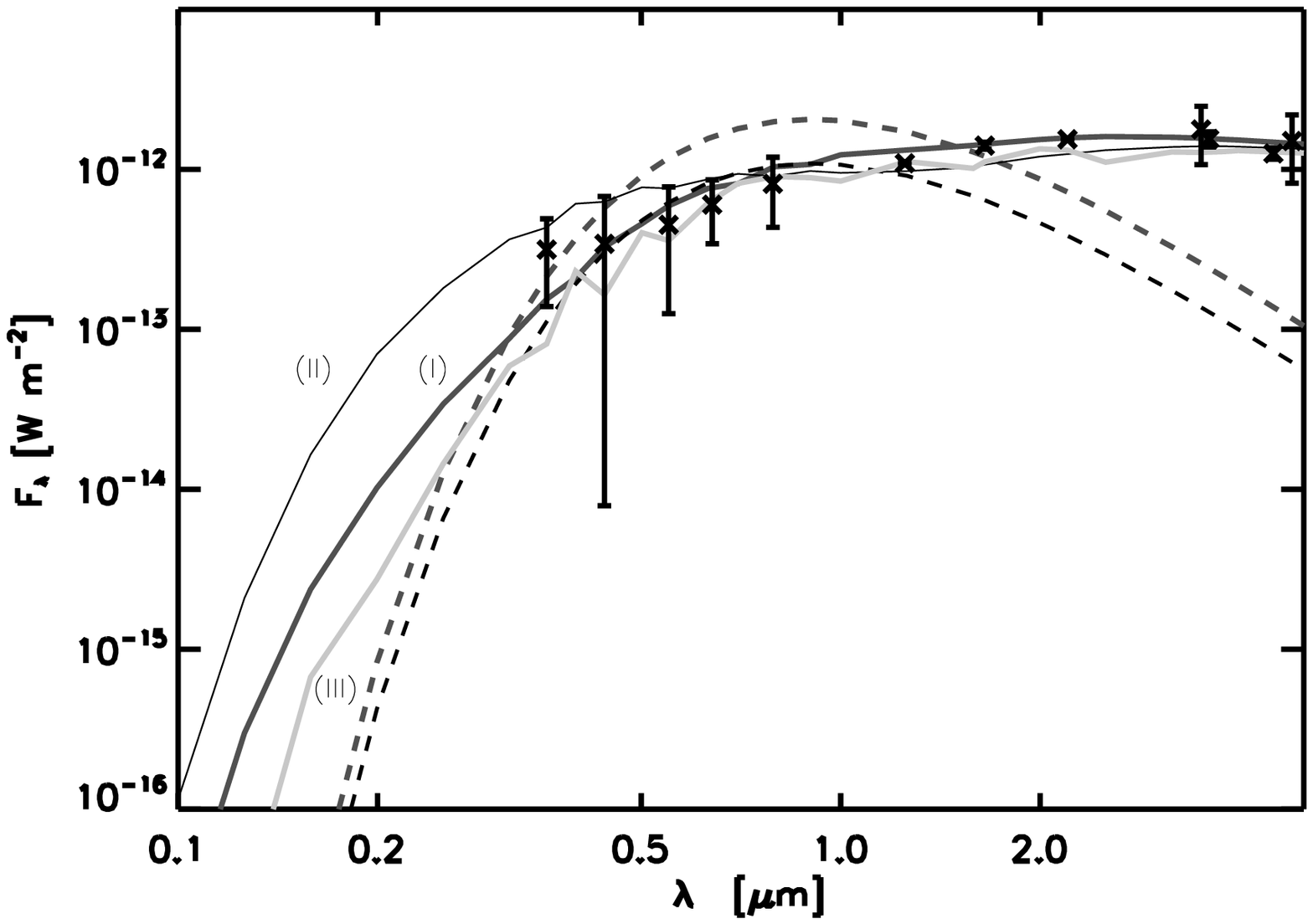}}\newline
    \resizebox{0.48\textwidth}{!}{\includegraphics{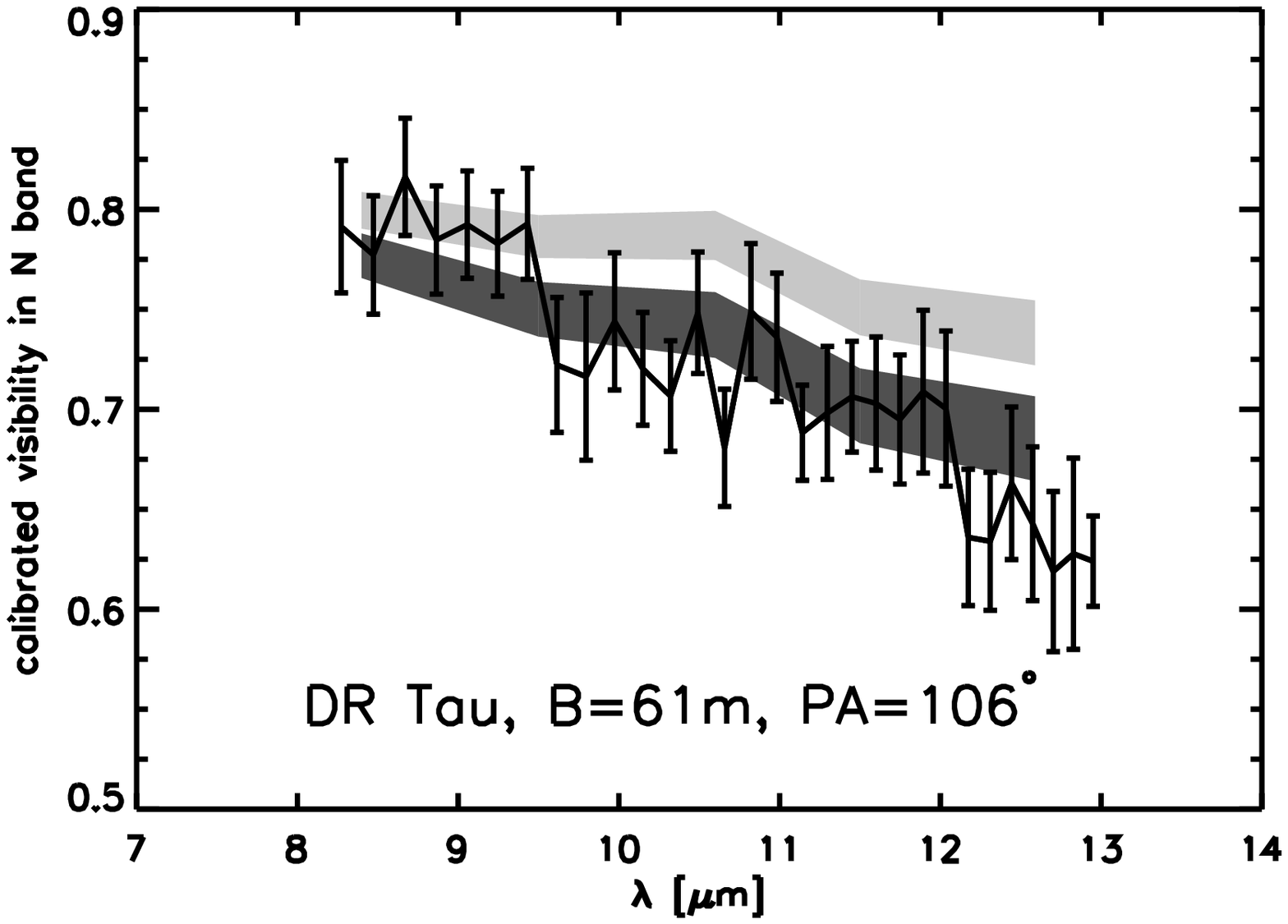}}  
    \resizebox{0.48\textwidth}{!}{\includegraphics{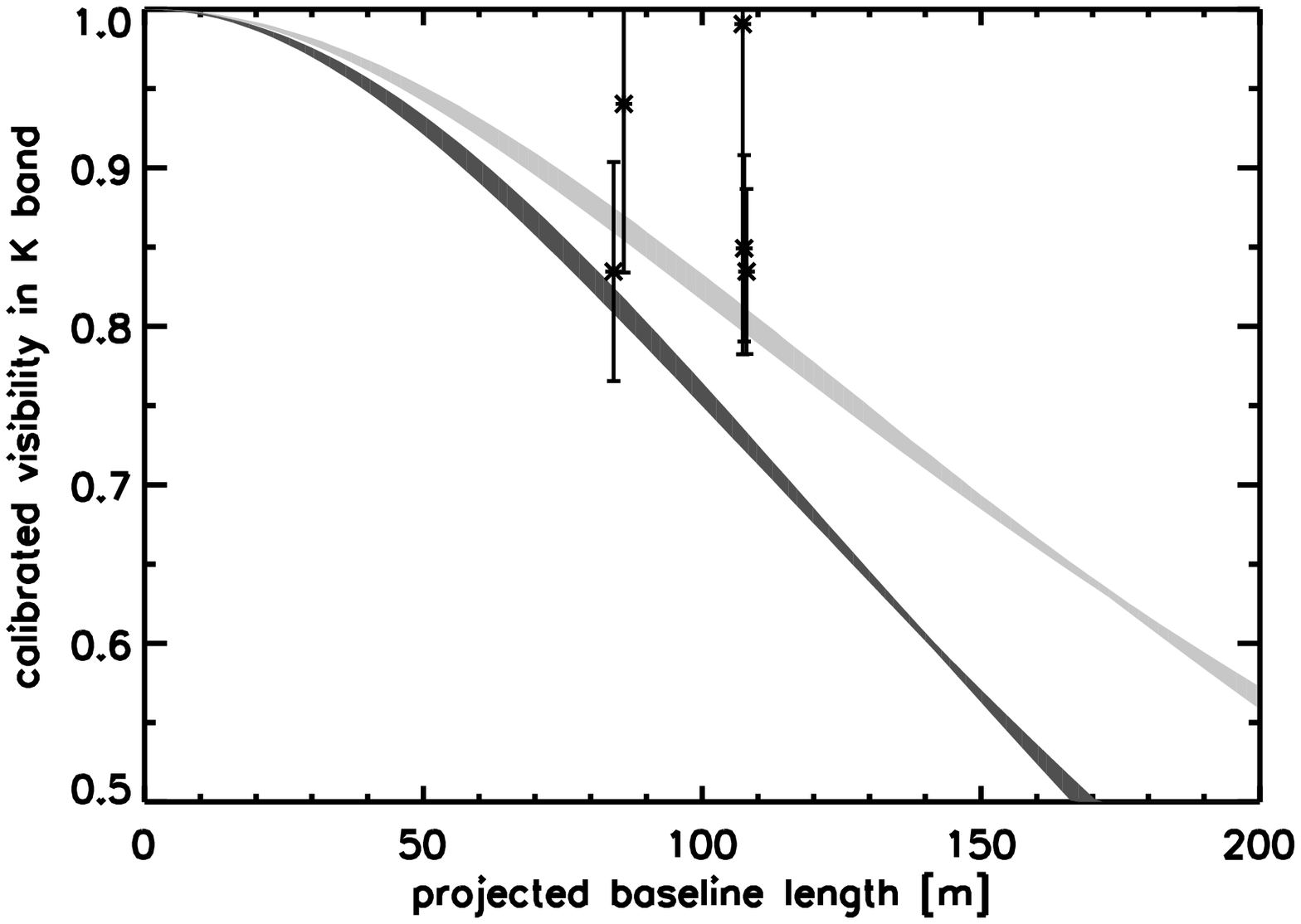}}
    \caption{SED and MIR visibilities for a projected baseline of 
      $B=60\,\mathrm{m}$ obtained from the measurements and our models (I), 
      (II), and (III) of DR\,Tau. The model parameters are listed in
      Table~\ref{table:properties-midisurvey} and 
      Table~\ref{table:properties-midisurveyII}. 
      {\it Upper row:} SED and an enlargement showing photometric data from
      the visible up to the NIR wavelength range. The curves are ascribed to
      the specific models by numbers. Real photometric data are plotted with
      error bars. The dashed lines represent the intrinsic stellar flux of
      models (I), (II), and (III). 
      {\it Lower-left figure:} The modeled visibilities were derived from the 
      corresponding model images for the wavelengths of $8.5\,\mathrm{\mu m}$,
      $9.5\,\mathrm{\mu m}$, $10.6\,\mathrm{\mu m}$, $11.5\,\mathrm{\mu m}$,
      and $12.5\,\mathrm{\mu m}$. The colored bars represent intervals that
      limit 
      the MIR visibilities $V(\lambda)$ for different position angles but the
      same inclination of the models (I: dark grey) and (III: light gray). For
      clarity, the visibilities obtained from model~(II) are not shown but
      they are located between the results of model~(I) and model~(III). The
      measured data are included with error bars (black, solid line). 
      {\it Lower right figure:} NIR visibilities as a function of the
      projected (effective) baseline length $B$ of an interferometer derived
      for a wavelength of $2.2\,\mathrm{\mu m}$ that result from model (III:
      light gray). Model (III) is the only model that fit all available data
      sets. The dark grey line represents the output of model (I) and (II),
      respectively. The 
      measured NIR visibilities obtained with the PTI
      (Akeson et al.~\cite{akeson}) are included with error bars.
      \newline } 
    \label{figure:drtau}
  \end{figure*}

  Considering the low gradient of the SED at wavelengths $\lambda \lessapprox 1\,\mathrm{\mu m}$
  (Fig.~\ref{figure:drtau}), it is evident that the (measured) SED of 
  DR\,Tau is represented by a black body with the stellar temperature of 
  $T_\mathrm{\star}=4050\,\mathrm{K}$ {\it and} an additional radiation source.  
  The accretion process is the other strong influence on the radiation emitted
  in this wavelength range. In addition to  
  the stellar luminosity and the accretion rate, the parameters of the model~(I) of our approach 
  confirm the values derived in the 
  previous study of Akeson et al.~(\cite{akeson}) using long-baseline NIR
  interferometry from the Palomar Testbed
  Interferometer (PTI). However, they found a stellar 
  luminosity of $L_\mathrm{\star}=0.9\,\mathrm{L_{\odot}}$, which is a factor
  of $\sim$$2$ lower than the value found in our
  model. Simultaneously, the accretion rate of their model, i.\/e., $\dot{M} = 8 
  \times 10^{-8}\,\mathrm{M_{\odot}yr^{-1}}$, is four times higher than the
  value determined in our first model. However, SED and MIR  
  vi\-sibilities can also be reproduced by a model~(II) by considering the stellar
  luminosity and the accretion rate found by Akeson et
  al.~(\cite{akeson}) while keeping all other model parameters constant. Both
  models similarly fit MIR visibilities and SED to
  within the error bars. We conclude that the intrinsic stellar luminosity and the
  accretion luminosity cannot 
  be disentangled considering SED and MIR visibilities, only. The derived
  accretion luminosity $L_\mathrm{acc}$ depends on the applied accretion rate
  $\dot{M}$: the model (I) has an 
  accretion luminosity of $L_\mathrm{acc}=0.19\,L_\mathrm{\odot}$ for $\dot{M} = 2 
  \times 10^{-8}\,\mathrm{M_{\odot}yr^{-1}}$ while an
  accretion luminosity of $L_\mathrm{acc} = 1.0\,L_\mathrm{\odot}$ for $\dot{M} = 8 
  \times 10^{-8}\,\mathrm{M_{\odot}yr^{-1}}$ is emitted
  by the model (II). The sum of the intrinsic stellar luminosity
  $L_\mathrm{\star}$ and the accretion luminosity $L_\mathrm{acc}$ in both
  models equals $\sim$$1.9\,\mathrm{L_\mathrm{sun}}$. Finally, we mention
  that Mohanty et al.~(\cite{mohanty}) derived an accretion rate of
  $\dot{M}=2.2 \times 10^{-7}\,\mathrm{M_{\odot}yr^{-1}}$ analysing Ca\,II
  lines. Robitaille et al.~(\cite{robitaille}) determined a lower and upper  
  limit to $\dot{M}$, i.\/e., $6.9 \times 10^{-9}\,\mathrm{M_{\odot}yr^{-1}}
  \leqq \dot{M} \leqq 1.0 \times 10^{-6}\,\mathrm{M_{\odot}yr^{-1}}$ assuming
  the SED, only. 
  If the NIR visibility data 
  obtained by Akeson et al.~(\cite{akeson}) is simultaneously taken into 
  account, we 
  find that the disks of models~(I) and (II) appear too spatially resolved,
  resulting in too low NIR visibilities. Only an additional reduction in 
  the inner disk radius to $R_\mathrm{in}=0.05\,\mathrm{AU}$
  improves our fit to the NIR visibility data in the model (II). A similar 
  improvement cannot be reached in model (I) by reducing its inner disk
  radius. Our model of DR\,Tau with 
  $L_\mathrm{\star}=0.9\,\mathrm{L_{\odot}}$, $\dot{M} = 8 
  \times 10^{-8}\,\mathrm{M_{\odot}yr^{-1}}$, and 
  $R_\mathrm{in}=0.05\,\mathrm{AU}$ is called
  model (III). The lower right panel in Fig.~\ref{figure:drtau} shows 
  measurements and model output for the 
  NIR visibility of model (III). We note that the fit to the MIR visibility data 
  at wavelengths of $12.5\,\mathrm{\mu m}$
  becomes poorer in model (III) and deviates from the measured data by $<7$\%  
  as the gradient of the modeled MIR visibility curve decreases. We summarize that 
  model (III) is the only model that reproduces all available data sets apart from 
  slight deviations in the MIR visibilities at $\sim$$12.5\,\mathrm{\mu
    m}$.  

  As we mentioned in Sect.~\ref{section:modeling}, the sublimation
  radius is approximated considering a stellar temperature of
  $T_\mathrm{\star} = 4000\,\mathrm{K}$ and the absorption
  coefficients of the dust composition that we used in our two-layer disk model. This
  approach is only an approximation because we assume an optically thin disk in
  determining of the sublimation radius $R_\mathrm{in}$. The true
  sublimation radius $R_\mathrm{in}$ should be larger because the back radiation of
  adjacent grains is neglected. We obtain a sublimation radius of
  $R_\mathrm{sub} \approx 0.06\,\mathrm{AU}$ even for the dust for which
  $a_\mathrm{max}=1\,\mathrm{mm}$ because of a low number of large
  grains with $a>0.25\,\mathrm{\mu m}$. Since the inner radius
  $R_\mathrm{in}$ of the modeled disk is slightly smaller than the sublimation
  radius, this value of $R_\mathrm{sub}$ could imply that 
  larger grains exist 
  in the disk, in the innermost disk in particular, than the MRN grain-size
  distribution predicts (Mathis et al.~\cite{mrn}). This result is also found
  in Sect.~\ref{section:dust composition} and predicted by Isella \&
  Natta~(\cite{isella}). However, the hint for larger grains is only weak since the
  large error bars of the NIR data still allow a slightly larger inner-disk
  radius $R_\mathrm{in}$ than we have assumed in model (III).

  As the disk crosses the line of sight of the observer with increasing
  inclination angle, the NIR flux at $\sim$$2.2\,\mathrm{\mu
    m}$ of the model of DR\,Tau decreases by $38\%$ in steps of $\Delta
  \vartheta = 5\degr$ for small inclinations.  
  Therefore, the inclination angle can be determined with high accuracy. If the measured visual 
  extinction  
  $A_\mathrm{V} \approx 1.6\,\mathrm{mag}$ (Muzerolle et al.~\cite{muzerolle}) can
  be ascribed exclusively to the  
  circumstellar disk around DR\,Tau that crosses the line of sight of the
  observer, an inclination of $\vartheta \approx 20\degr$ can  
  be derived. If the extinction is also caused by interstellar media outside
  the system, which is not considered in the model, the derived inclination
  angle is correspondingly smaller. 

  Our observation of DR\,Tau with MIDI achieved a theoretical spatial 
  resolution of $\la 17\,\mathrm{mas}$ ($\sim$$2\,\mathrm{AU}$). To this 
  limit, the visibility data do not provide any evidence
  of a companion with a brightness ratio close to unity, which would
  have produced a characteristic sinusoidal signature in the visibility (e.\/g.,
  Schegerer et al.~\cite{schegerer}). We note that the detection of a close
  binary is difficult when the binary is oriented nearly orthogonal to the
    baseline vector of our interferometric observations. 
  
  Our finding confirms the results of former speckle-interferometric
  measurements in the NIR range (Ghez et al.~\cite{ghez}), where a theoretical 
  spatial resolution of $53\,\mathrm{mas}$ could be achieved. Two
  previously unpublished lunar occultation data sets of high quality ($SNR$ 
  $\approx 20-40$) are available for DR\,Tau from the $3.5\,\mathrm{m}$ 
  telescope in Calar Alto, the first one recorded in November 1997 with an
  InSb fast photometer and the second one in September 1999 with the IR array 
  OmegaCass in fast readout for a $32 \time 32$ pixel window. A detailed  
  analysis allows us to conclude that the source appeared point-like with an
  upper limit of $(2.70\pm0.65)\,\mathrm{mas}$, 
  i.\/e., $(0.4\pm0.1)\,\mathrm{AU}$ and $(5.0\pm0.4)\,\mathrm{mas}$, 
  i.\/e., $(0.7\pm0.06)\,\mathrm{AU}$ in 
  both data sets. Upper limits can also be set to the flux because of a 
  possible circumstellar emission, and they are $\lessapprox 5$\% over
  $160\,\mathrm{mas}$ and $\lessapprox 2.5$\% over $420\,\mathrm{mas}$, for
  both 
  observations. 

  \subsection{GW\,Ori}\label{section:gwori} 
  \begin{figure*}[!tb] 
    \center
    \resizebox{0.48\textwidth}{!}{\includegraphics{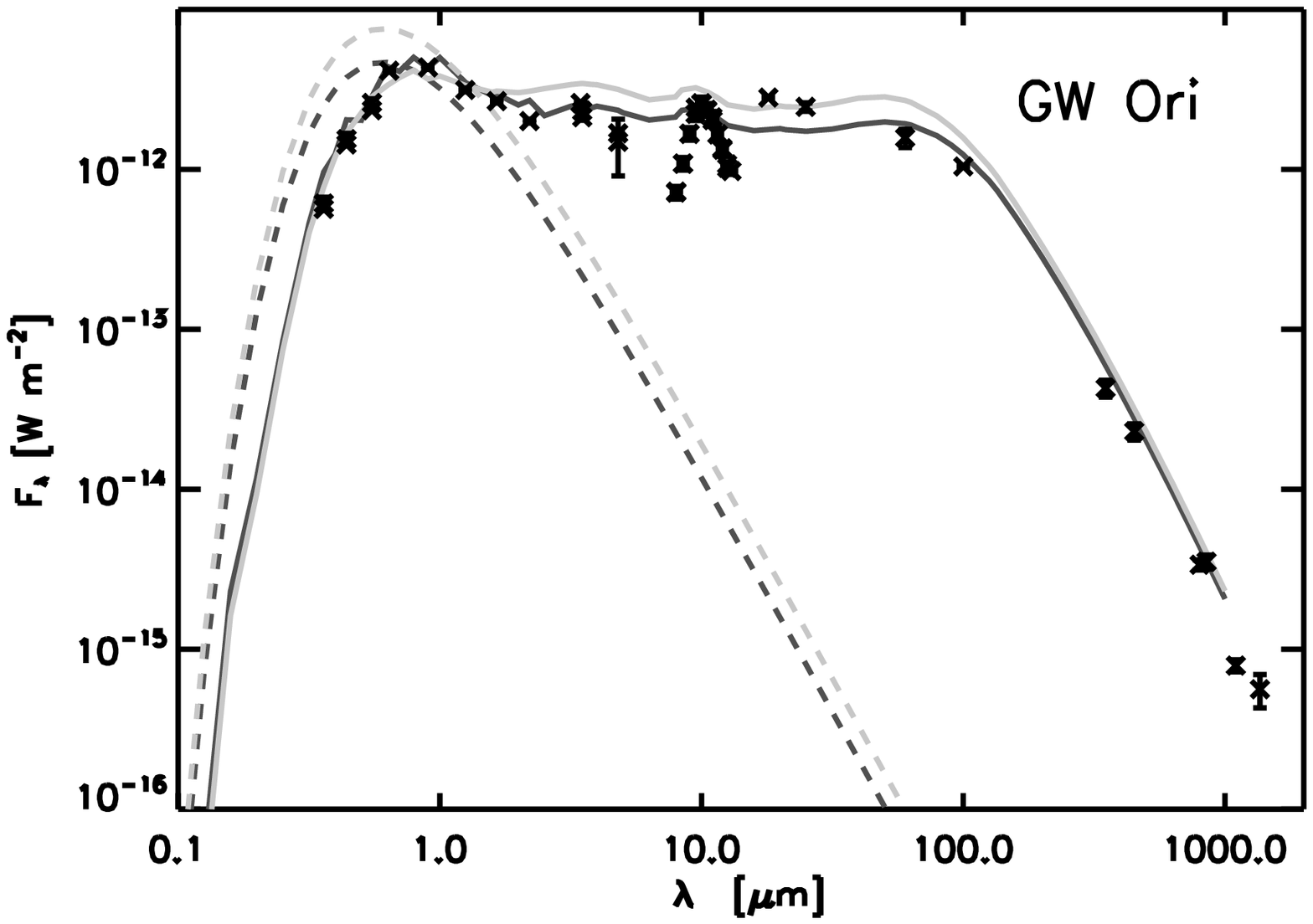}}\newline 
    \resizebox{0.48\textwidth}{!}{\includegraphics{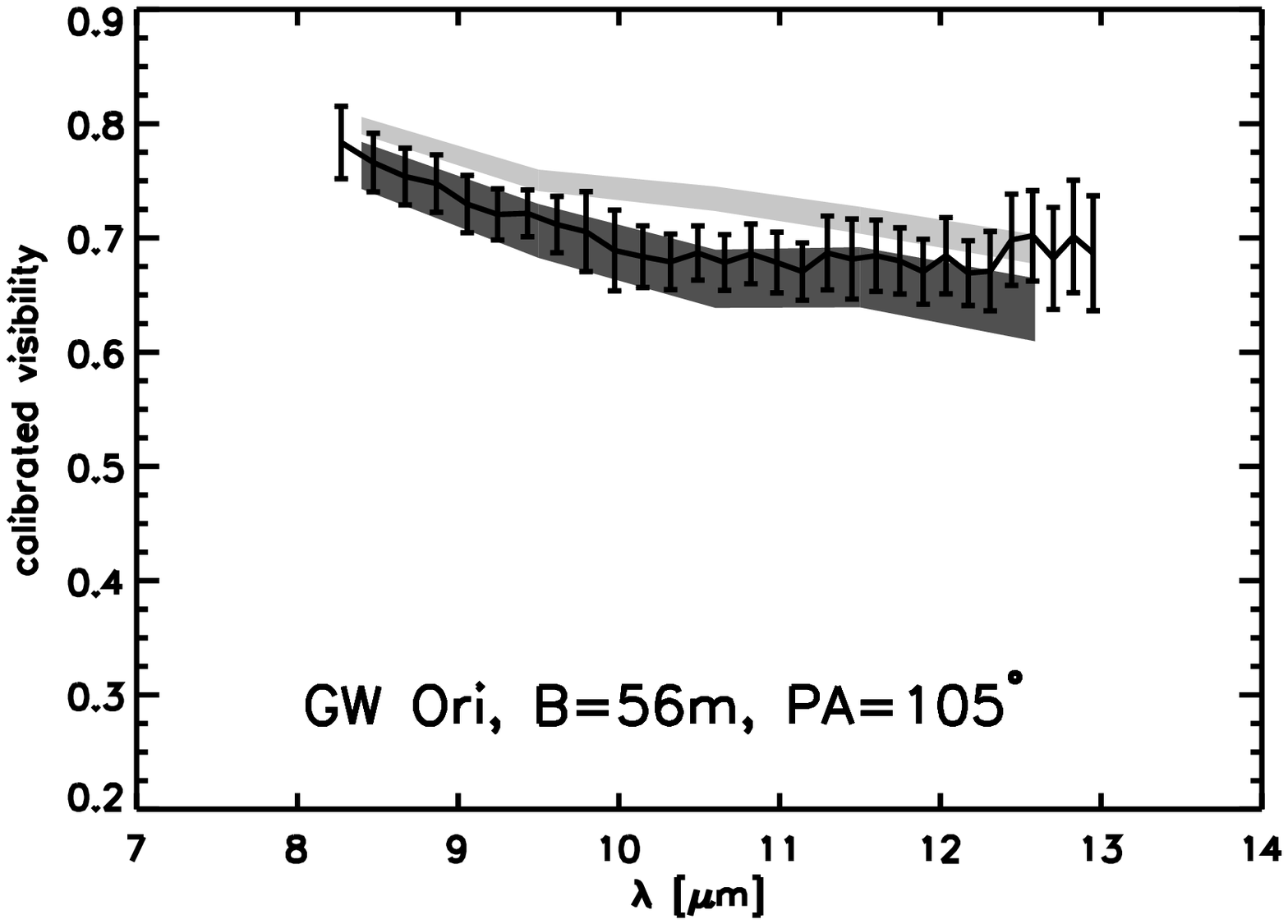}}
    \resizebox{0.48\textwidth}{!}{\includegraphics{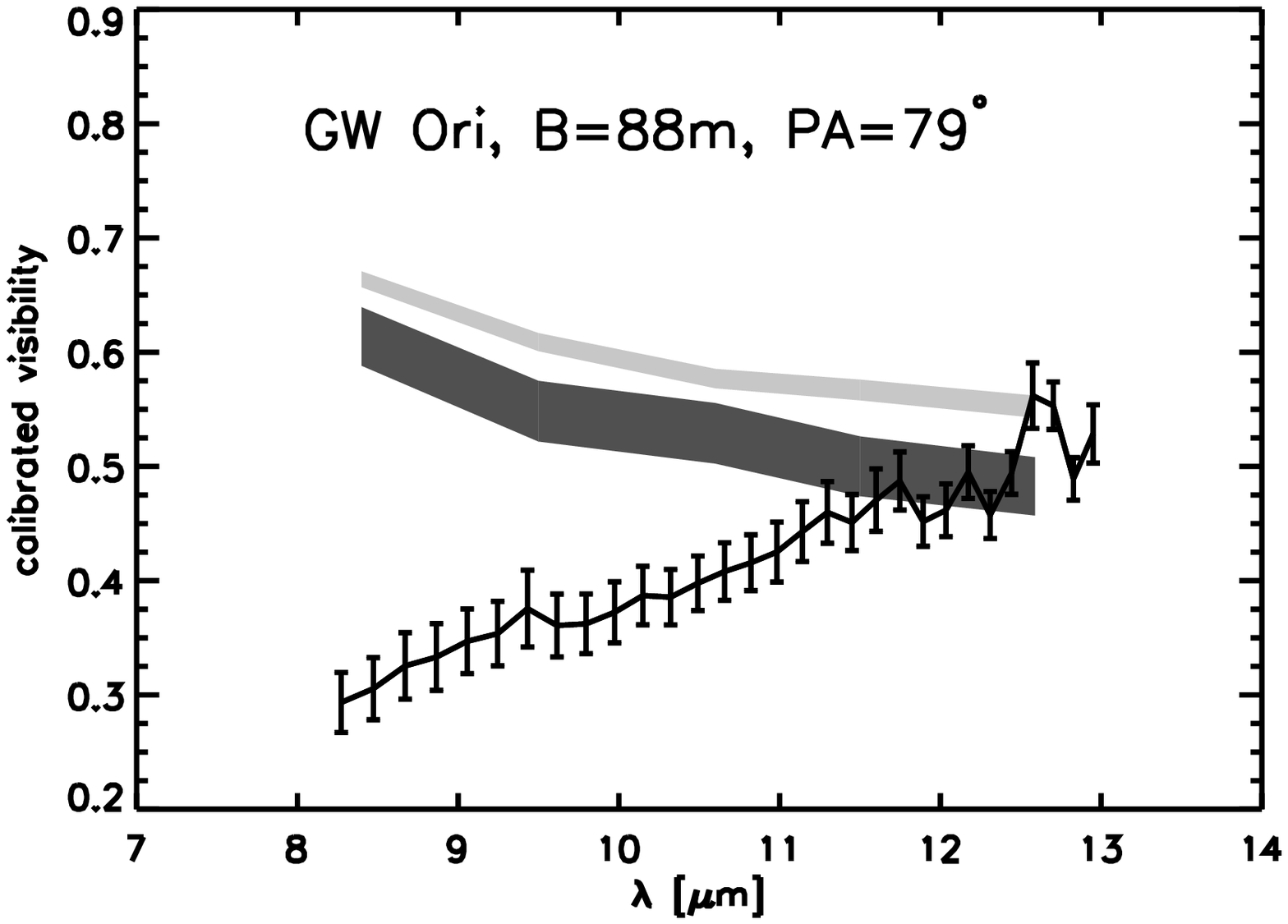}} 
    \caption{SED and MIR visibilities for the projected baselines of 
      $B=56\,\mathrm{m}$ and $B=88\,\mathrm{m}$ obtained from the measurements
      and our model of GW\,Ori. The results of two different models are plotted   
      that differ in the stellar luminosity (dark gray:
      $L_{\star}=40\,\mathrm{L_{\odot}}$, light gray: $L_{\star} =
      62\,\mathrm{L_{\odot}}$). The increasing visibility of the longest
      baseline $B=87\,\mathrm{m}$, which cannot be reproduced by our modeling
      approach, could point to a bright, truncated density structure in
      the inner disk region ($< 5\,\mathrm{AU}$), although different scenarios
      are also possible. } 
    \label{figure:gwori}
  \end{figure*}

  As former millimeter measurements have shown (Mathieu et
  al.~\cite{mathieuII}) and can be confirmed by this  
  study, the circumstellar disk around GW\,Ori is very massive,
  i.\/e., $M_\mathrm{disk}=1.0\,\mathrm{M_{\odot}}$.  
  Furthermore, compared to other objects in this study, the disk of
  GW\,Ori has the largest outer  
  radius $R_\mathrm{out}$. Besides the luminosity, all other stellar parameters
  and the accretion rate found in this study are consistent with the result of 
  Calvet et al.~(\cite{calvet}), who analyzed the UV spectrum of this
  source. The accretion luminosity is $L_\mathrm{acc} =  
  5\,\mathrm{L_\mathrm{\odot}}$ in our model. If an intrinsic stellar
  luminosity of $L_\mathrm{\star} = 62\,\mathrm{L_\mathrm{\odot}}$ is  
  used, as was determined by Calvet et al.~(\cite{calvet}), the resulting
  flux in K band would exceed the  
  photometric measurement by $20\,\%$. In Fig.~\ref{figure:gwori},
  the SED and MIR visibilities of a model  
  with a stellar luminosity of $L_\mathrm{\star} =
  62\,\mathrm{L_\mathrm{\odot}}$ is also shown (model~(II): gray  
  curves/visibility bars). Although both the SED and the visibilities for
  $B=56\,\mathrm{m}$  
  can be reproduced, 
  no model could be found that reproduces simultaneously the visibilities of a
  baseline of $B=88\,\mathrm{m}$. For  
  spectroscopic measurements, Mathieu et al.~(\cite{mathieu}) found that
  GW\,Ori has a stellar companion at a  
  projected separation $a_\mathrm{sep}$ between $a_\mathrm{sep}=1.08\,\mathrm{AU}$ and 
  $a_\mathrm{sep}=1.18\,\mathrm{AU}$ and of stellar mass between $M_\star = 
  0.5\,\mathrm{M_{\odot}}$ and $M_\star = 1\,\mathrm{M_{\odot}}$. According to
  Mathieu et al.~(\cite{mathieu}),  
  this companion creates a dust-free ring in the circumstellar 
  disk between $r=0.17\,\mathrm{a_\mathrm{sep}}$ and $r=3.0\,\mathrm{a_\mathrm{sep}}$. 
  By means of a theoretical investigation, Artymowicz \&
  Lubow~(\cite{artymowicz}) determined an inner radius of  
  $\sim$$2.3\,\mathrm{a_\mathrm{sep}}$ for the circumbinary disk of the
  system. They claimed that each component  
  has its own 
  circumstellar disk of outer radii of $0.46\,\mathrm{a_\mathrm{sep}}$
  and $0.20\,\mathrm{a_\mathrm{sep}}$. 
  
  In our approach, the inner disk radius equals the sublimation 
  radius ($R_\mathrm{in}=R_\mathrm{sub}$). We do not consider a dust-free
  inner gap as proposed by Mathieu et  
  al.~(\cite{mathieu}) and Artymowicz \& Lubow~(\cite{artymowicz}). However, in
  another model a dust-free,  
  ring-shaped gap is cut in our disk model. We therefore assumed the same disk parameters
  $\beta$ and  
  $h_\mathrm{100}$ for the circumbinary
  and circumstellar disk. The disk gap extends from the radius
  $0.5\,\mathrm{AU}$ to $3.0\,\mathrm{AU}$.  
  In the latter model, we obtained the following results (not shown in the figure):
  the NIR flux obtained from this model  
  decreases by $70\%$ and cannot fit the NIR range of the SED
  anymore. Although the modeled vi\-si\-bilities $V(\lambda)$ for
  the long baseline  
  $B=87\,\mathrm{m}$ corresponds to a 
  positive gradient, the visibilities for the short baseline $B=56\,\mathrm{m}$
  decrease by 
  $20\%$ across the complete N band. Any modifications of the radii  
  that limit the disk gap could not even improve our model considering the 
  measurements. 

  We assume that the disk gap is probably neither dust-free nor
  ring-shaped. In a theoretical study, it was shown that material  
  streams between the circumbinary and the circumstellar disks guarantee the
  accretion onto each component  
  (G\"unther \& Kley~\cite{guentherII}). Even the stellar companion that is
  not implemented in our  
  approach could affect the MIR visibilities and SED.  A strong concentration of MIR
  intensity in the innermost disk regions that could not be spatially resolved with MIDI
  and an abrupt truncation of the MIR intensity distribution potentially
  caused by a disk gap result in the increase of the MIR 
  visibility on long baselines.  

  To reveal the innermost disk region of GW\,Ori, more complicated modeling  
  approaches as well as additional interferometric observations are
  essential. Measurements with the NIR  
  interferometer AMBER at the VLTI, for instance, could provide information about the 
  small-scale structures of the innermost regions that potentially deviate from axial symmetry. 

  \subsection{HD\,72106\,B}\label{section:hd72106}
  HD\,72106\,B is the least well-studied object in our sample. In our mo\-deling
  approach, we use a luminosity 
  of $L_\mathrm{\star}=28\,\mathrm{L_{\odot}}$. This value results directly 
  from a fit by a Planck function of  
  temperature $T_\mathrm{\star}=9500\,\mathrm{K}$ (Sch\"utz et
  al.~\cite{schuetz}) to the photometric  
  measurements in U-, V- and R-band. Two properties make HD\,72106\,B a
  special object in our sample:  
  
  First, with an age of $10\,\mathrm{Myr}$, HD\,72106\,B is a relatively old T\,Tauri
  object with weak infrared 
  excess. The equivalent width of the H$\alpha$ line that is smaller than $10\,$\,{\AA}
  (Vieira et al.~\cite{vieira}) is a sign of weak accretion. Our model of
  HD\,72106\,B consists of a passive disk without considering any accretion
  effects. For several T\,Tauri
  objects of similar age (e.\/g., TW\,Hya; s. Calvet et
  al.~\cite{calvetII}) an inner, dust-free gap in the range of
  $\lessapprox 1\,\mathrm{AU}$, i.\/e., larger than the sublimation radius
  could be detected. However, because the SED and MIR
  visibilities of HD\,72106\,B can be simulated by assuming that the inner edge of
  the (dust) disk equals the sublimation radius, a shift in the
  inner-disk edge to larger radii is not obvious.  
  
\begin{figure*}[!tb]
    \center
    \resizebox{0.48\textwidth}{!}{\includegraphics{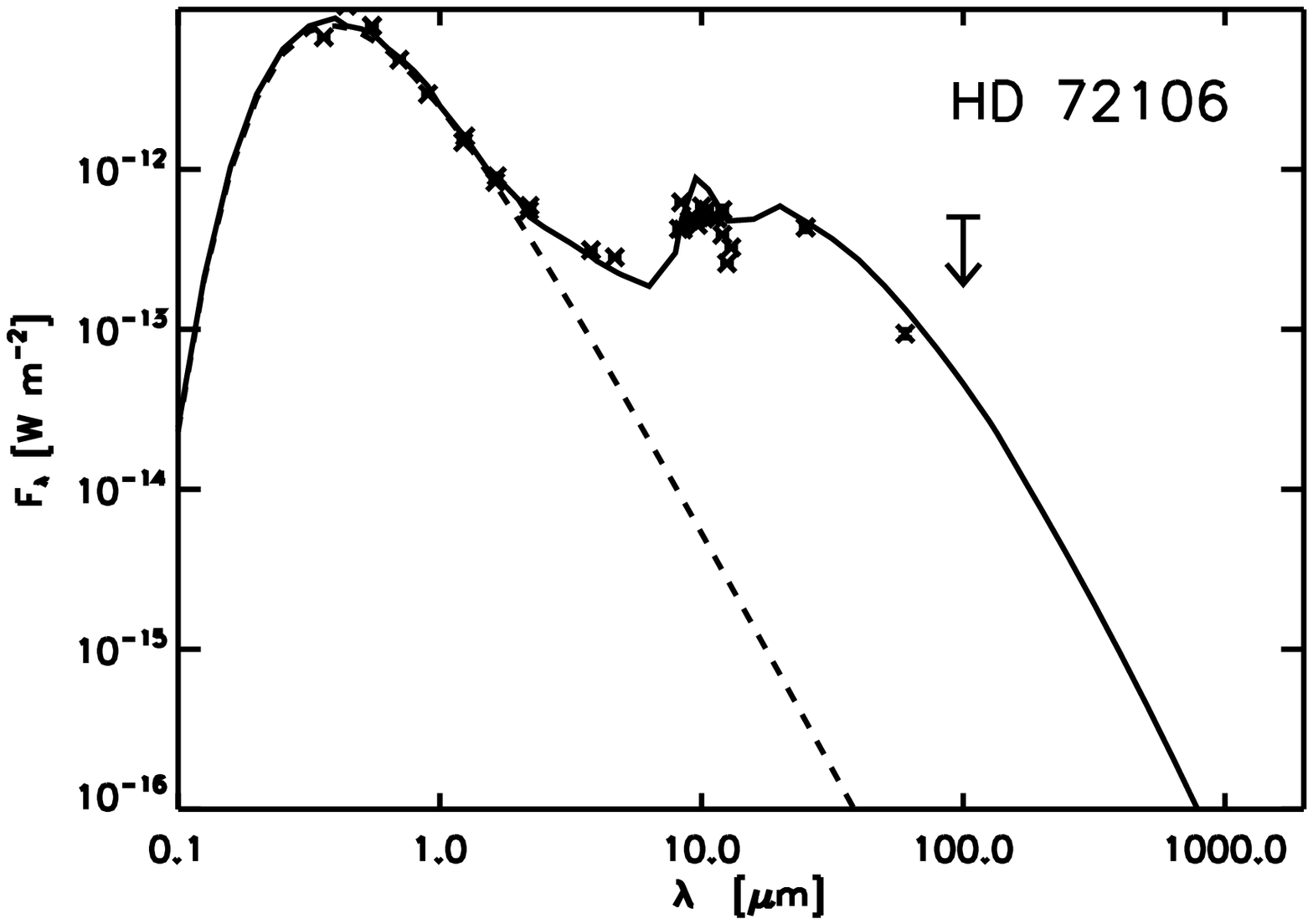}}\newline
    \resizebox{0.48\textwidth}{!}{\includegraphics{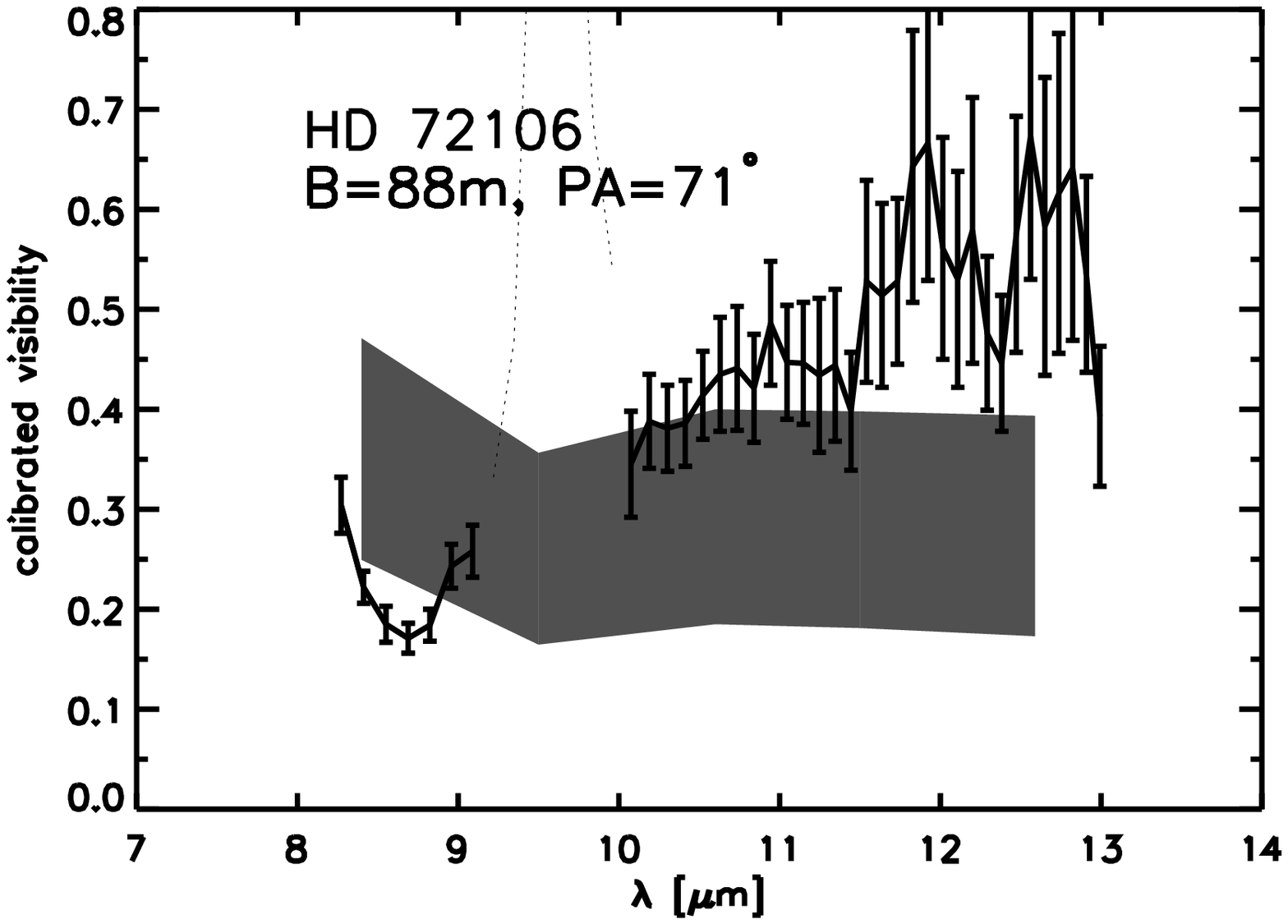}}
    \resizebox{0.48\textwidth}{!}{\includegraphics{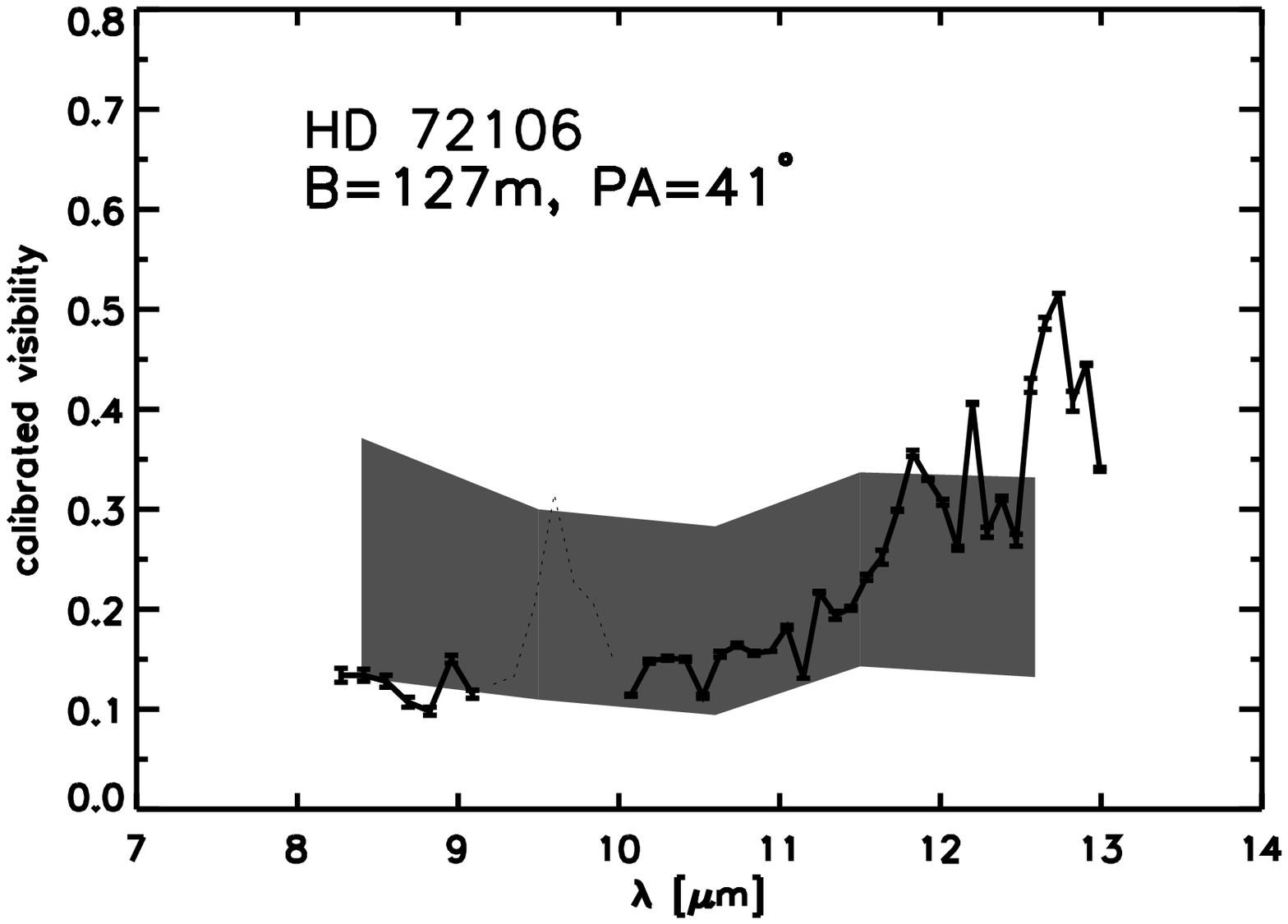}}
    \caption{SED and MIR visibilities for the projected baselines of
      $B=88\,\mathrm{m}$ and $B=127\,\mathrm{m}$  
      that results from the measurements and our model of HD\,72106\,B. Dotted
      curves represent remnants of the  
      telluric ozone band that could not be eliminated by the data reduction.}
    \label{figure:hd72106}
  \end{figure*}

  Second, HD\,72106\,B is the infrared companion of HD\,72106\,A with a projected
  distance of  
  $280\,\mathrm{AU}$. HD\,72106\,A has already developed into a main-sequence
  star (Wade et al.~\cite{wade}). The  
  main component is probably responsible for the truncation of the outer disk
  around the B component considering an outer radius of $40\,\mathrm{AU}$. The
  small outer radius of the disk finally produces a strong decrease in the
  FIR flux  
  (s. Fig~\ref{figure:hd72106}). 

  \subsection{RU\,Lup}\label{section:rulup}
  \begin{figure*}[!tb]
    \centering
    \resizebox{0.48\textwidth}{!}{\includegraphics{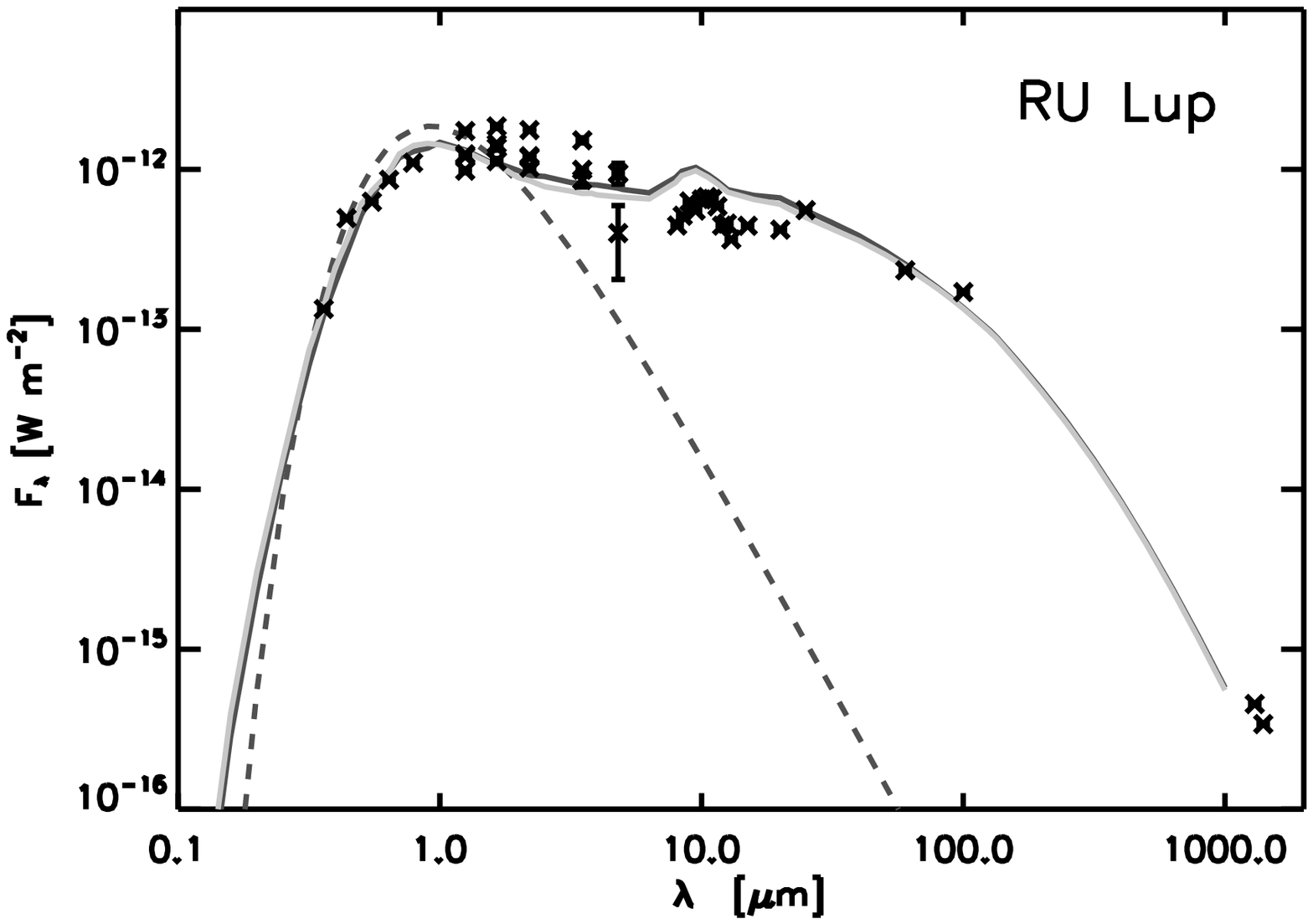}}\newline
    \resizebox{0.48\textwidth}{!}{\includegraphics{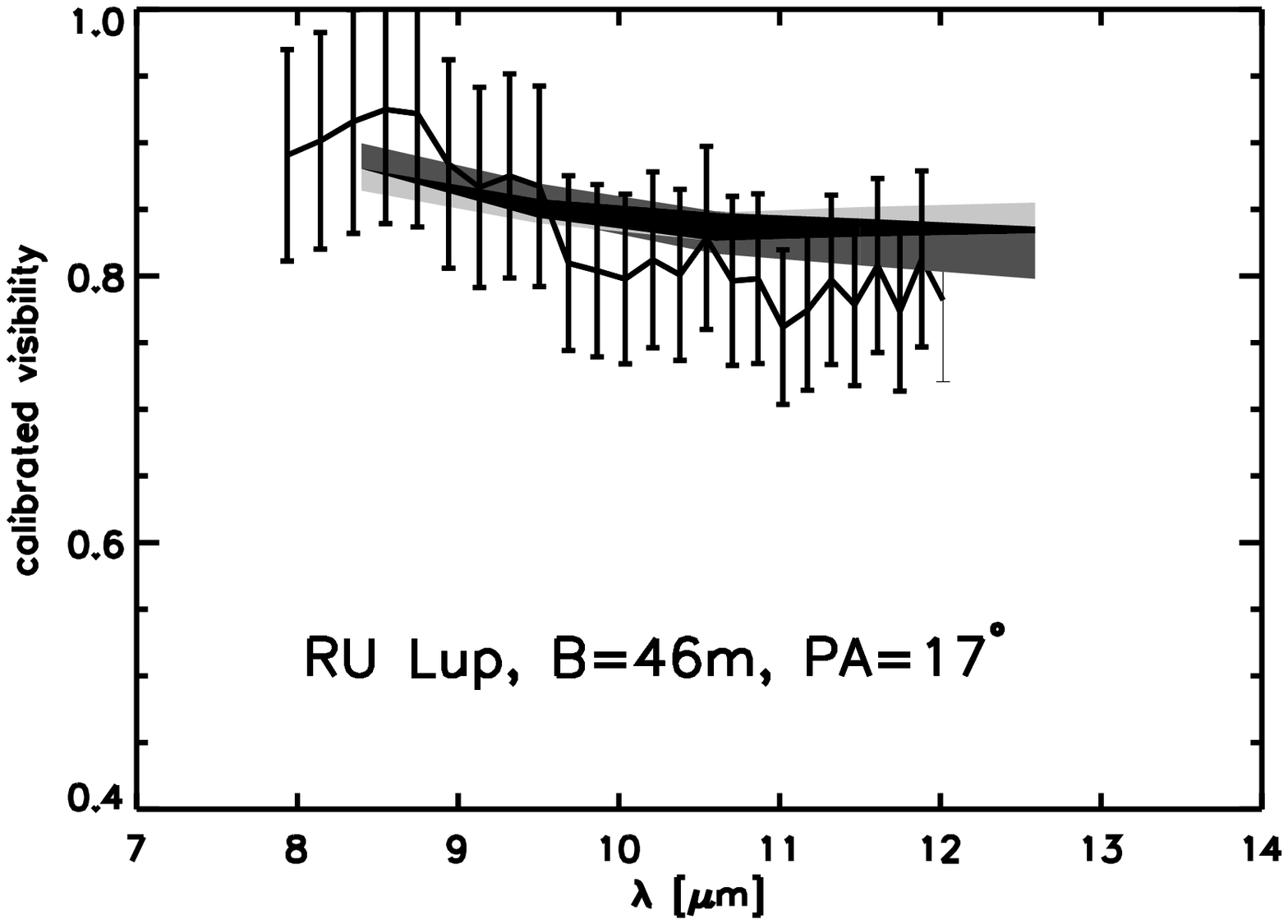}}
    \resizebox{0.48\textwidth}{!}{\includegraphics{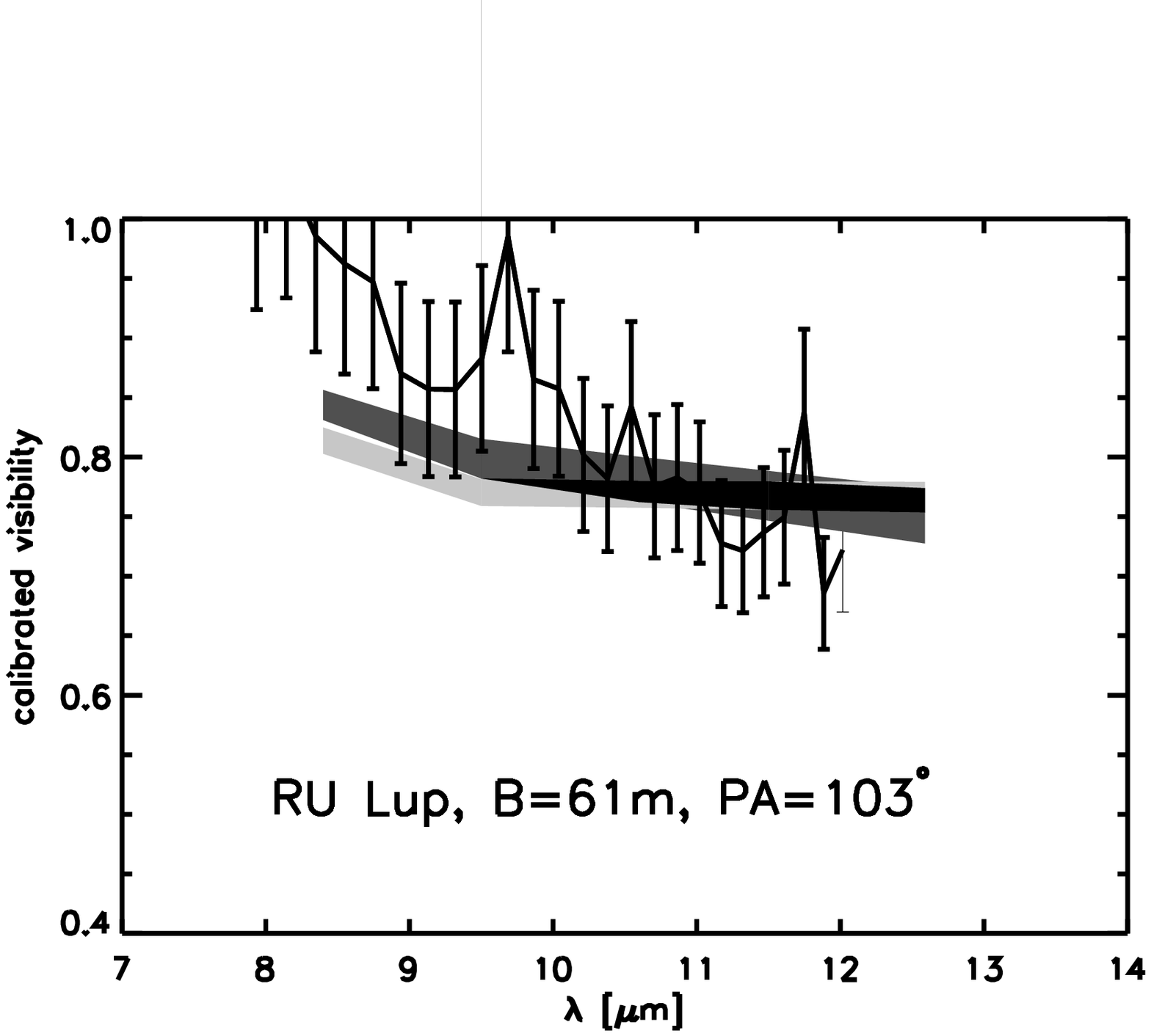}}\newline
    \resizebox{0.48\textwidth}{!}{\includegraphics{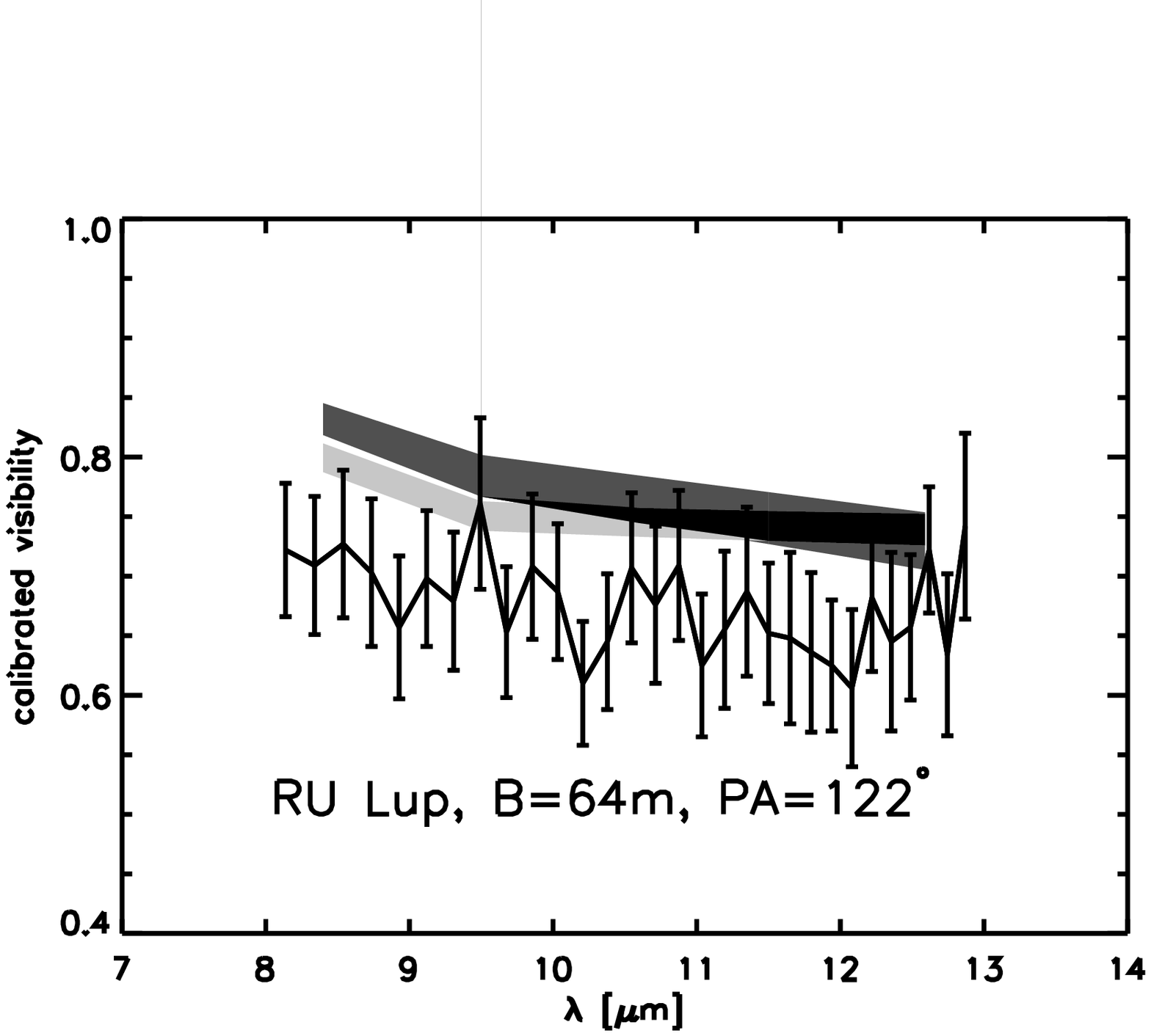}}
    \resizebox{0.48\textwidth}{!}{\includegraphics{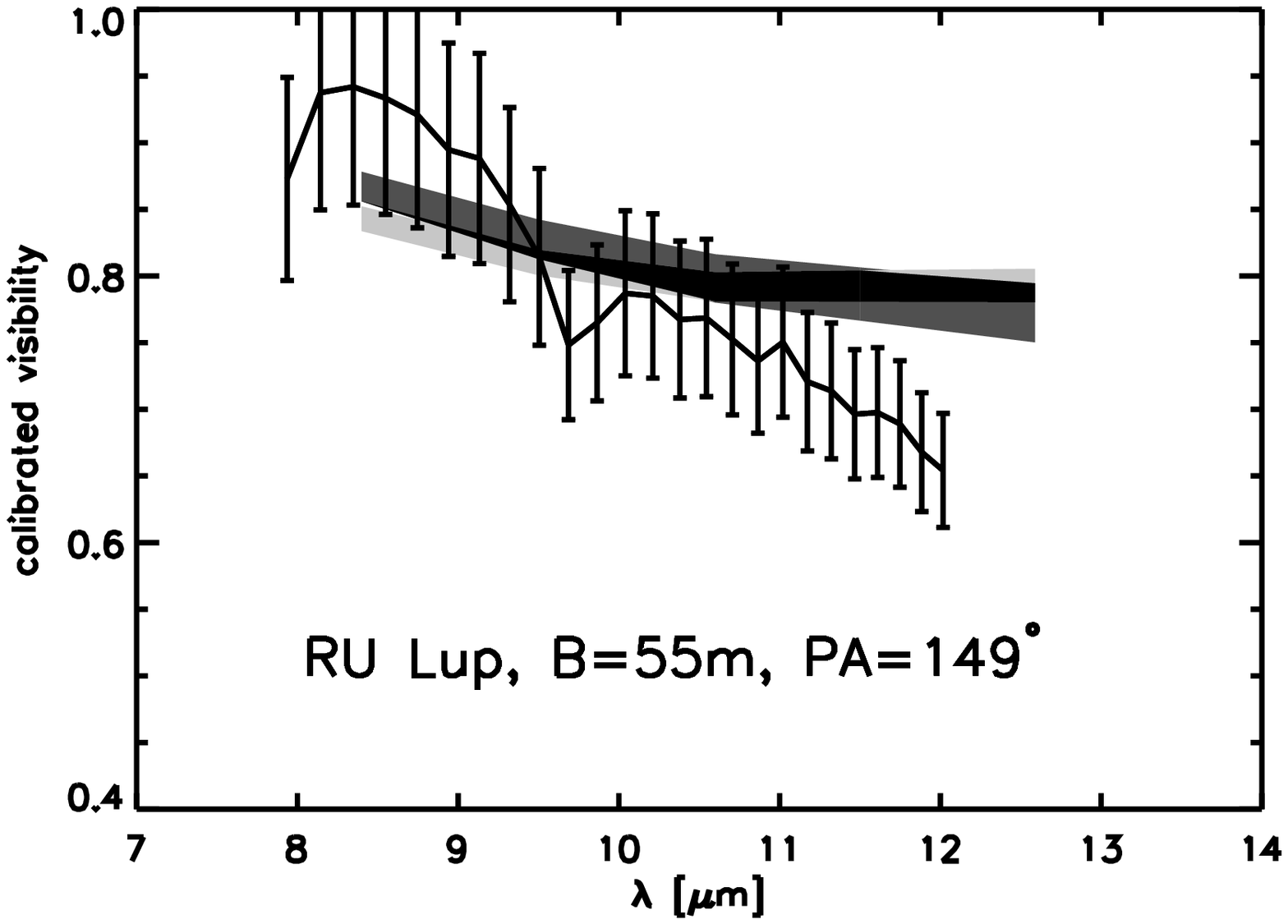}}
    \caption{SED and MIR visibilities for the projected baselines of
      $B=46\,\mathrm{m}$, $B=55\,\mathrm{m}$,  
      $B=61\,\mathrm{m}$, and $B=64\,\mathrm{m}$ obtained from the
      measurements and our model of RU\,Lup. For  
      comparison, the light gray and dark gray colored curves/bars are results from two
      different models with and without disk gap, respectively. The black
      region of the visibility bars represents an intersection. The gap -if
      present- stretches from $2\,\mathrm{AU}$ to $4\,\mathrm{AU}$ according to Takami
      et al.~(\cite{takamiII}). }
    \label{figure:rulup}
  \end{figure*}

  As Lamzin et al.~(\cite{lamzin}) has shown, RU\,Lup was at least
  temporarily a strongly accreting object. By  
  analyzing its SED, they derived an accretion rate of $\dot{M} = 3 \times
  10^{-7}\,\mathrm{M_{\odot}yr^{-1}}$. In this context, Reipurth et
  al.~(\cite{reipurthII}) found that the H$\alpha$ emission has broad wings of
  up to $900\,\mathrm{km s^{-1}}$ in width. Furthermore, the H$\alpha$ emission of RU\,Lup
  shows one of the broadest equivalent widths among T\,Tauri stars
  (Giovannelli et al.~\cite{giovannelli}). However, by considering SED and MIR
  visibilities, we can exclude such a high accretion rate $\dot{M}$.  
  We determine a maximum value of $\dot{M} = 1 \times
  10^{-8}\,\mathrm{M_{\odot}yr^{-1}}$ with an accretion  
  luminosity of only $L_\mathrm{acc} = 0.1\,\mathrm{L_{\odot}}$. Herczeg et
  al.~(\cite{herczeg}) found that $\dot{M} = (5  
  \pm 2) \times 10^{-8}\,\mathrm{M_{\odot}yr^{-1}}$ which supports our
  result. A higher accretion rate produces 
  a decrease in the MIR visibility because the irradiation of the disk
  increases. Simultaneously, the NIR and MIR  
  flux would exceed the photometric measurements. Figure~\ref{figure:rulup}
  represents our best-fit model with  
  $\dot{M}=1 \times 10^{-8}\,\mathrm{M_{\odot}yr^{-1}}$ (black curve/visibility
  bars). 
  The stellar luminosity of 
  $L_\mathrm{\star} = 1.27\,\mathrm{L_{\odot}}$ corresponds to a value 
  derived by  
  Gras-Vel\'azquez \& Ray~(\cite{gras-velazquez}). It represents the average
  of previous measurements (Herczeg et  
  al.~\cite{herczeg}: $0.49\,\mathrm{L_{\odot}}$; N\"urnberger et
  al.~\cite{nuernberger}:  
  $2.2\,\mathrm{L_{\odot}}$). A model that simulates the flux in the N band,
  i.\/e., the silicate feature, could
  not be found. 
  The relative 
  deviation in the MIR range is $37\%$. The visibilities obtained on
  the baselines $B=61\,\mathrm{m}$ and $B=65\,\mathrm{m}$ and for similar
  position angles significantly differ, which could be due to a measurement error. 

  Based on spectro-astrometric observations, Takami et al.~(\cite{takamiII})
  suggested that an inner gap in the circumstellar disk of RU\,Lup exists
  between $\sim$$2\,\mathrm{AU}$ and $\sim$$4\,\mathrm{AU}$ ($20$-$30\,\mathrm{mas}$)
  as justified by the following arguments. Because of the positional
  displacement of the extended wings, the emission of forbidden
  lines such as [OI] and [SII] in the spectra of RU\,Lup is assumed to arise
  in a collimated large-scale outflow. The apparent lack of redshifted line components
  results from an obscuration of the dorsal outflow 
  by the circumstellar
  disk (Eisl\"offel et al.~\cite{eisloeffel}). Takami et
  al.~(\cite{takamiII}) found that the H$\alpha$ line in the spectrum of
  RU\,Lup, which is evident predominatly in the innermost regions, shows both
  blueshifted {\it and} redshifted wings. This finding is 
  explained by an inner disk gap that allows regions behind
  the disk to be observed. The gap could be evoked by an unseen (planetary)
  companion. Since the wings of the H$\alpha$ line shows the same positional
  displacement as the forbidden lines, Takami et
  al.~(\cite{takamiII}) suggested that the H$\alpha$ line originates predominantly in
  the outflow. 

  Similar to our modeling approach of GW\,Ori, we defined a dust-free gap in
  the disk model that is presented in Fig.~\ref{figure:rulup}
  (Tables~\ref{table:properties-midisurvey},\ref{table:properties-midisurveyII}).  
  The gap size was varied using values of 
  $R_\mathrm{in}$ of up to $3\,\mathrm{AU}$ for the inner gap radius, and
  $3\,\mathrm{AU}$ and $5\,\mathrm{AU}$ for the outer gap radius. An outflow
  was not considered in our model. The resulting
  changes can be summarized as follows: as shown in
  Fig.~\ref{figure:rulup}, a gap lowers the NIR flux by $\sim$$7$\%. We note that
  the NIR flux is highly variable during a year (Giovannelli et
  al.~\cite{giovannelli}). Previous photometric measurements
  (Giovannelli et al.~\cite{giovannelli}) would be consistent with lower NIR
  fluxes. A disk gap as used here would even lower the MIR
  flux and slightly improve our fit to the N band. The MIR
  visibility simultaneously decreases at wavelengths $<10.5\,\mathrm{\mu m}$
  as the relative MIR 
  flux contribution of outer disk regions in relation increases. For longer wavelengths
  $>10.5\,\mathrm{\mu m}$, 
  the disk becomes more compact as the visibilities slightly increase. Our visibility
  measurements are inconsistent with a disk gap and an inner radius
  of $r=R_\mathrm{in}$, only, but a gap of between $2\,\mathrm{AU}$ and
  $4\,\mathrm{AU}$ fits the visibility data as well as the model without
  gap. 

  Although there are observational hints for a disk gap, we are unable to
  conclude whether the 
  model with a disk gap of inner radius $\gtrapprox 2\,\mathrm{AU}$ or a model
  disk without 
  gap provide a closer fit to the data. Further interferometric observations
  and simultaneous 
  photometric measurements in the NIR wavelength range would
  allow a final confirmation or rejection of the presence of a gap in the
  disk of RU\,Lup. 
    
  \subsection{HBC\,639}\label{section:hbc639}
  \begin{figure*}[!tb]
    \center
    \resizebox{0.48\textwidth}{!}{\includegraphics{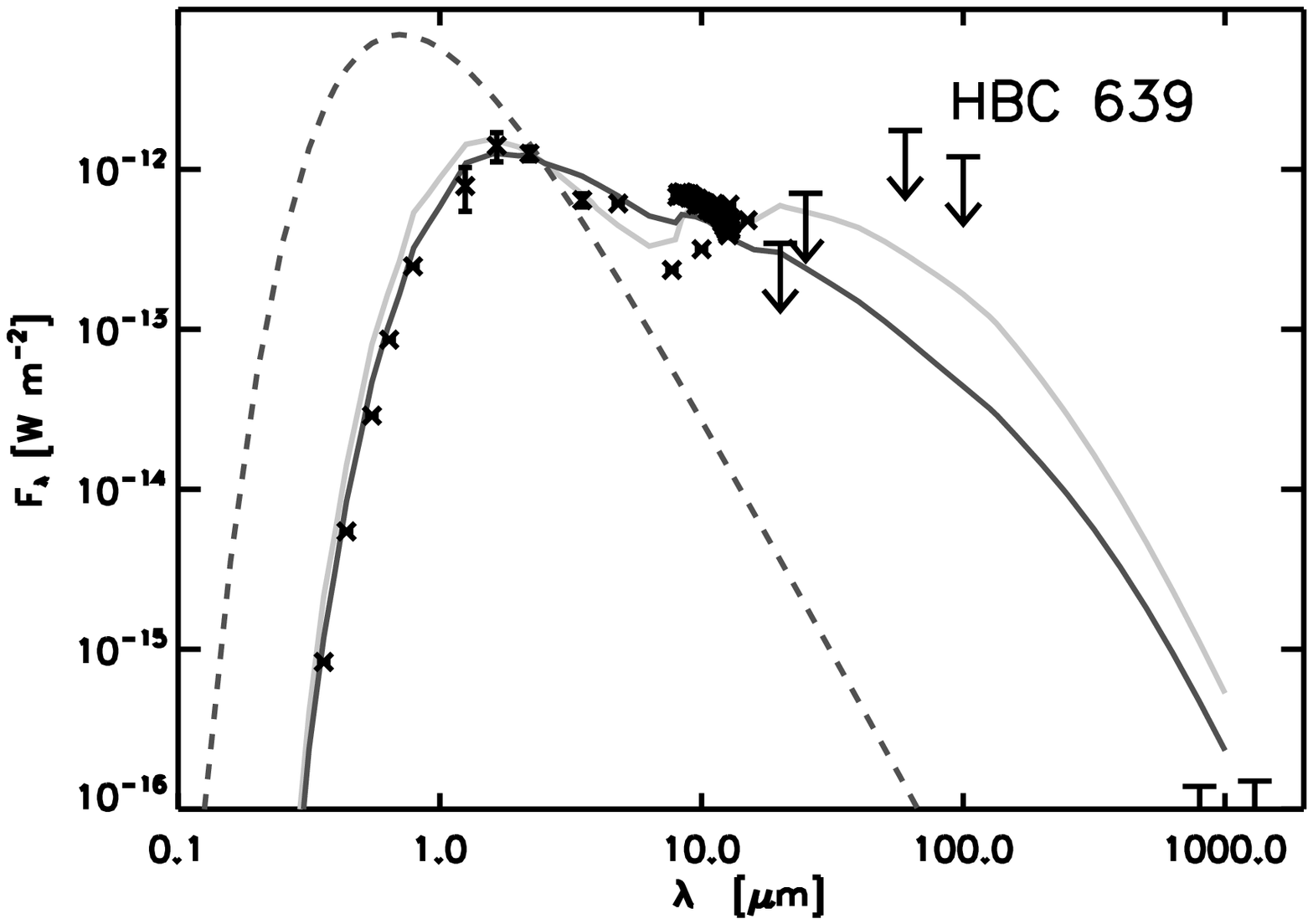}}
    \resizebox{0.48\textwidth}{!}{\includegraphics{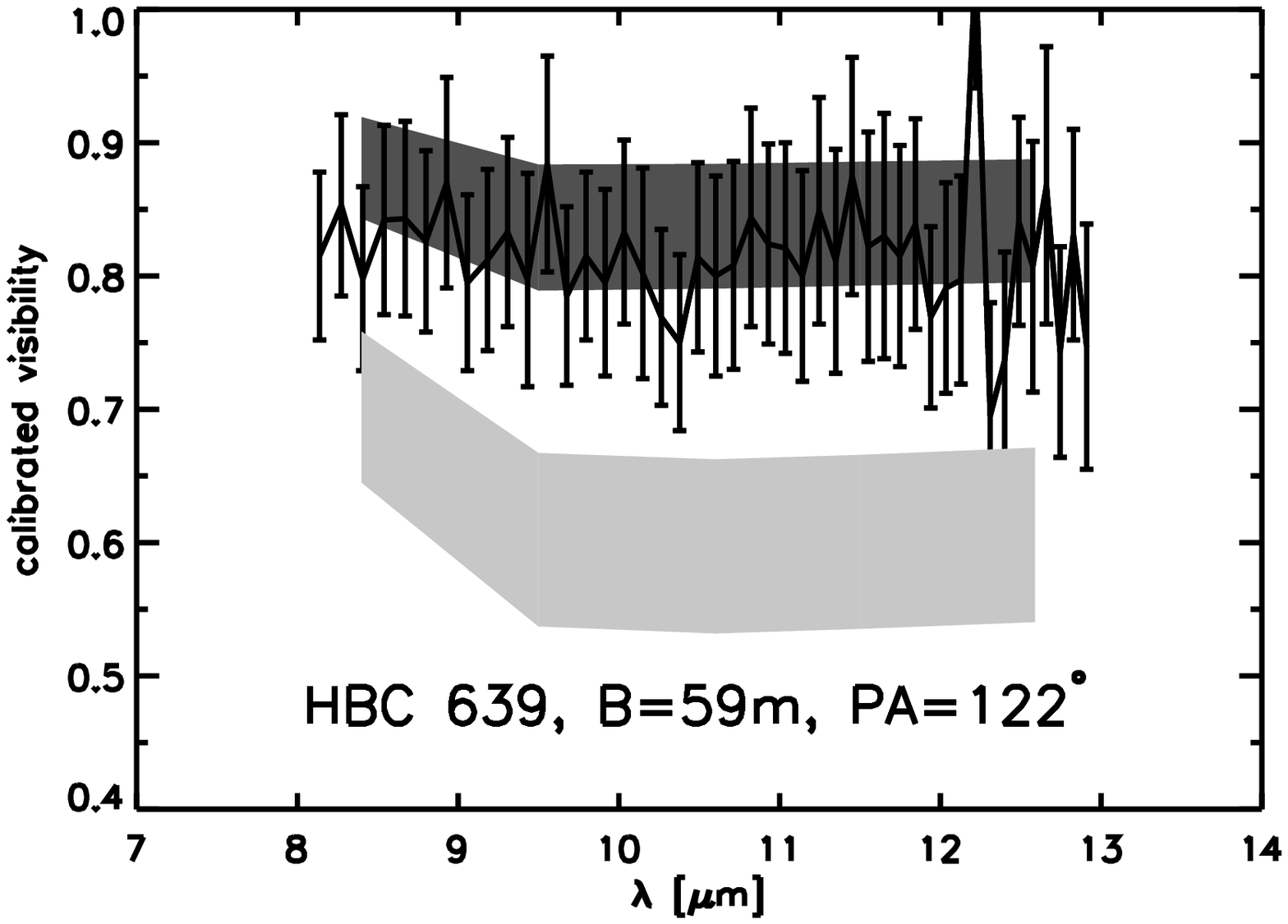}}\newline
    \resizebox{0.48\textwidth}{!}{\includegraphics{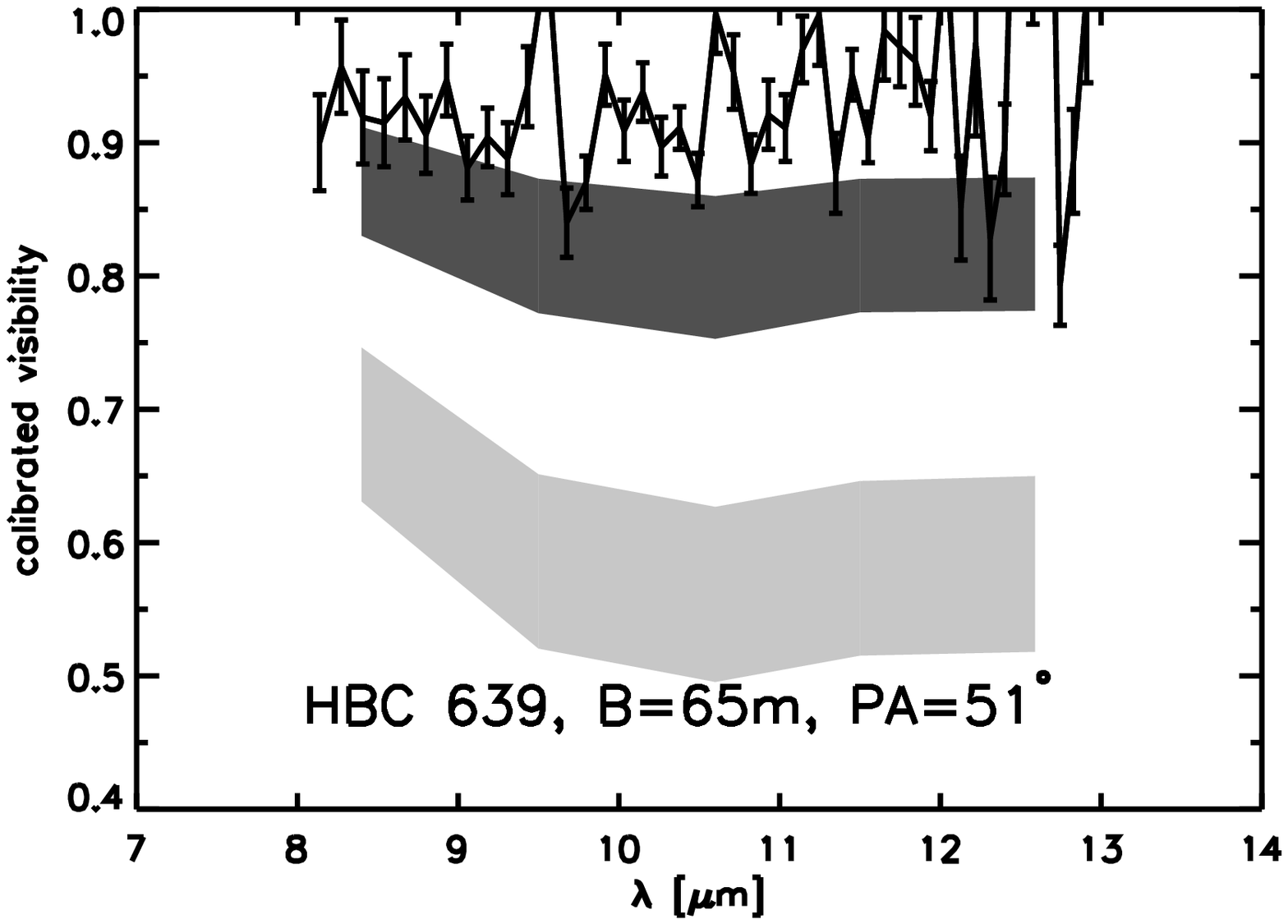}}
    \resizebox{0.48\textwidth}{!}{\includegraphics{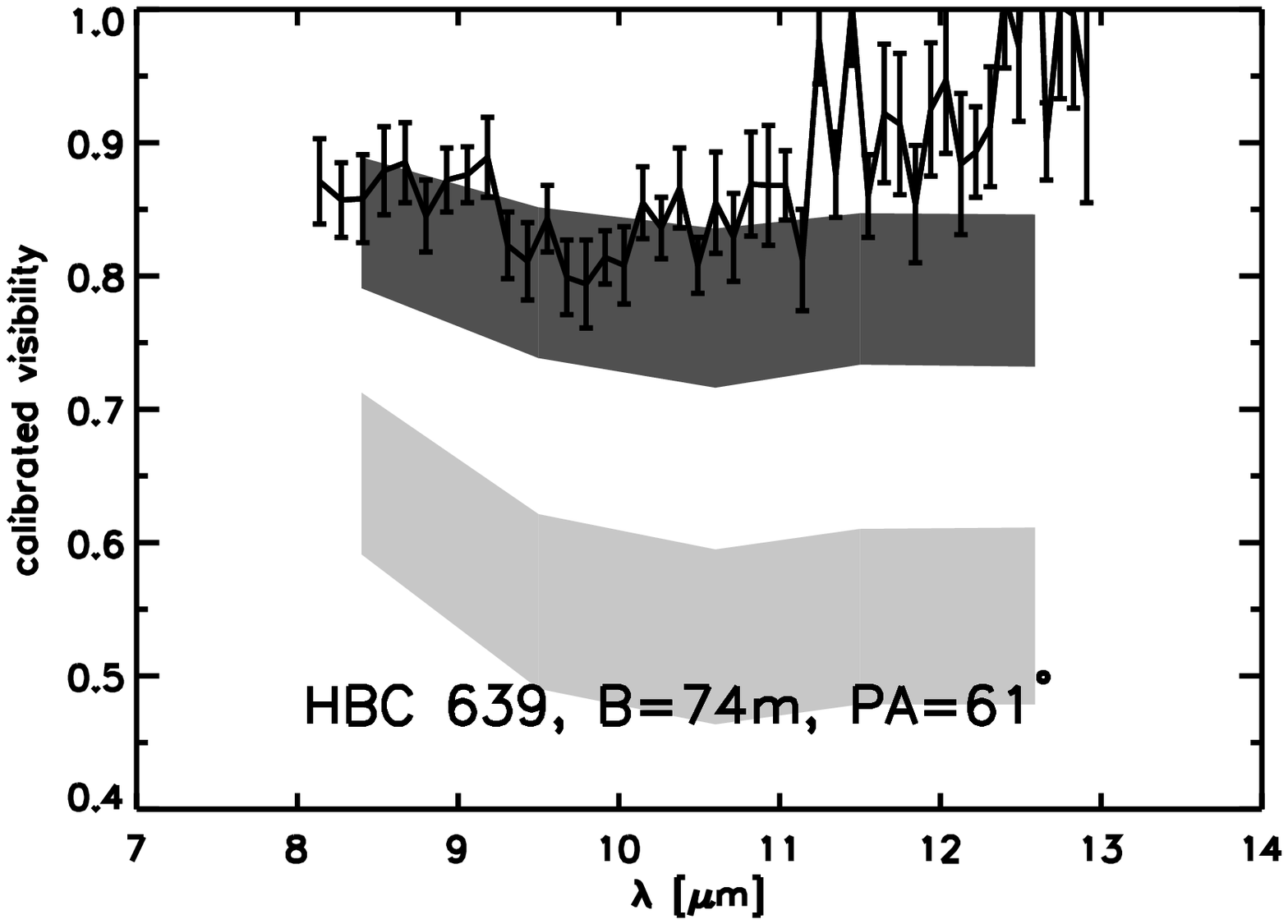}}\newline
    \caption{SED and MIR visibility for the projected baselines of
      $B=59\,\mathrm{m}$, $B=65\,\mathrm{m}$,  
      and $B=74\,\mathrm{m}$ obtained from the observations and the model of
      HBC\,639. We assume that the photometric 
      measurements in the FIR range are ascribed to the infrared
      companion. The light gray curves/bars  
      represent a model with a disk parameter
      $\beta=1.1$ instead of using $\beta =  
      1.0$ (grey colored curves/bars).} 
    \label{figure:hbc}
  \end{figure*}

  Apart from HD\,72106\,B, the source HBC\,639 is the only object where
  accretion has not   been considered. Although HBC\,639 has an age of between
  $1$ and $3\,$million
  years (Gras-Vel\'azquez \&  
  Ray~\cite{gras-velazquez}), we can neglect the derived accretion rate (Prato
  et al.~\cite{prato}), for which we estimate an upper limit of
  $\dot{M} < 1 \times 10^{-8}\,\mathrm{M_{\odot}yr^{-1}}$. Since the  
  equivalence width of the H$\alpha$ line is smaller than $10\,$\AA, HBC\,639
  is not a classical T\,Tauri star  
  but belongs to the class of weak-line T\,Tauri stars. 
  
  The parameter combination listed in
  Table~\ref{table:properties-midisurveyII} could reproduce  
  the 
  MIR visibilities. However, this model produces a strong decline in the FIR
  flux and is therefore unable to reproduce 
  the photometric upper limits in the FIR 
  range. HBC\,639 has an infrared 
  companion at a projected distance of $320\,\mathrm{AU}$
  (s. Appendix~\ref{appendix}). The strong decline
  in the FIR flux could be explained by the presence of an infrared
  companion that truncates the outer-disk region of the main
  component. In the latter case, the measured flux in the FIR (s.  
  Fig.~\ref{figure:hbc}) can mainly be ascribed to this infrared
  companion. The companion is again a binary, as  
  well, and 
  deeply embedded in a circumbinary envelope. 
  
  A larger value for the disk parameter $\beta$ results in a more flared
  disk. Outer disk regions can be more  
  effectively heated and the intensity distribution in the MIR range therefore
  decreases less strongly. A  
  consequence is a decrease in the MIR visibilities. 
  Fig.~\ref{figure:hbc} also shows the results of a model with  
  $\beta=1.1$ for which none of the other parameters were modified. We note
  that the  
  visibilities decrease by at least $10\%$ for all baselines and the FIR flux 
  increases by almost a factor of $2$.  
  
  \subsection{S\,CrA\,N}\label{section:scraa}
  \begin{figure*}[!tb]
    \center
    \resizebox{0.48\textwidth}{!}{\includegraphics{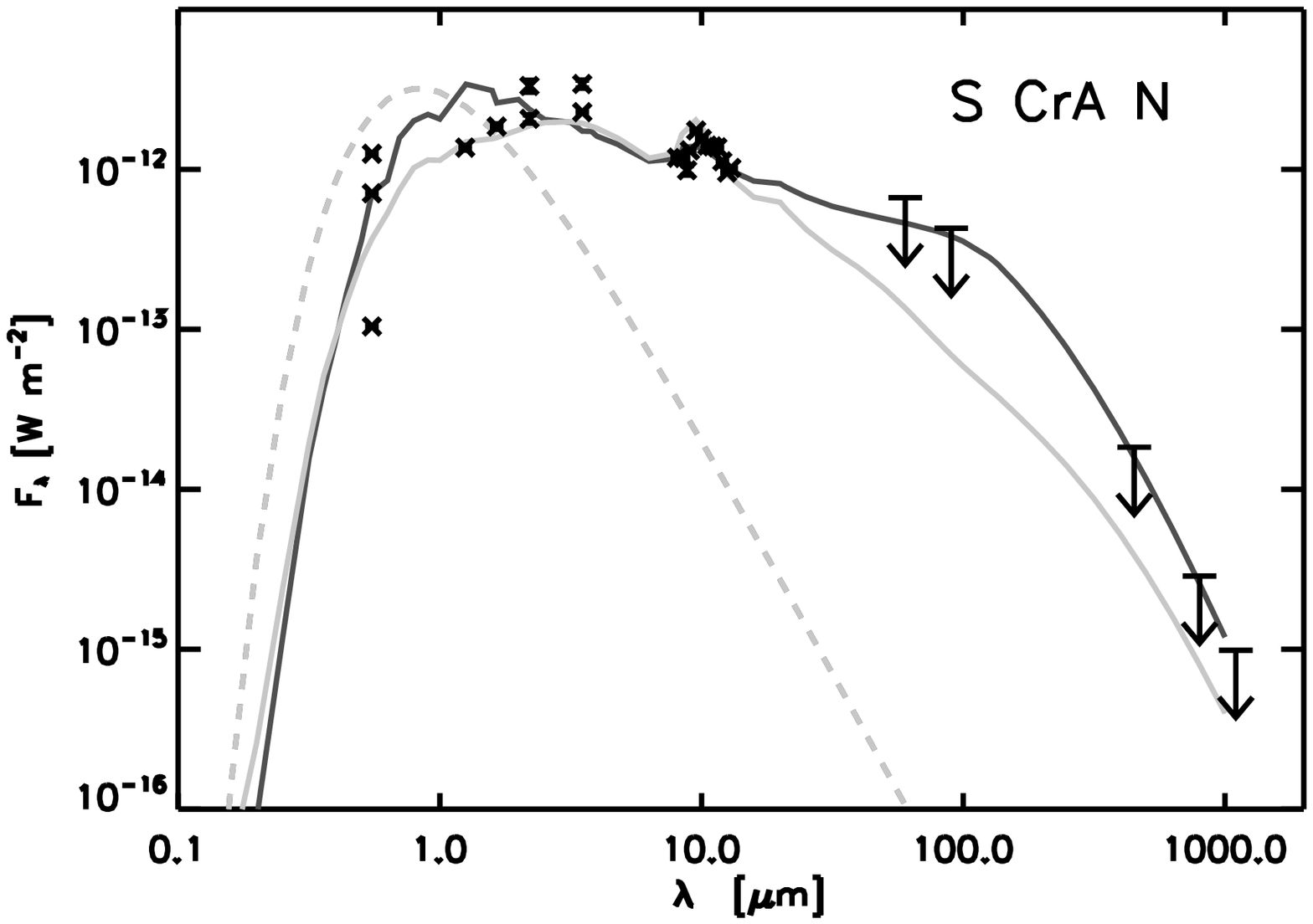}}
    \resizebox{0.48\textwidth}{!}{\includegraphics{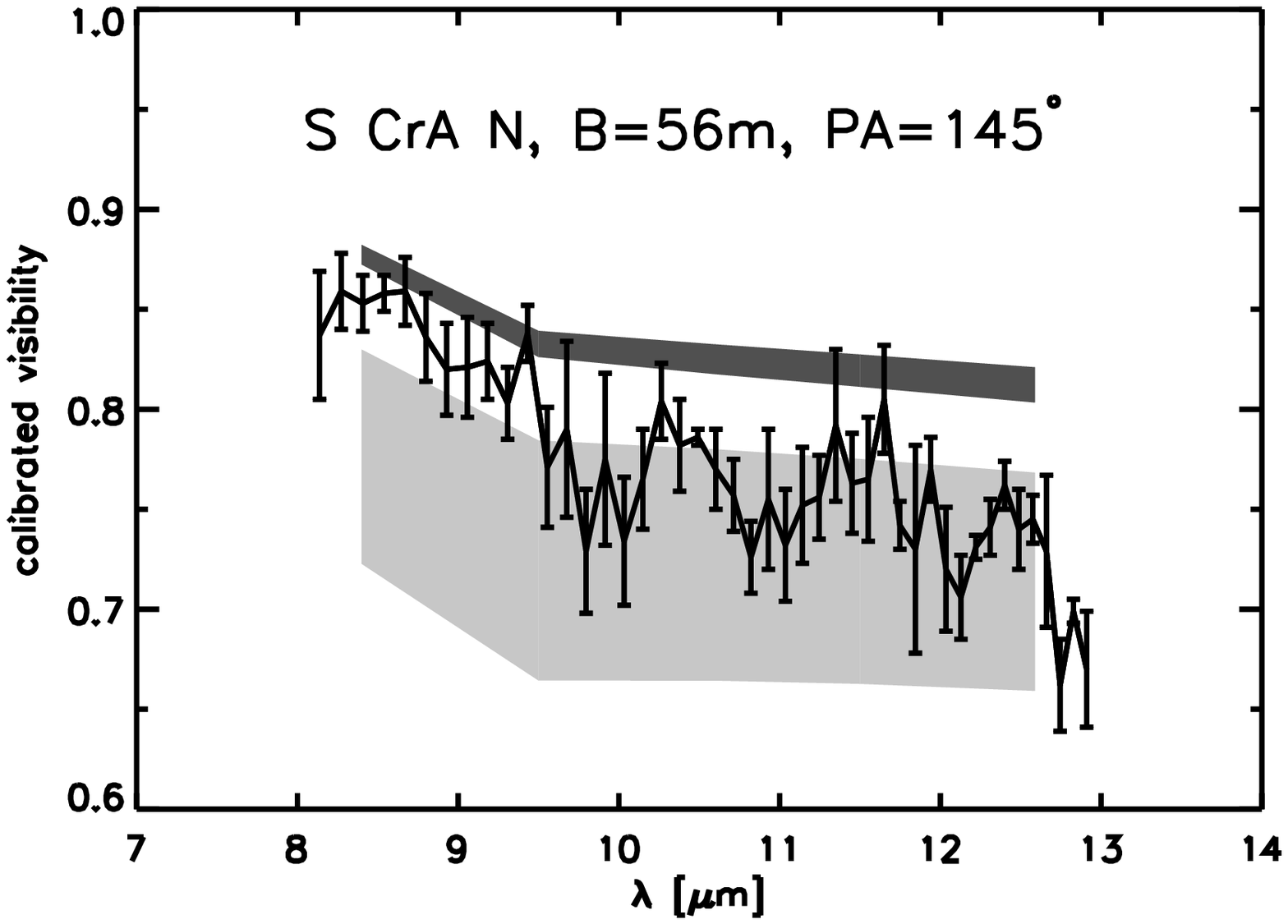}}\newline
    \resizebox{0.48\textwidth}{!}{\includegraphics{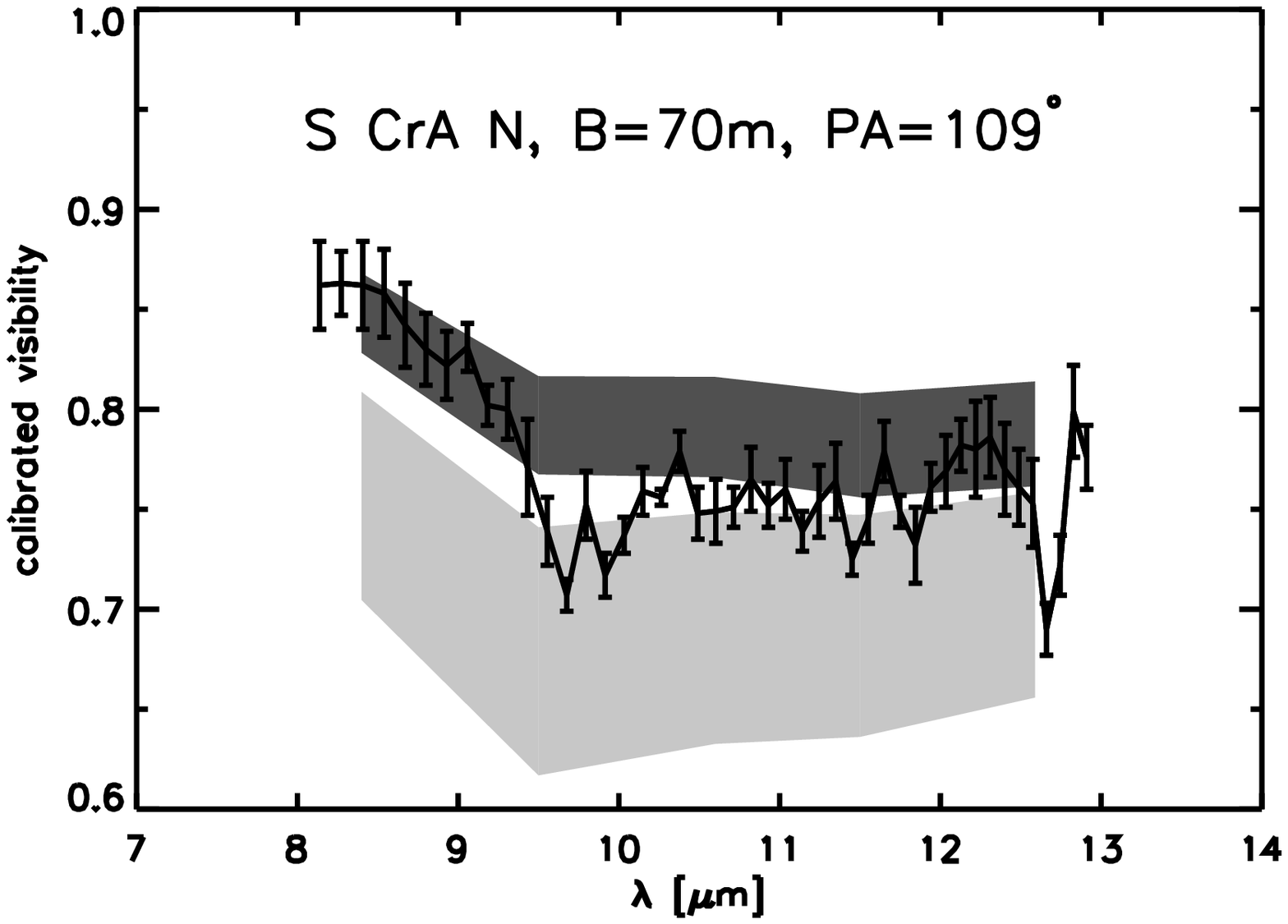}}
    \caption{SED and MIR visibilities for the projected baselines of
      $B=56\,\mathrm{m}$, and $B=70\,\mathrm{m}$ obtained from the
      observations and the model of S\,CrA\,N (dark grey line/colored
      bars). A second model with $\beta=0.90$, and
      $h_\mathrm{100}=12\,\mathrm{AU}$ (light gray lines/colored bars) can even
      improve the reproduction of the NIR fluxes.}
    \label{figure:scraa}
  \end{figure*} 
  Only the photometric measurements in the NIR and MIR range can exclusively 
  be ascribed to the Northern  
  component of this binary. All other photometric measurements could not spatially resolve
  the individual components of the binary system. The visual extinction  
  $A_\mathrm{V}$ and the effective stellar temperature $T_\mathrm{\star}$
  used in the model are within  
  the $1\,\sigma$-deviations ($1.0\,\mathrm{mag}$ and $400\,\mathrm{K}$) of
  the previously derived results of Prato et  
  al.~(\cite{prato}). The NIR flux that  
  results from our model deviates by $60\%$ from the measurements
  (s. Fig.~\ref{figure:scraa}). The 
  accretion luminosity is equal to $L_\mathrm{acc} =  
  0.7\,\mathrm{L_\mathrm{\odot}}$. 
  
  A second model for S\,CrA\,N (grey curve/bars in Fig.~\ref{figure:scraa}) with $\beta=0.90$ and  
  $h_\mathrm{100}=12\,\mathrm{AU}$ 
  does not reproduce the visibility
  measurements as well as the first model (deviations from the measurements
  are   
  $6\%$) but the SED for $\lambda < 14\,\mathrm{\mu m}$ can be 
  simulated more accurately. In contrast to HBC\,639, the FIR  
  flux must then originate in the Southern component. Furthermore, to 
  reproduce the visual wavelength  
  range, the second model still allows an inclination angle of up to
  $\vartheta \lessapprox 45\degr$, while only lower inclinations  
  angles of $\vartheta < 10\degr$ can be used in the first model. 
    
  \subsection{S\,CrA\,S}\label{section:scrab}
\begin{figure*}[!t]
    \center 
    \resizebox{0.48\textwidth}{!}{\includegraphics{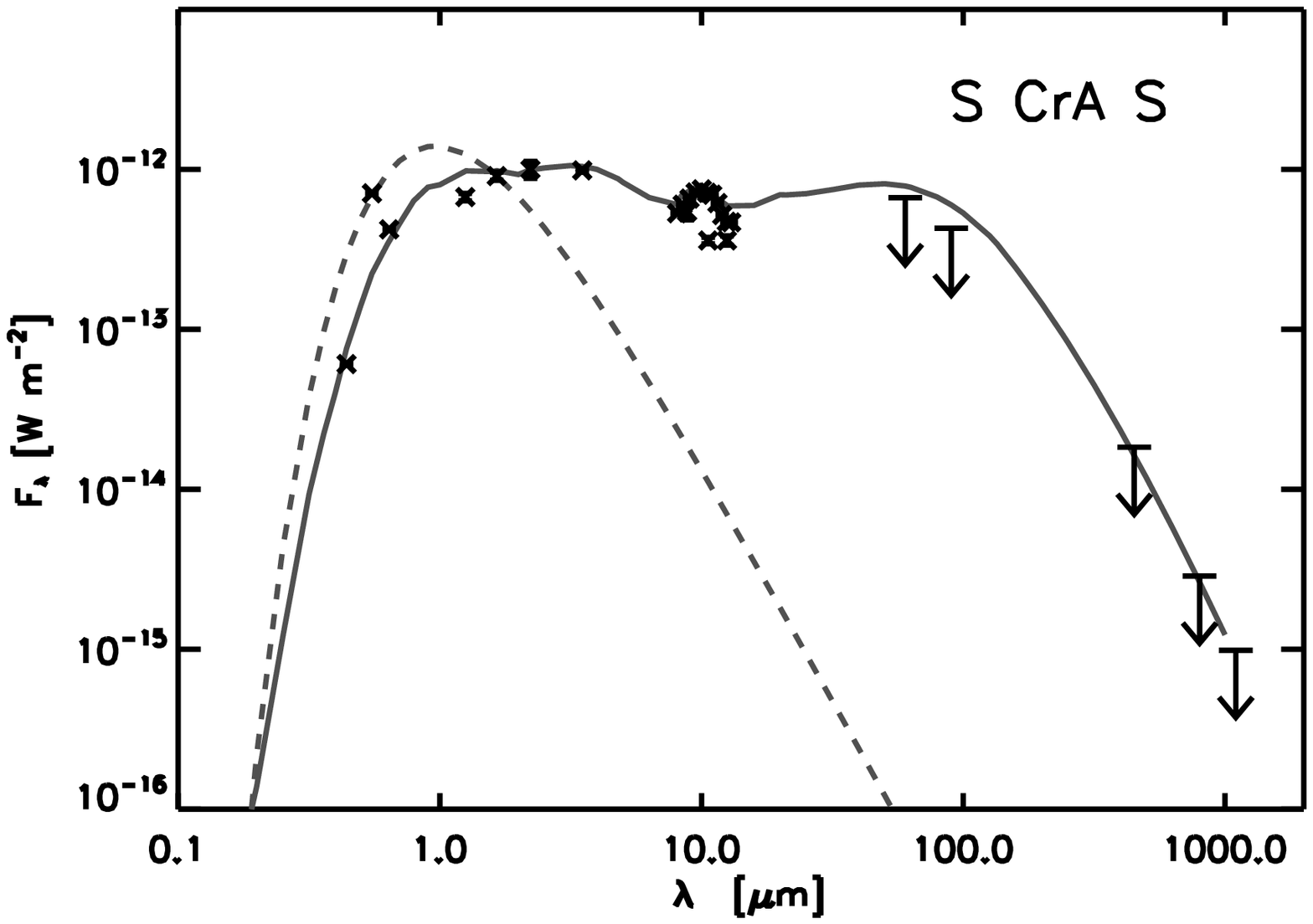}}
    \resizebox{0.48\textwidth}{!}{\includegraphics{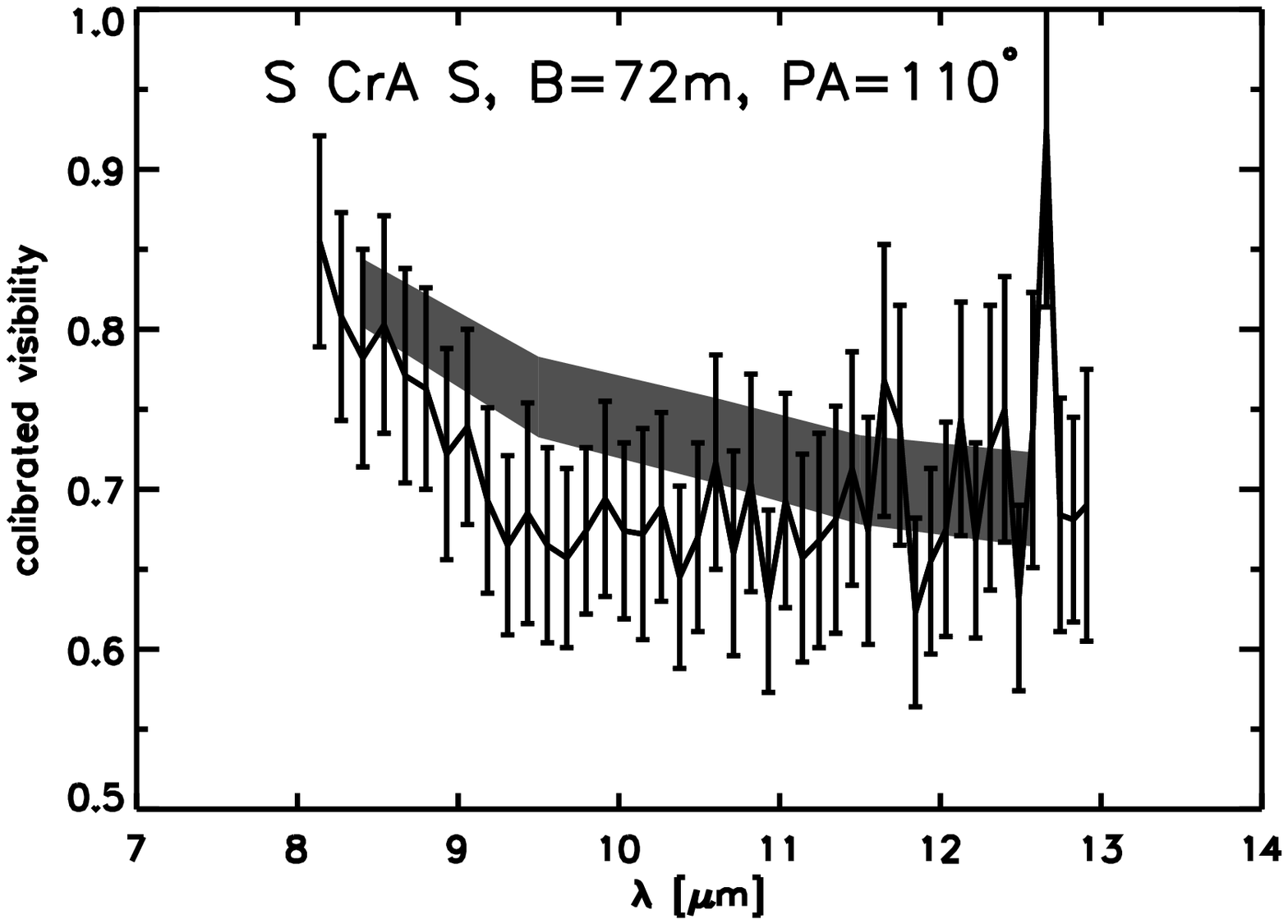}}
    \caption{SED and MIR visibilities for a projected baseline of
      $B=72\,\mathrm{m}$ obtained from the  
      measurements and our best model for S\,CrA\,S.}
    \label{figure:scrab}
  \end{figure*}

  The Southern component of S\,CrA is the fainter component in the NIR and MIR
  wavelength range. An  
  investigation of the Br$\gamma$ line showed that this component is as active
  as the Northern component (Prato \& Simon~\cite{pratoII}). A value for the
  accretion rate had not yet been derived. The accretion luminosity in  
  our model is $L_\mathrm{acc}=0.2\,\mathrm{L_\mathrm{\odot}}$. The
  corresponding accretion rate of $\dot{M} = 4 \times
  10^{-8}\,\mathrm{M_{\odot}yr^{-1}}$ is already an upper limit because  
  higher values would produce an increase in the disk irradiation and
  a decrease in the visibilities. If accretion is not considered in our model, the MIR visibilities
  increase by only $2\%$. Figure~\ref{figure:scrab}
  represents our best-fit model. 
  
  \section{Radial gradient of the dust composition in circumstellar disks
    around T\,Tauri stars}\label{section:dust composition}
  For each single interferometric
  observation with MIDI and for each single telescope, one uncorrelated N band
  spectrum 
  can be obtained. For each interferometric observation, we also obtained a
  correlated spectrum 
  reflecting the flux emitted by regions that are not spatially resolved by
  the interferometer. An increasing effective baseline length $B$ of the
  interferometer results in a higher resolution. The uncorrelated, i.\/e.,
  single-dish spectra as 
  well as the correlated spectra contain spectral contributions of the N band
  from the entire disk, while the contributions from the hotter
  and brighter, i.\/e., inner regions increase with increasing
  effective baseline length $B$ assuming a homogeneous, axial-symmetric disk. 
  It was shown by Schegerer et al.~(\cite{schegerer}) that the local silicate
  dust composition in the circumstellar disk around RY\,Tau depends on the
  radial distance from the star, i.\/e., the relative contribution of  
  crystalline and large dust grains increases towards the
  star.\footnote{In this study,
    crystalline and large grains are called evolved dust grains.} Small and
  large dust grains are assumed to have a radius of
  $0.1\,\mathrm{\mu m}$ and $1.5\,\mathrm{\mu m}$, respectively. In the following, we  
  investige whether a corresponding correlation between radial location and dust
  composition can also be found in the T\,Tauri stars of our sample. 

  The (averaged) uncorrelated spectra and the correlated
  spectra are shown in Fig.~\ref{figure:midi-10umfit}. 
  All uncorrelated and correlated spectra exhibit the silicate emission
  feature apart from the
  correlated spectra
  of HD\,72106 and HBC\,639 that were obtained with the longest
  interferometric baseline. The uncorrelated
  spectra obtained during different nights differ
  in absolute flux scale by $4$\% for S\,CrA\,N and up to $35$\% for GW\,Ori, while
  the shape of the spectra does not change. We assume that the variation in
  the absolute flux scale of a spectrum is caused by an erroneous photometric calibration
  during data reduction. Przygodda~(\cite{przygoddaII}) indeed confirmed
  that the scale factor of the photometric calibration can vary by more than
  $15$\% during a single night. If available, spectra acquired by 
  the Thermal Infrared Multi Mode Instrument\,2 (TIMMI\,2) at
  ESO's observatory La\,Silla (Przygodda et al.~\cite{przygodda}; Schegerer et 
  al.~\cite{schegererI}) are also plotted in
  Figs.~\ref{figure:midi-10umfit} and \ref{figure:midi-10umfit2}. The
  spectra obtained
  with TIMMI\,2 can be used in the following comparison. With respect to the
  TIMMI\,2 data, the shapes of the
  uncorrelated MIDI spectra are largely preserved for GW\,Ori (deviation:
  $6$\%), RU\,Lup ($9$\%), S\,CrA\,N ($18$\%), and S\,CrA\,S ($25$\%) and
  differ only in absolute scale. However, the shapes of the spectra of
  DR\,Tau, and of HBC\,639 in particular, differ significantly in a way 
  that could be due to an intrinsically temporal variation. We 
  note that the TIMMI\,2 spectra were obtained $\sim$$3\,$years before our
  MIDI observations. However, the difference could also be caused by a doubling
  of the spatial resolution power using a $8.2\,\mathrm{m}$-single-dish of
  MIDI in contrast to the $3.6\,\mathrm{m}$-telescope of TIMMI\,2. Therefore,
  observations with a single-dish of the VLTI allows us to observe exclusively
  more central regions of the source, where more evolved grains are assumed to
  exist. We should mention that a
  spatial resolution of $\sim$$43\,\mathrm{AU}$ at a distance of
  $\sim$$140\,\mathrm{pc}$ can theoretically be achieved with a
  single-dish observation with the VLT. 
  With respect to the TIMMI\,2 spectrum, the flattening of the MIDI spectrum of
  HBC\,639 could originate from an increase in the mass contribution of
  large silicate grains in the more central regions considering their
  specific absorption efficiency $\kappa$ (Dorschner et al.~\cite{dorschner}). 
  Our latter assumption is also confirmed by the correlated spectra of
  the objects HD\,72106\,B and HBC\,639 obtained with the longest
  interferometric baseline. These spectra are flat 
  and do not show any silicate emission band. We refer to a
  study of Min et al.~(\cite{min}), where it was found that the silicate band
  disappears if the averaged dust grain size exceeds $\sim$$4\,\mathrm{\mu m}$. 

  The origin of the evolved dust contributing to the
  TIMMI\,2 spectra can be derived by considering the following comparison with the
  MIDI spectra. A black body
  $B(\nu,T)$ that represents the putative underlying continuum (Schegerer et 
  al.~\cite{schegererI}) and is fitted to the extended wings of the emission
  features $F(\nu)$ of TIMMI\,2 as well as correlated and uncorrelated MIDI
  spectra, is subtracted from N band. We derive normalized features using: 
  \begin{eqnarray}
    \label{eq:normalize}
    \hfill{}
    F_\mathrm{norm}(\nu)= 1 + \frac{(F(\nu)-B(\nu,T))}{B(\nu,T)}.
    \hfill{}
  \end{eqnarray} 
  This normalization procedure preserves the shape of the emission
  feature (Przygodda et al.~\cite{przygodda}). According to Schegerer et
  al.~(\cite{schegererI}), the contributions of amorphous and
  crystalline dust to the TIMMI\,2 spectra are determined using a $\chi^2$
  fitting routine, where the absorption efficiencies $\kappa$ of dust are
  linearly combined. In the context of the latter findings, we
  additionally subtract the contribution of small amorphous dust from the
  normalized TIMMI\,2 spectra and compare the resulting spectra with the
  normalized uncorrelated and correlated MIDI spectra.\footnote{For simplicity,
    the modified TIMMI\,2 spectra after normalization {\it and} after subtraction of the
    contribution of small amorphous grains are called normalized TIMMI\,2
    spectra.} In
  Figs.~\ref{figure:10um_abzug} and \ref{figure:10um_abzug2}, we only 
  compare the normalized TIMMI\,2 spectra with both the normalized
  uncorrelated and normalized correlated MIDI 
  spectra obtained with the longest effective baseline length. With
  respect to our results, all sources can be divided into three classes:
  \renewcommand{\labelenumi}{\roman{enumi}.}
  \begin{enumerate}    
  \item The normalized TIMMI\,2 spectra reproduce 
    well the normalized uncorrelated and normalized correlated MIDI
    spectra. The evolved grains that
    contribute to the TIMMI\,2 spectra originate therefore in more central
    regions. Since the normalized TIMMI\,2 spectra similarly correspond
    to the normalized uncorrelated {\it and} normalized correlated MIDI
    spectra, a further evolution of 
    dust in regions that MIDI is able to study, cannot be determined. DR\,Tau
    and S\,CrA\,N belong to this class. 
  \item The normalized TIMMI\,2 and MIDI
    spectra of HBC\,639 differ at $\sim$$10.8\,\mathrm{\mu m}$, where
    the crystalline compound enstatite generally shows a specific emission
    feature. If the remaining feature at $\sim$$10.8\,\mathrm{\mu m}$ in the
    normalized TIMMI\,2 spectrum can be ascribed to enstatite, our comparison suggests
    that enstatite has its origin in more outer disk regions. These outer
    regions can already be resolved with a single-dish of MIDI and do not
    appear in the MIDI spectra. S\,CrA\,S could also be ascribed to this class
    by considering its correlated, normalized spectra. This result is based on
    the assumption that the spectral 
    resolution power of MIDI is high enough to resolve the
    $\sim$$10.8\,\mathrm{\mu m}$ feature of enstatite 
    only in the inner regions. 
  \item An increasing flattening of the emission feature and an
    increasing amplitude of the forsterite feature at $11.3\,\mathrm{\mu m}$ with increasing
    baseline length $B$ suggest grain evolution towards more central regions
    (e.\/g., GW\,Ori).
    The comparison with the corresponding normalized TIMMI\,2 spectra shows, however,
    that the contribution of small amorphous dust grains, emitting a
    triangular emission feature with the maximum at $9.8\,\mathrm{\mu m}$, is
    still large even in the correlated spectrum that were obtained with the longest
    baselines. GW\,Ori, RU\,Lup, and S\,Cr\,A\,S belong to this class. 
  \end{enumerate}
  We note that all of these findings are based on the assumption that the underlying
  continuum of the emission feature can be represented by a single black body. 
  This procedure is circumstantially discussed in Schegerer et al.~(\cite{schegererI}).
 
  Considering a measurement error in the visibility of $\sigma \approx 0.1$ as well as the low signal-to-noise ratio of $\sim$$4$
  and the low spectral resolution reached, we avoid fitting the
  uncorrelated and correlated MIDI spectra using a linear combination of
  different-sized, amorphous and crystalline silicate grains. The latter
  procedure was
  presented by Schegerer et al.~(\cite{schegererI}) and Schegerer et
  al.~(\cite{schegerer}). 
  However, we conclude that the local silicate dust composition in the 
  circumstellar disks around the T\,Tauri objects of our sample 
  depends on the radial disk location. Features
  of crystalline and large silicate grains indeed originate predominantly in
  more central disk 
  regions of several AUs in size. However, the features of enstatite
  could originate in more outer regions, and the spectral contribution of
  small, non-evolved dust grains from the inner disk region could remain high. 

  \begin{figure*}[!tb]
    \centering
    \includegraphics[scale=0.29]{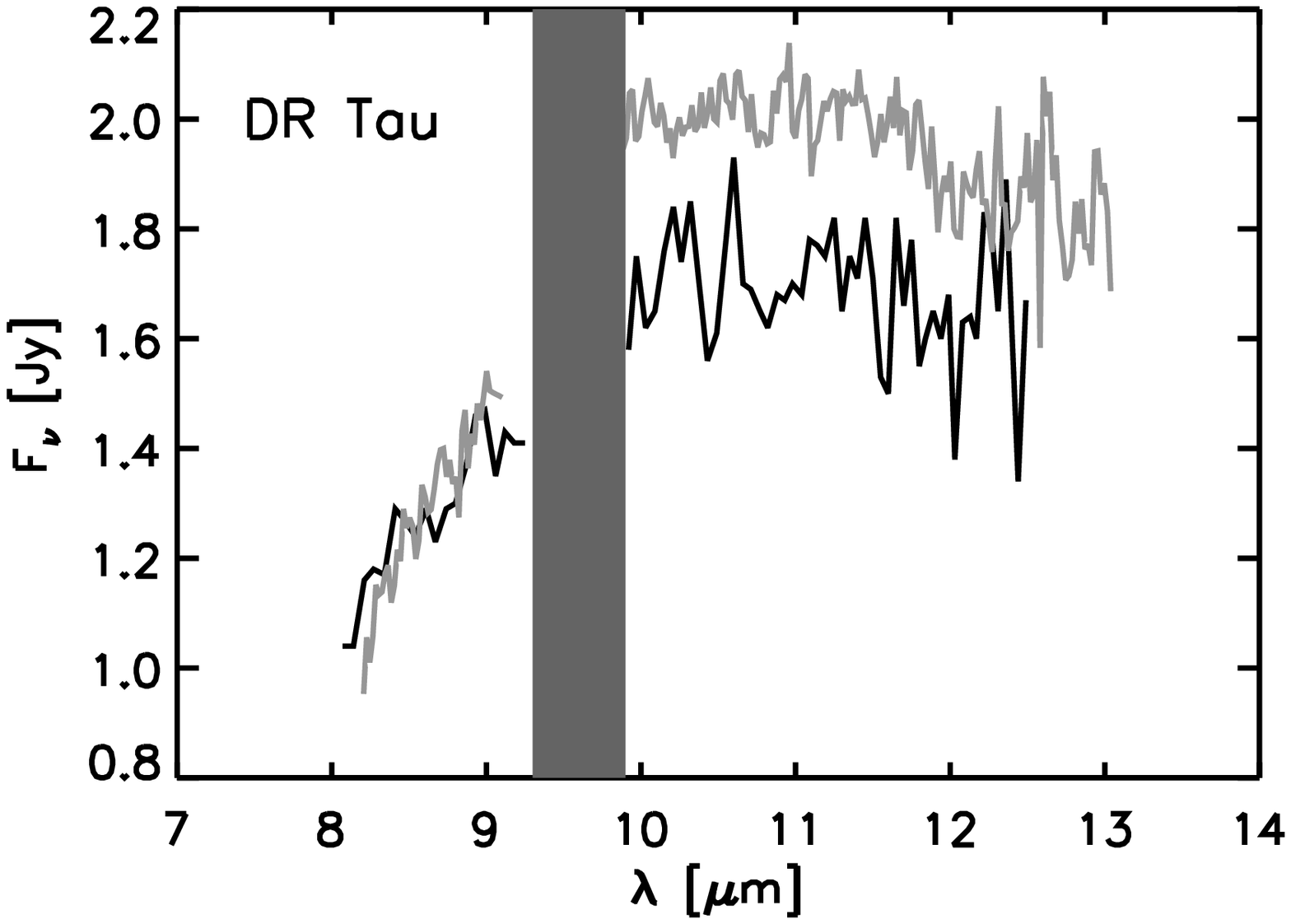}
    \includegraphics[scale=0.29]{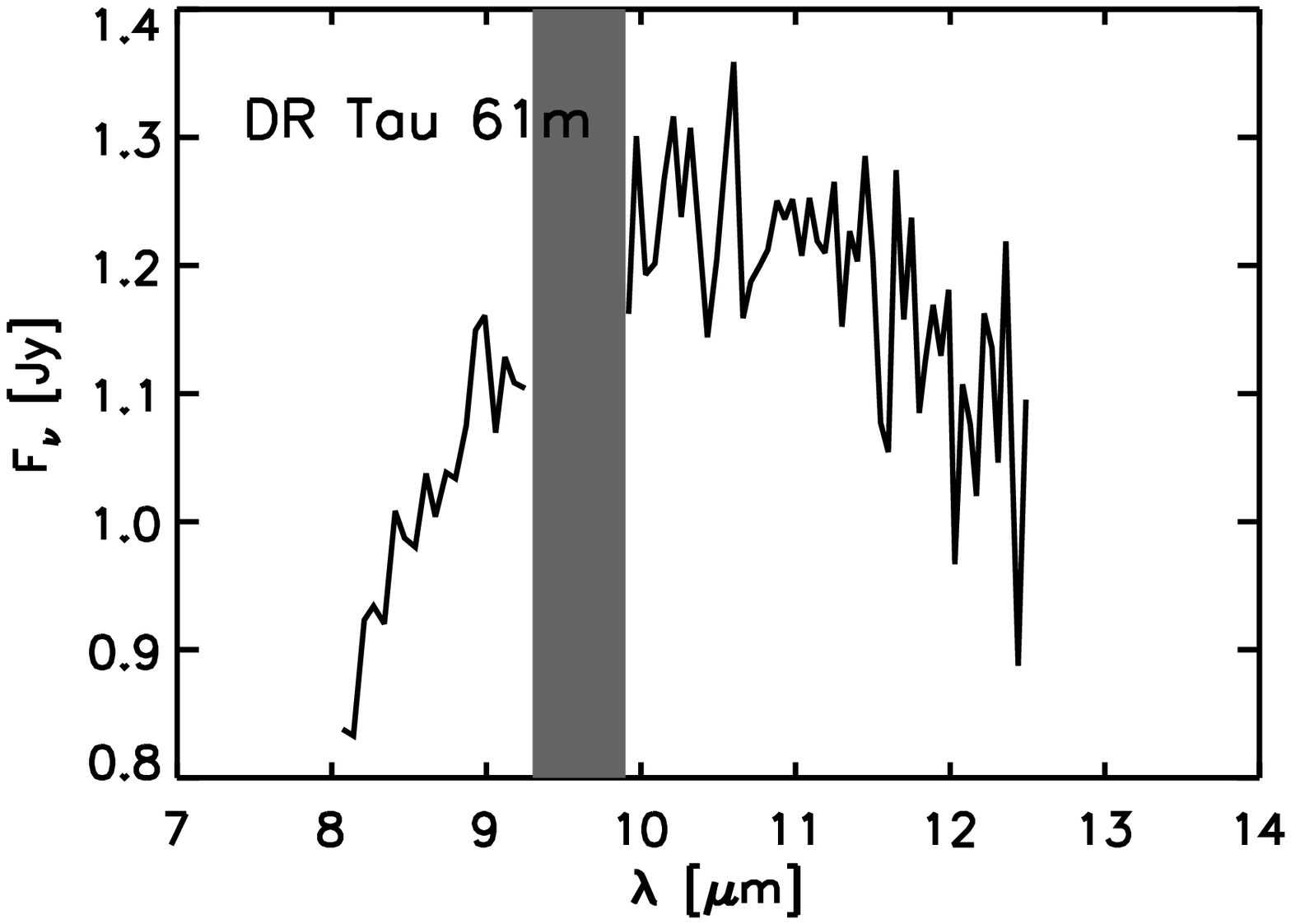}\newline
    \includegraphics[scale=0.29]{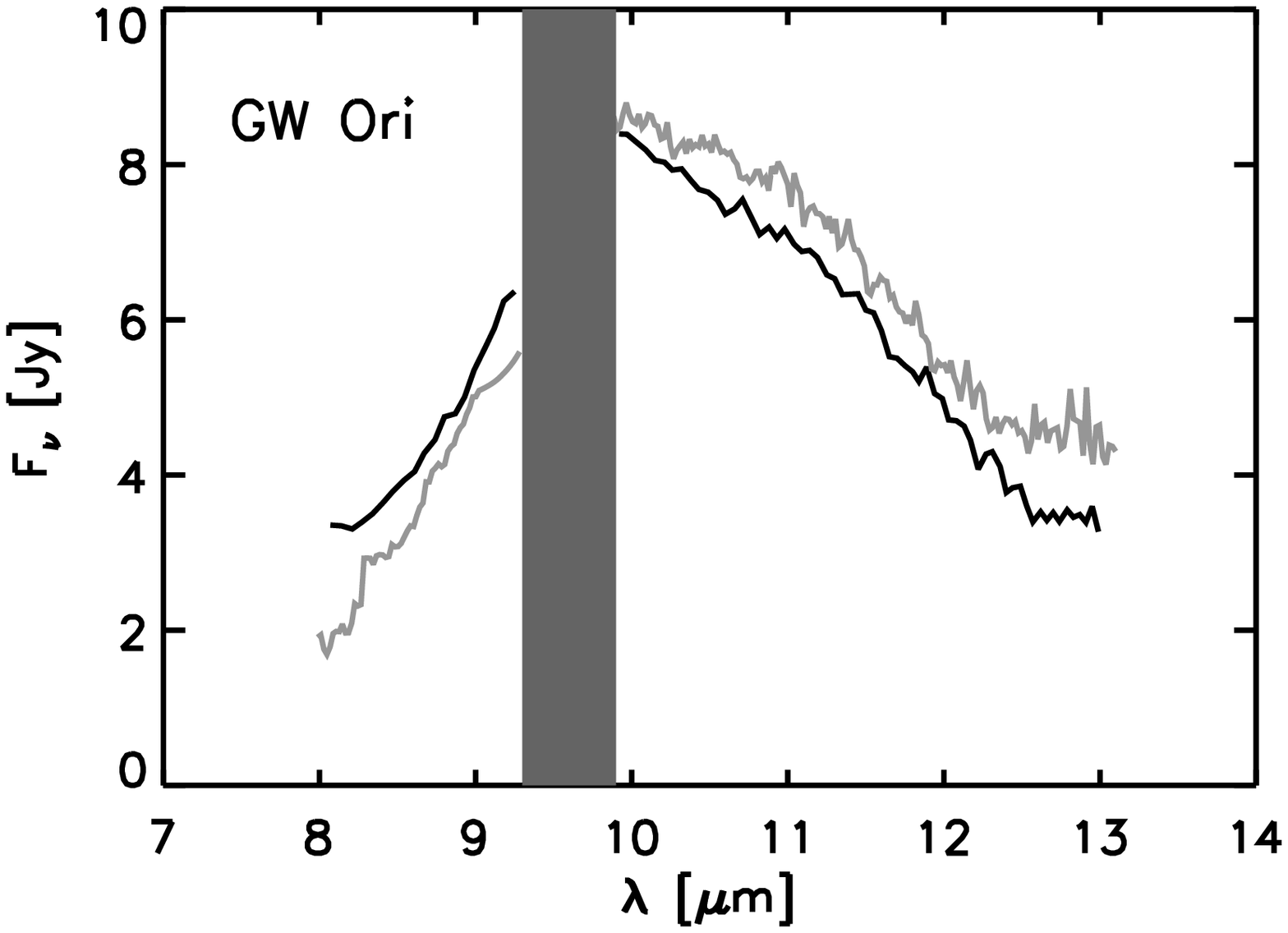}
    \includegraphics[scale=0.29]{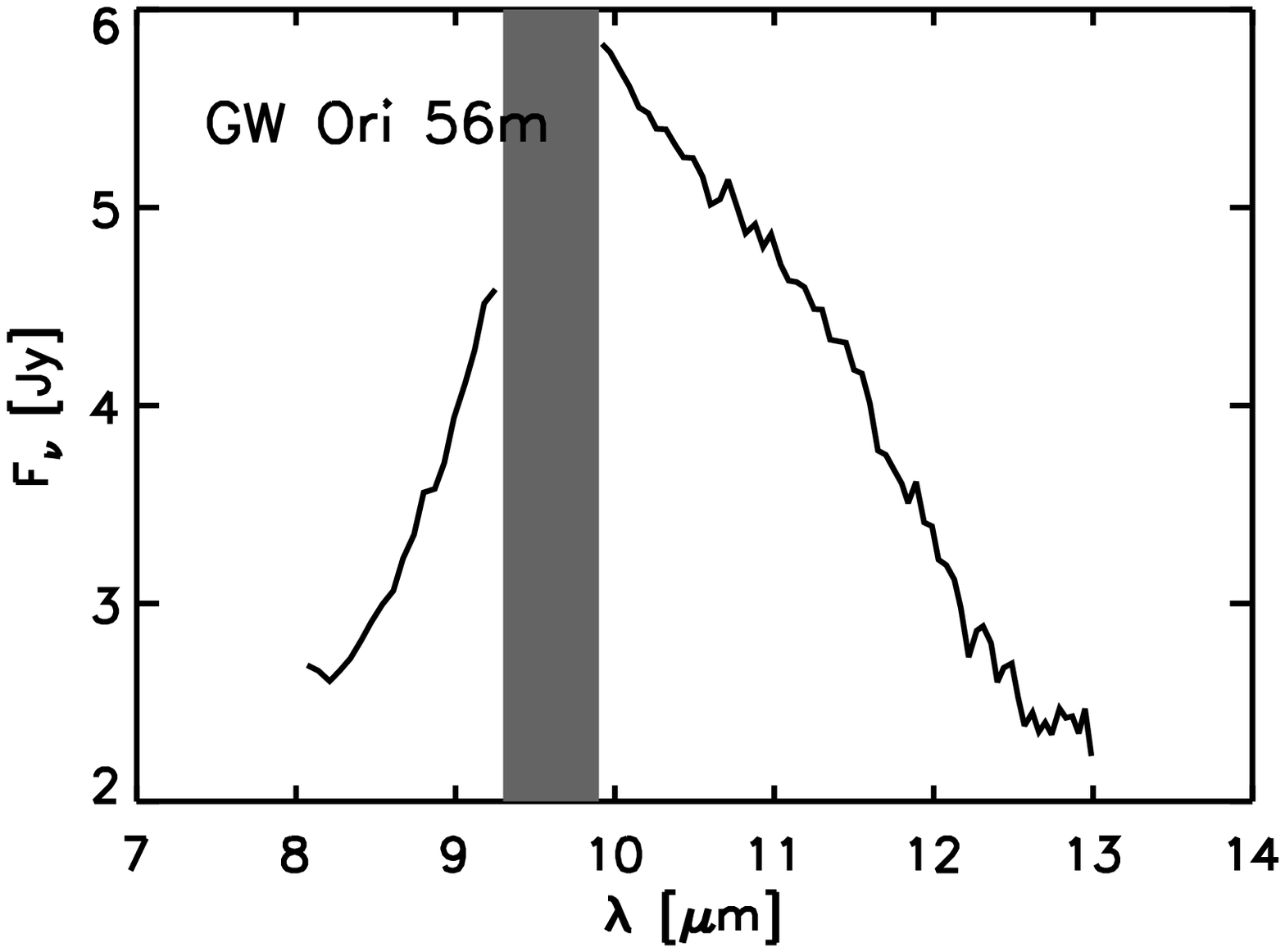}
    \includegraphics[scale=0.29]{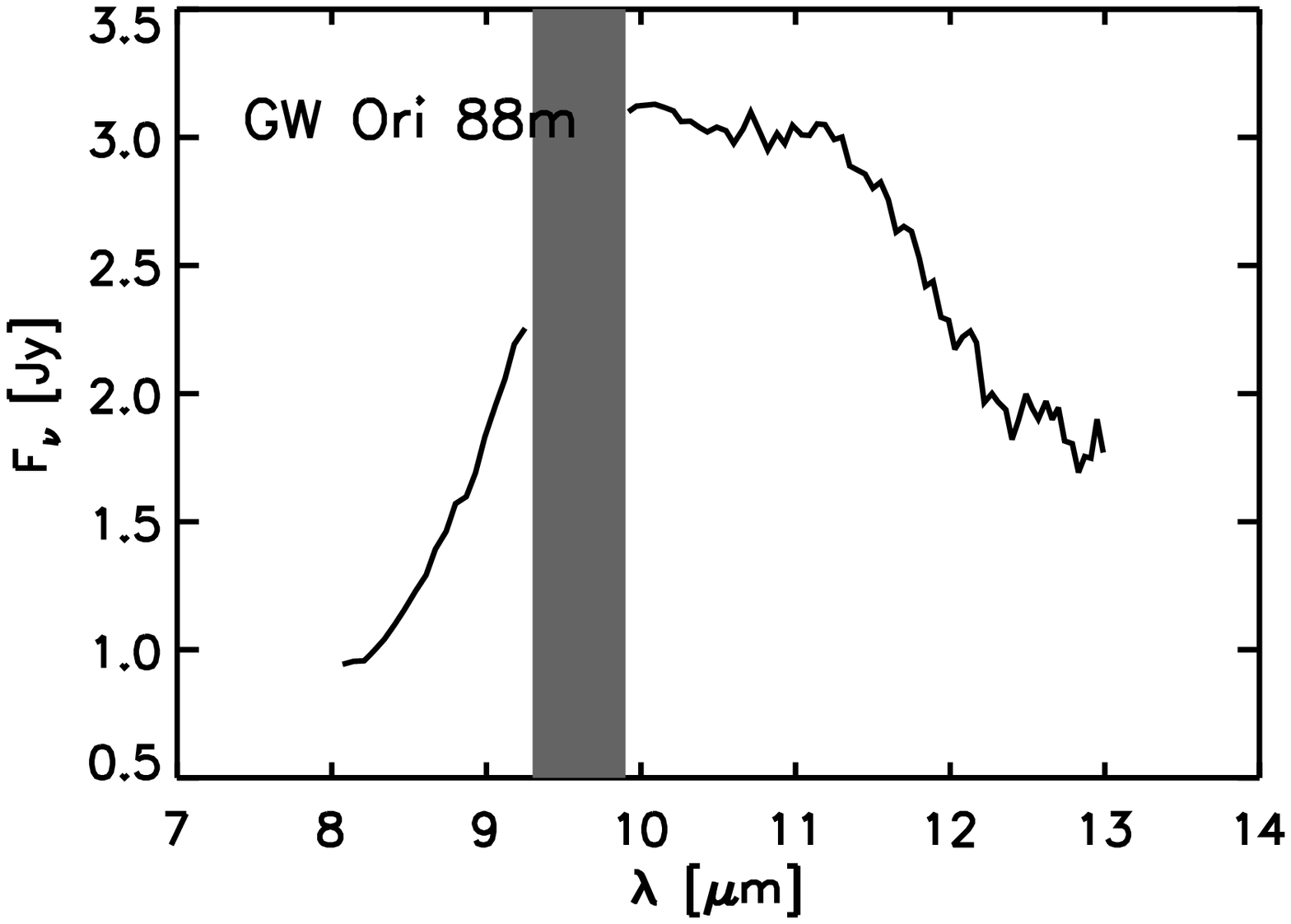}\newline
    \includegraphics[scale=0.29]{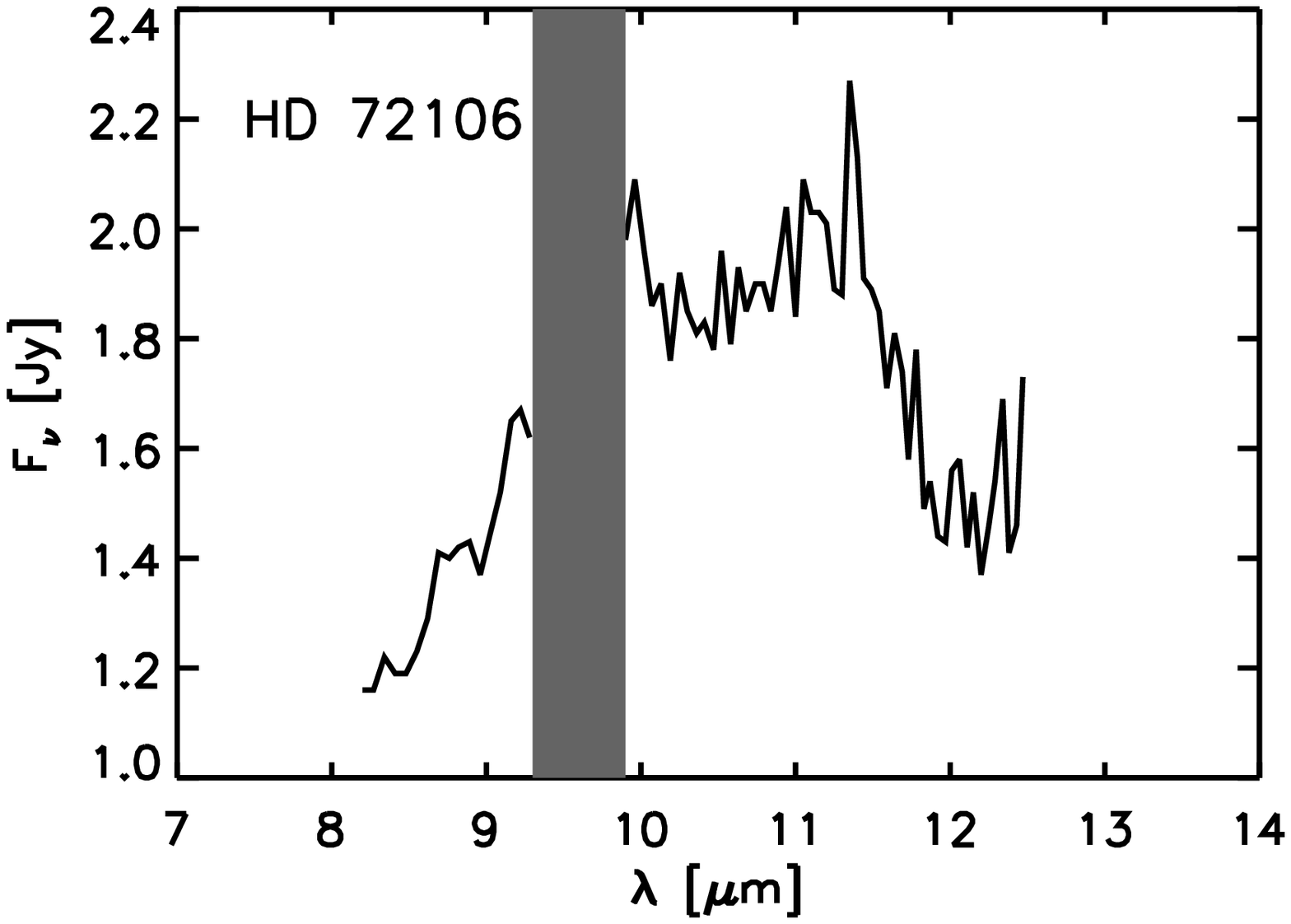}
    \includegraphics[scale=0.29]{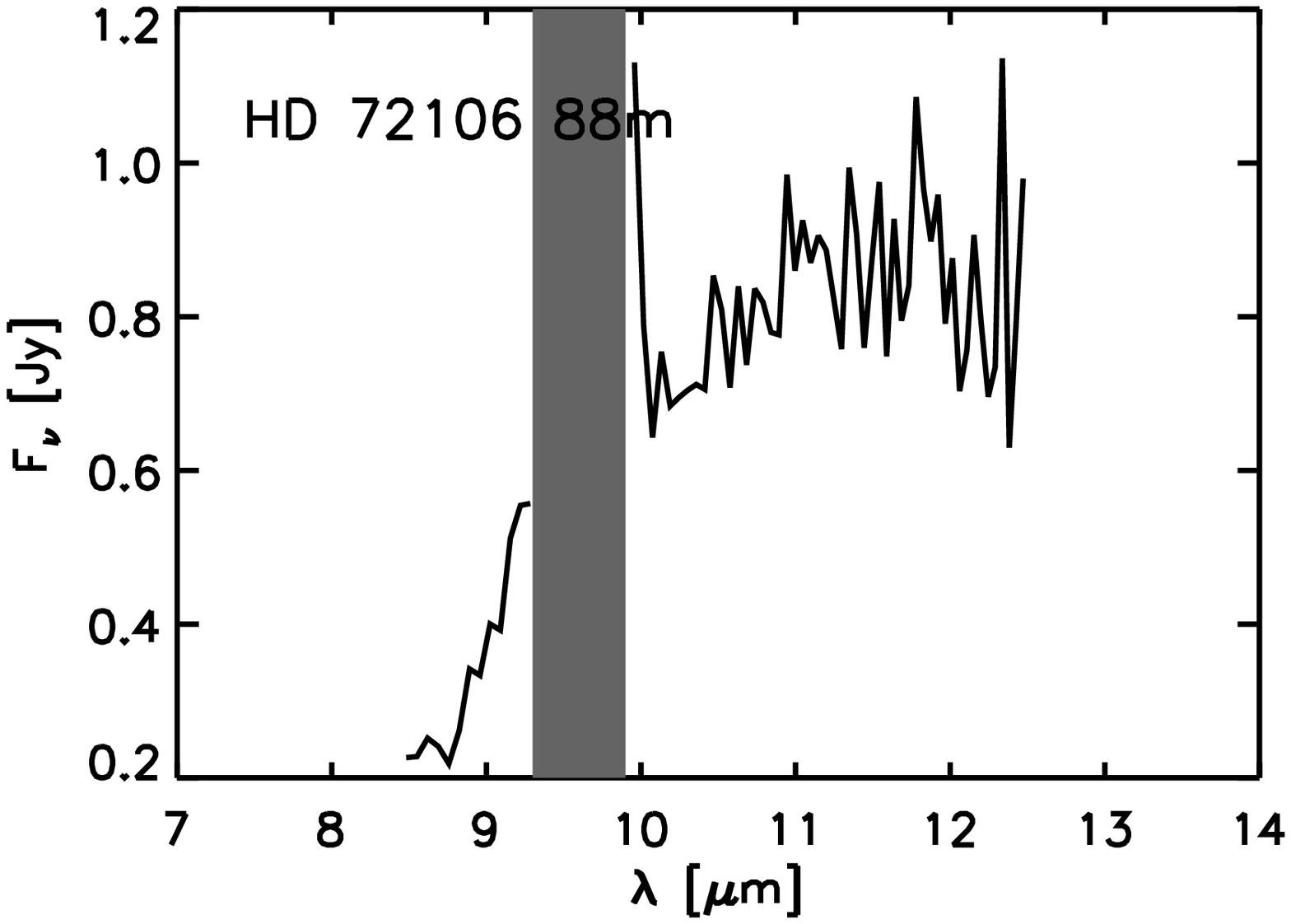}
    \includegraphics[scale=0.29]{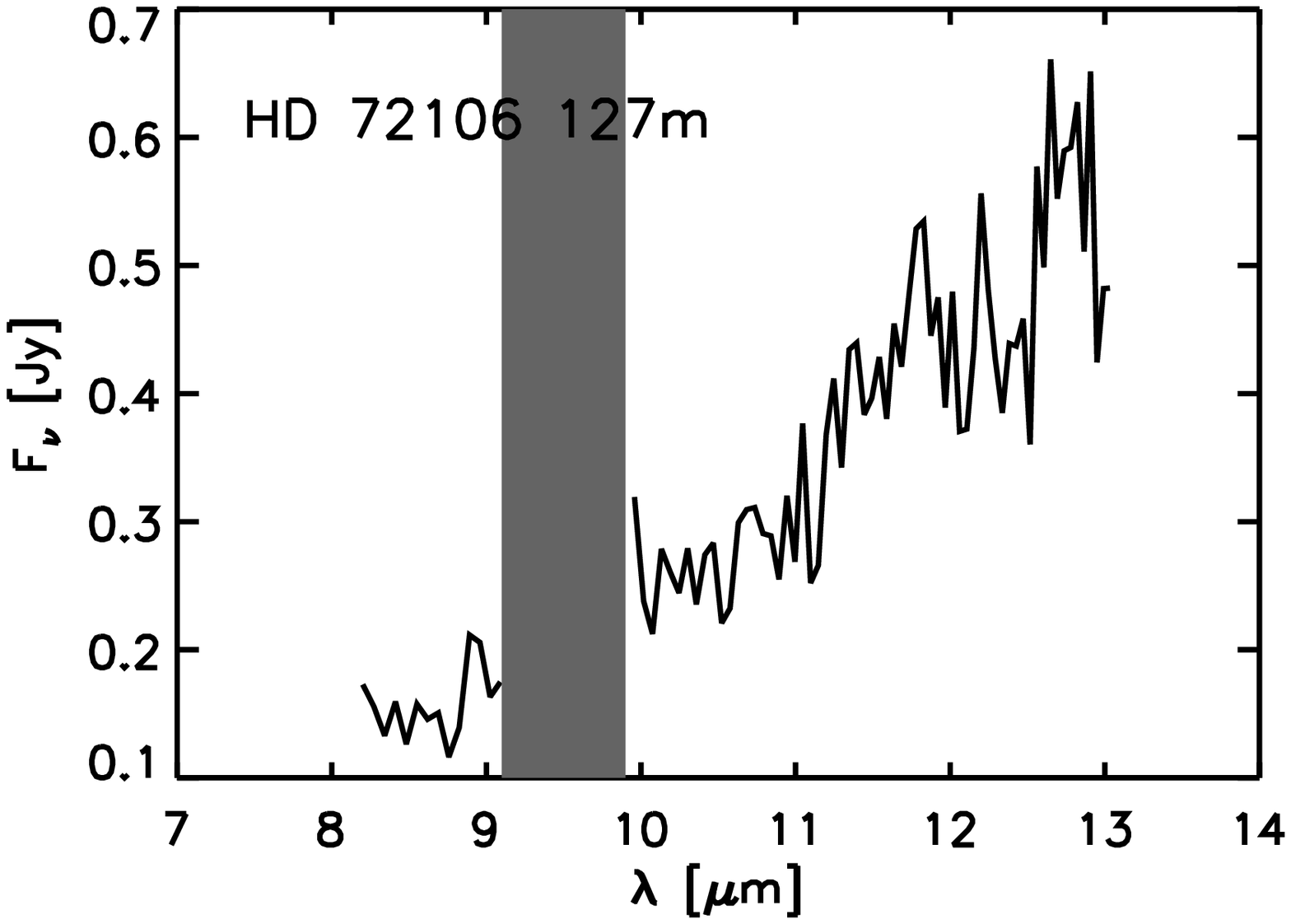}\newline
    \includegraphics[scale=0.29]{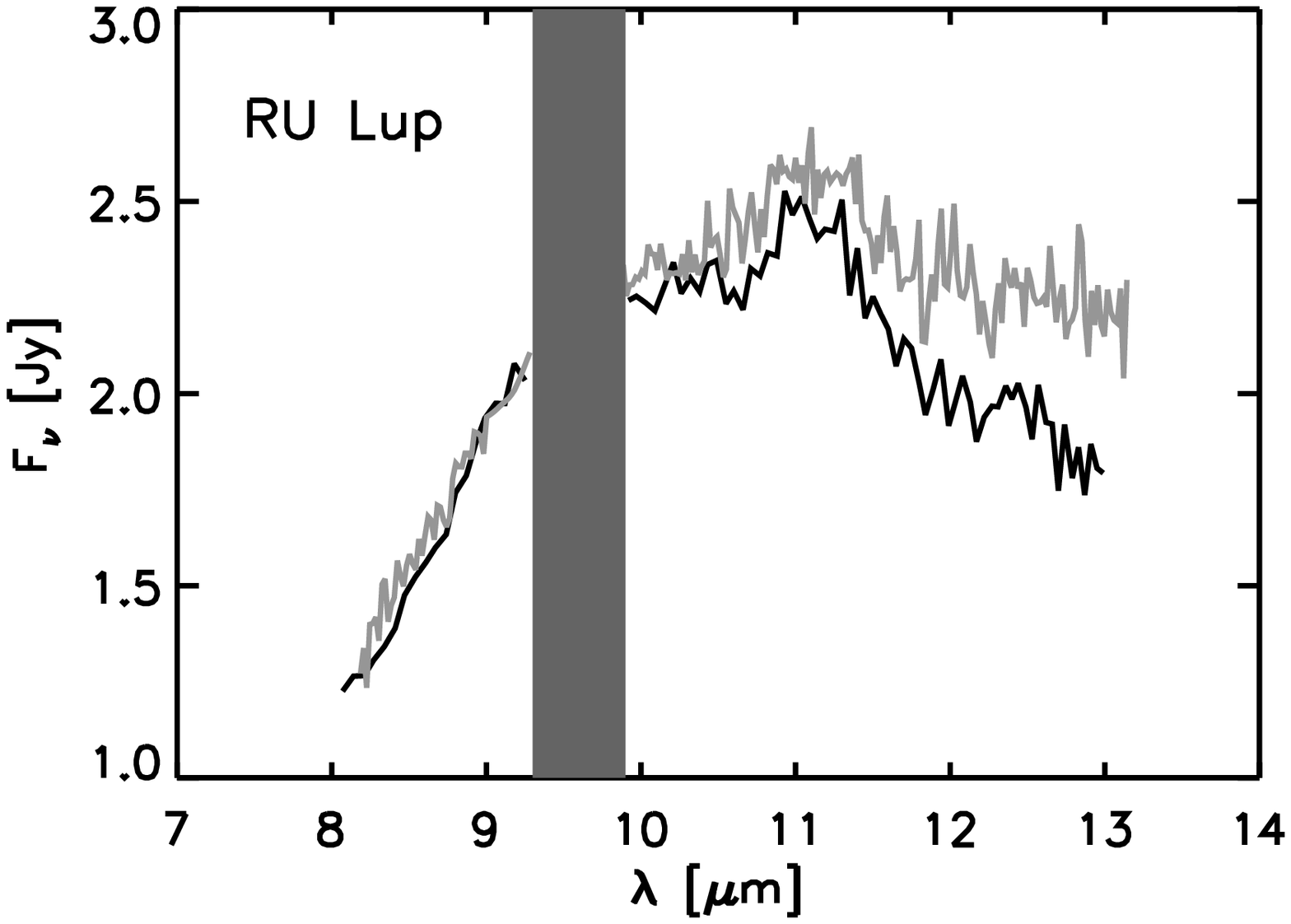}
    \includegraphics[scale=0.29]{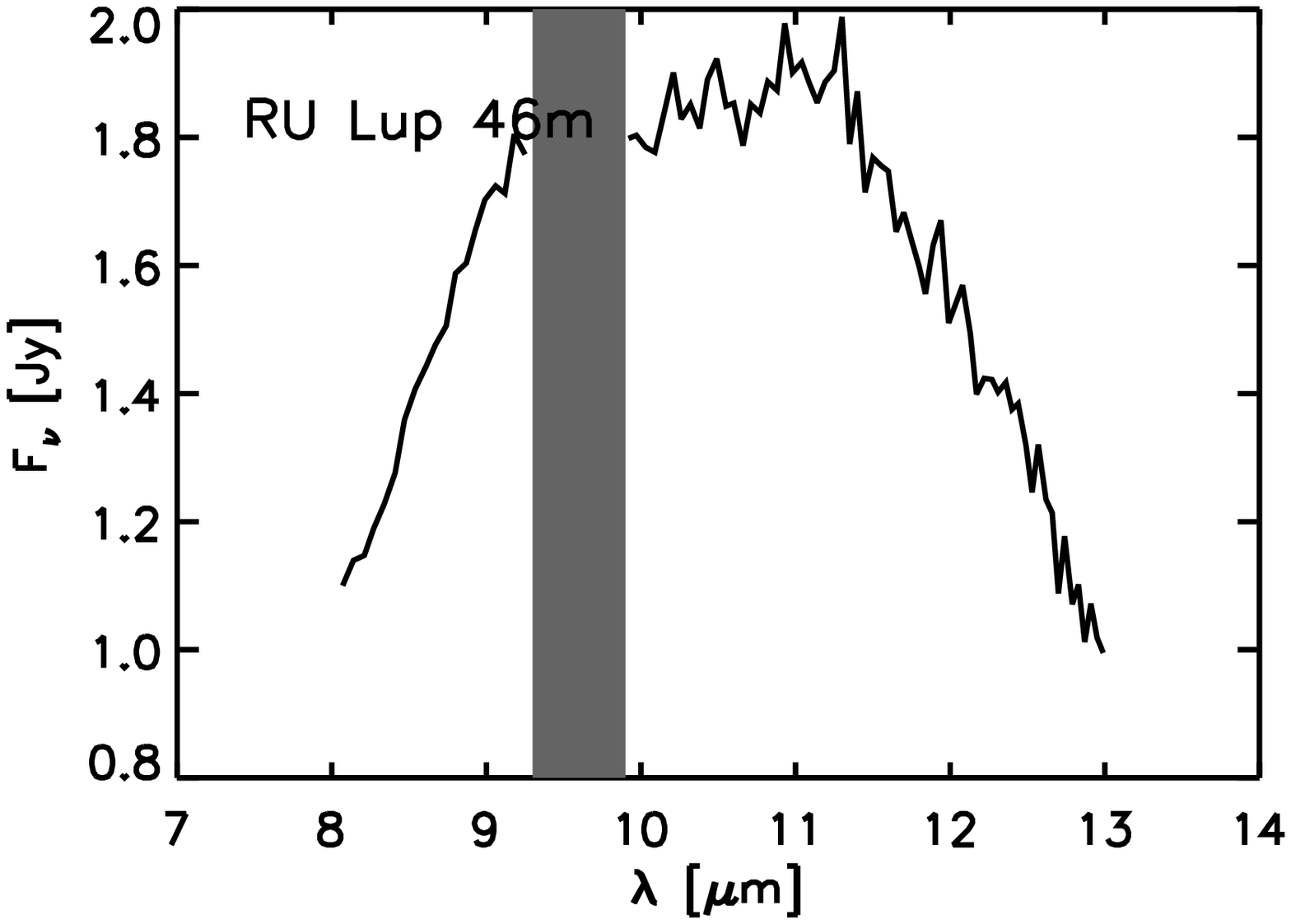}
    \includegraphics[scale=0.29]{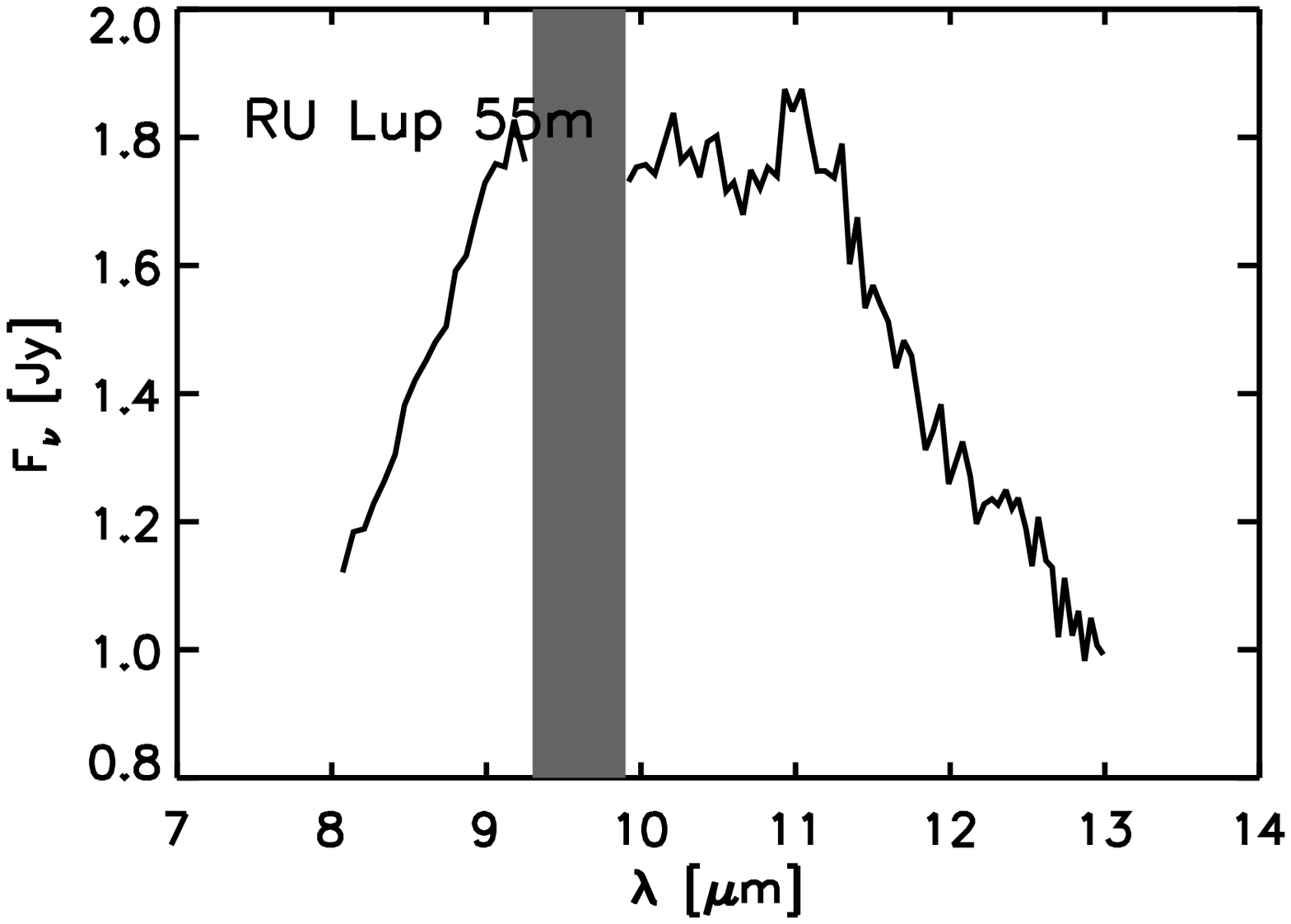}\newline
    \includegraphics[scale=0.29]{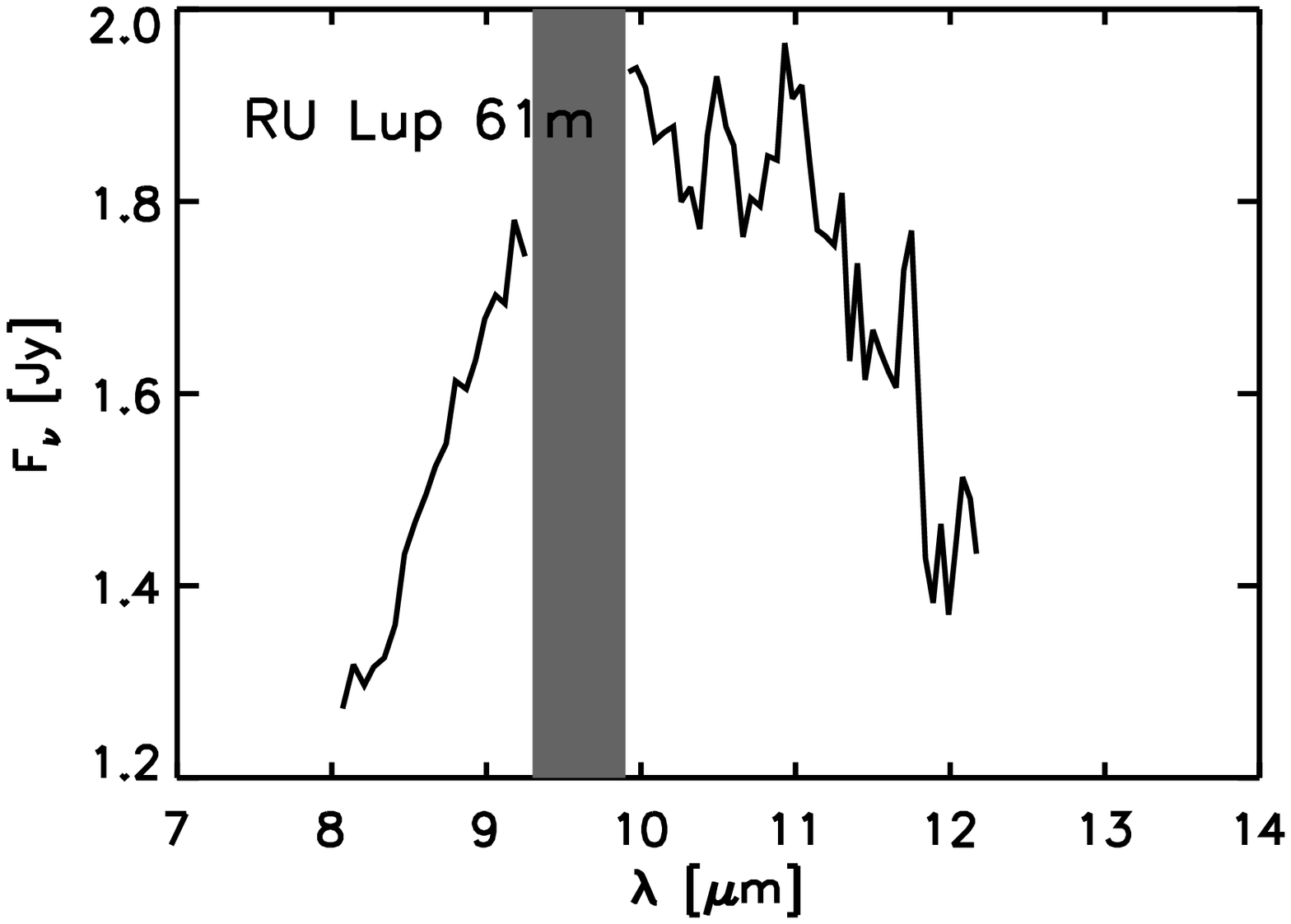}
    \includegraphics[scale=0.29]{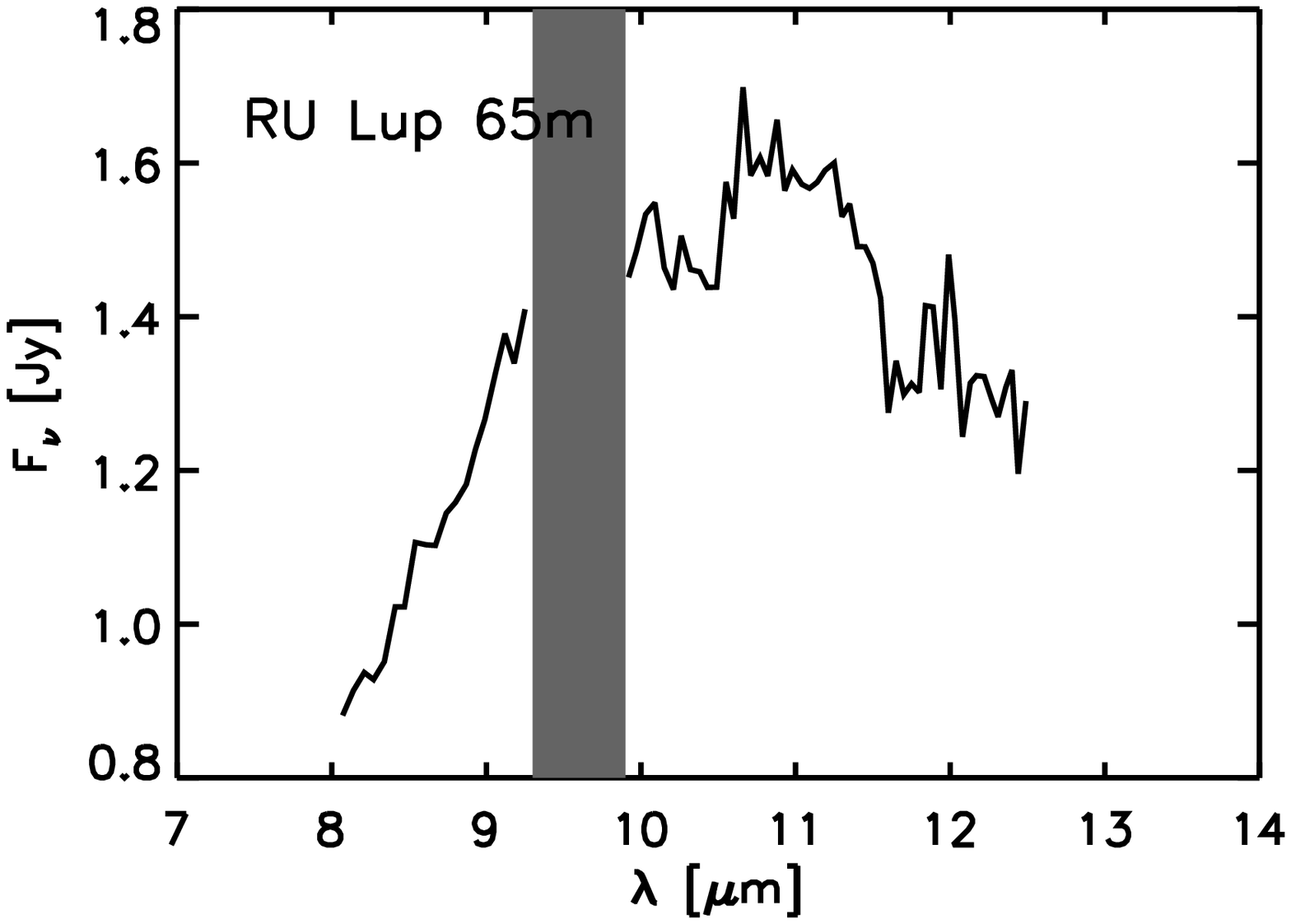}\newline
    \caption{Uncorrelated and correlated N band spectra obtained from the
      objects that were observed with MIDI  
      for this study (black lines). {\it Correlated} spectra are designated by
      the additional information about the baseline length $B$
      (Table~\ref{table:observation-midisurvey}). 
      Because of the remaining ozone band in  
      several spectra the wavelength interval from $\sim$$9.3\,\mathrm{\mu m}$
      up to $\sim$$9.9\,\mathrm{\mu m}$ is cut and underlaid with gray color. 
      The gray lines represent data formerly obtained with TIMMI\,2
      (Przygodda et al.~\cite{przygodda}; Schegerer et
      al.~\cite{schegererI}).}
    \label{figure:midi-10umfit}
  \end{figure*} 
  \begin{figure*}[!tb]
    \centering
    \includegraphics[scale=0.29]{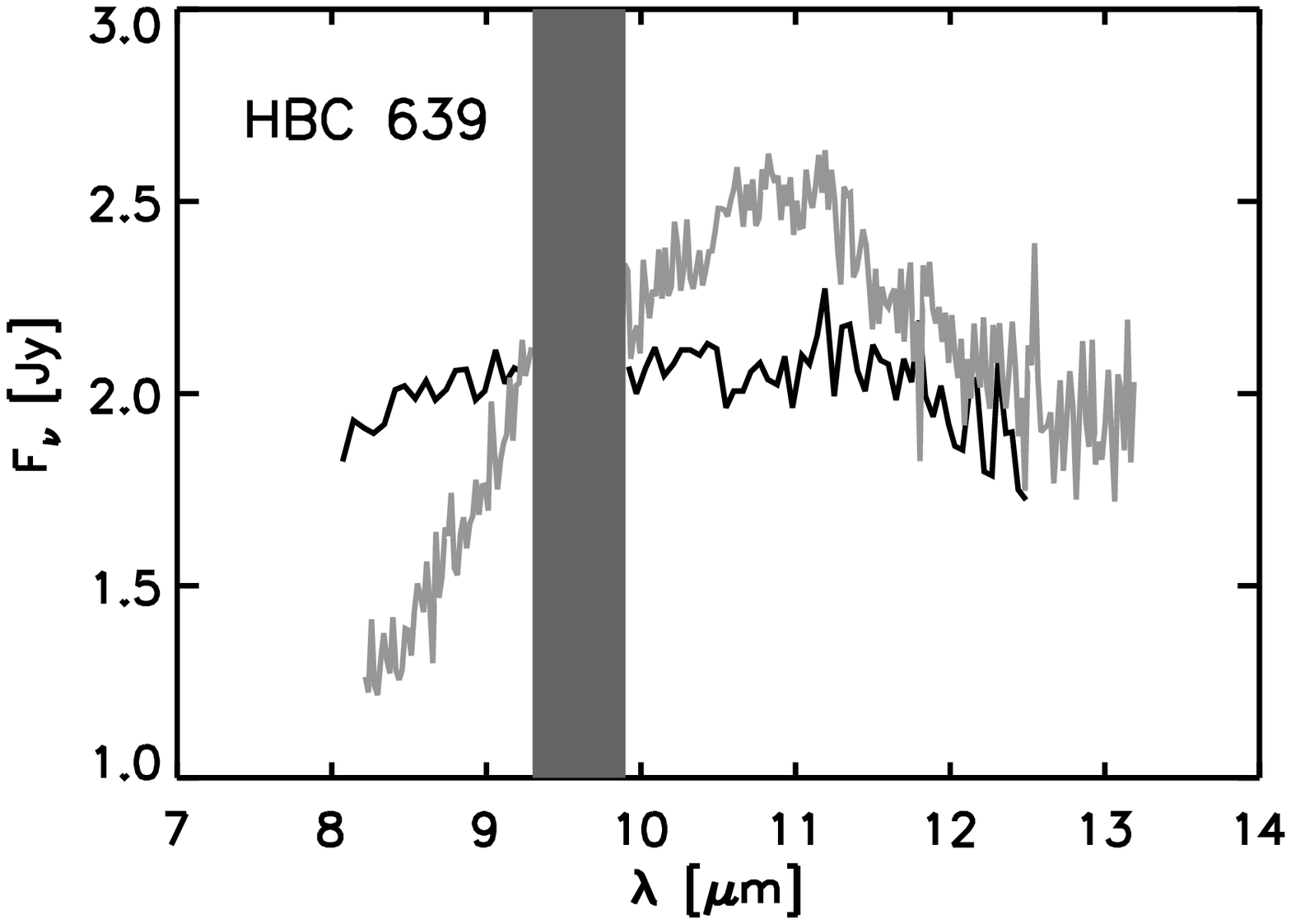}
    \includegraphics[scale=0.29]{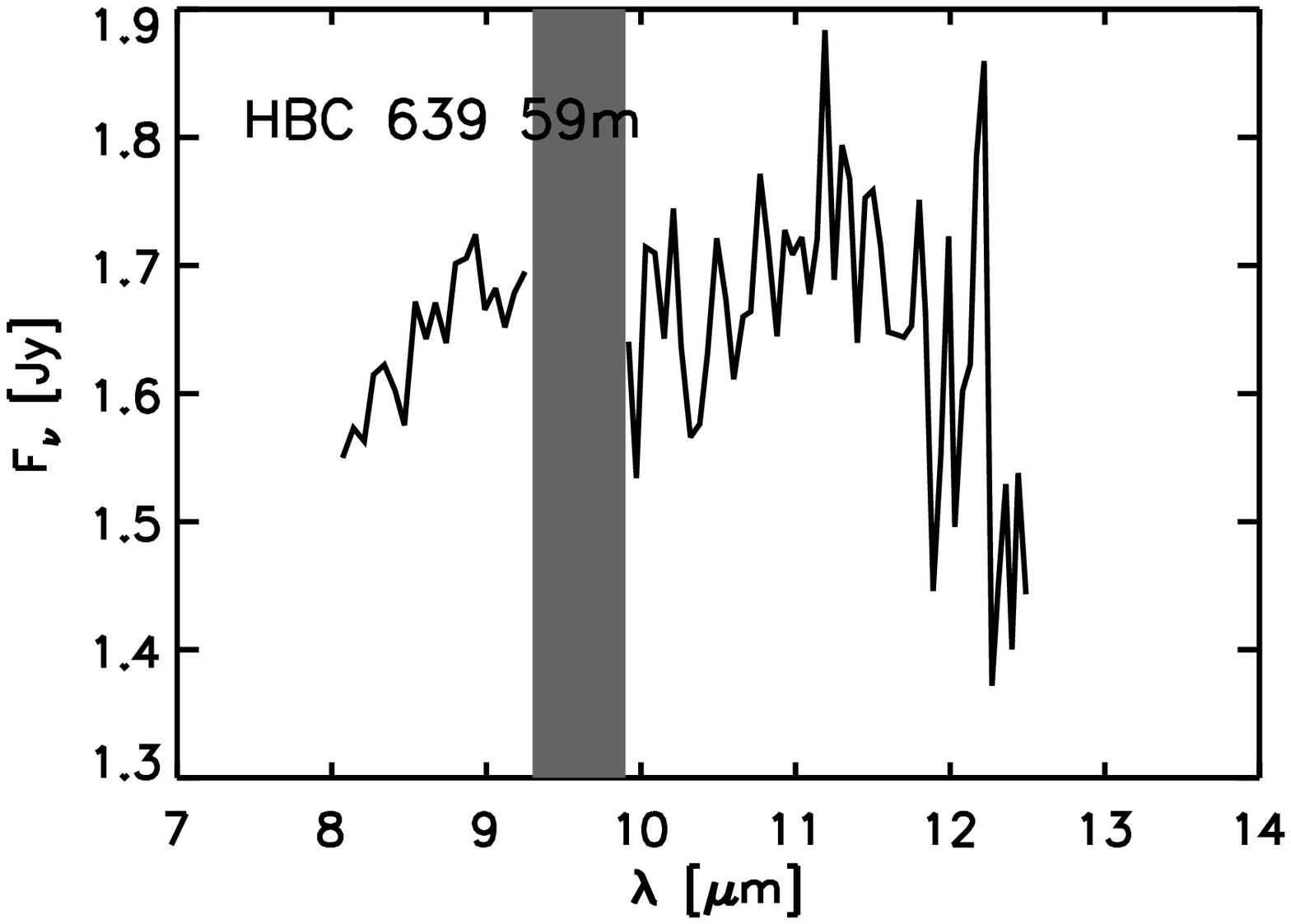}\newline
    \includegraphics[scale=0.29]{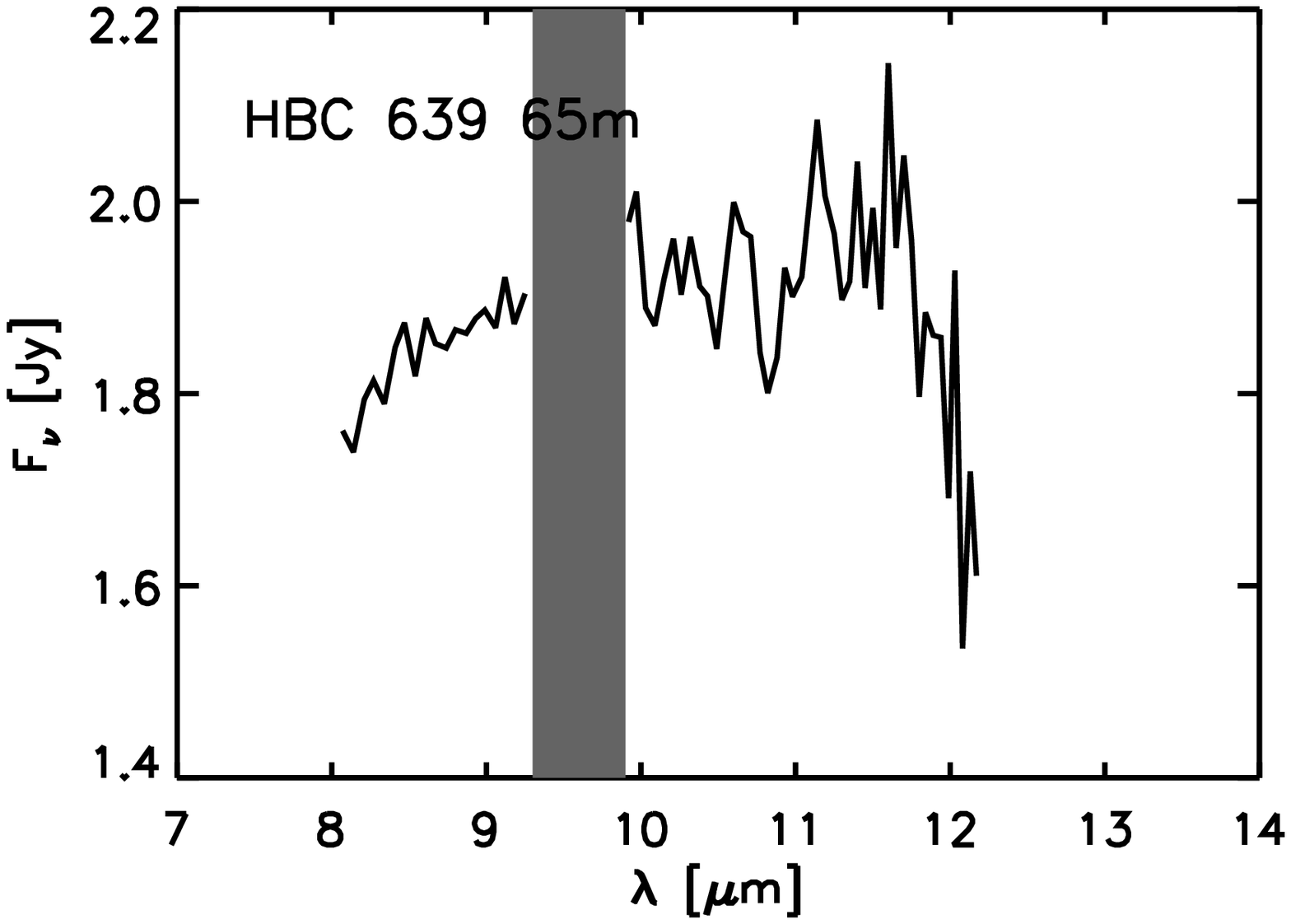}
    \includegraphics[scale=0.29]{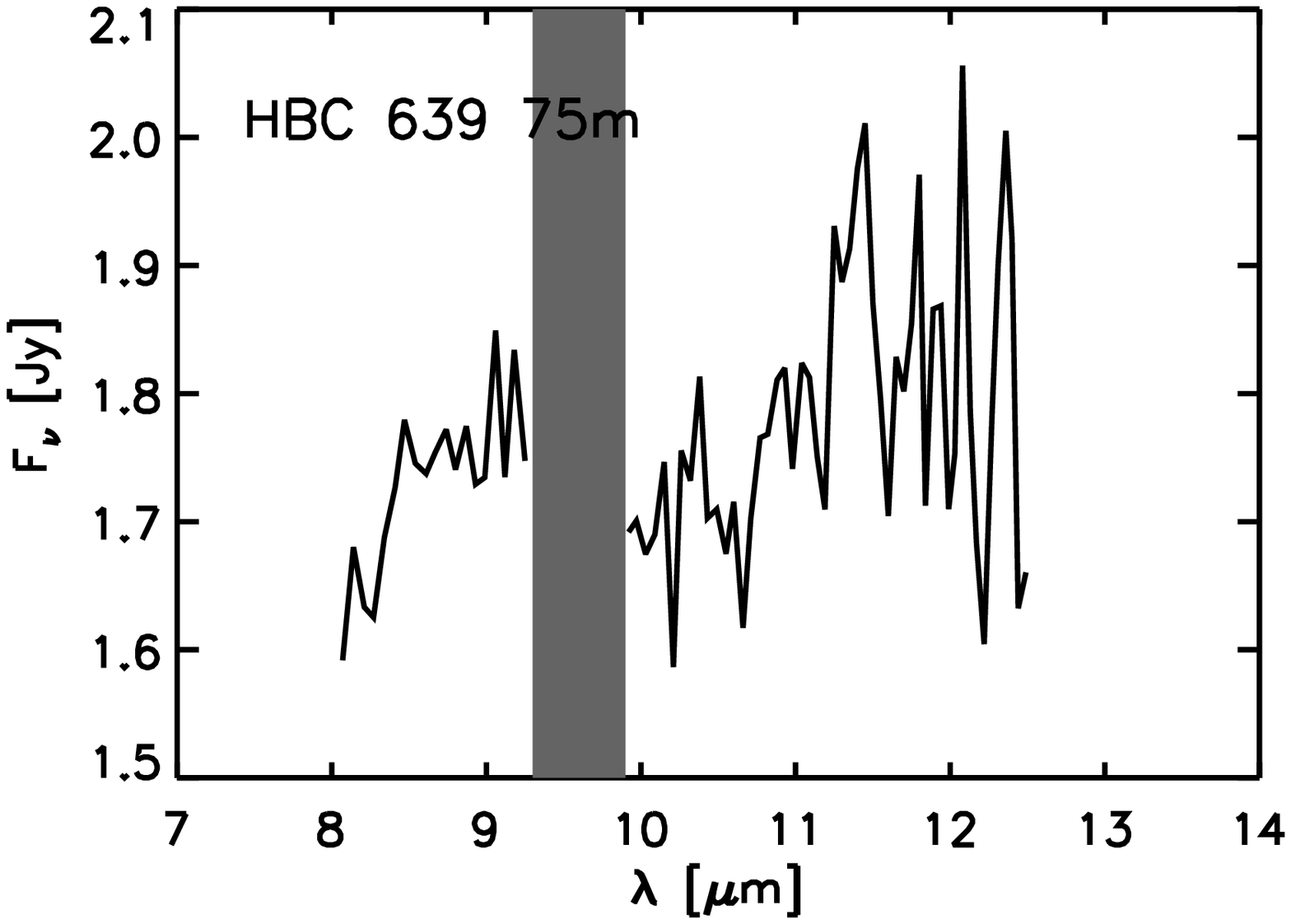}\newline
    \includegraphics[scale=0.29]{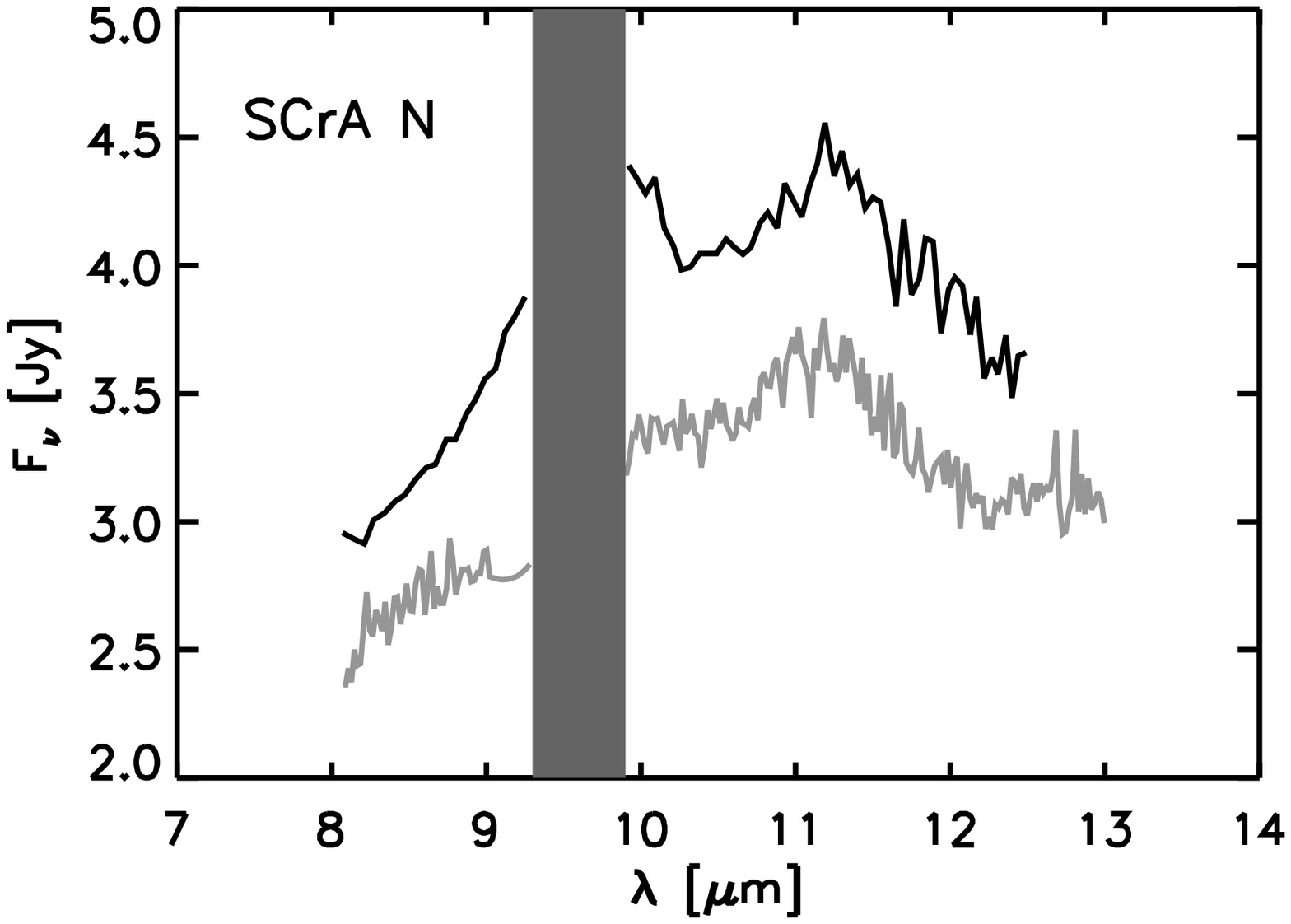}
    \includegraphics[scale=0.29]{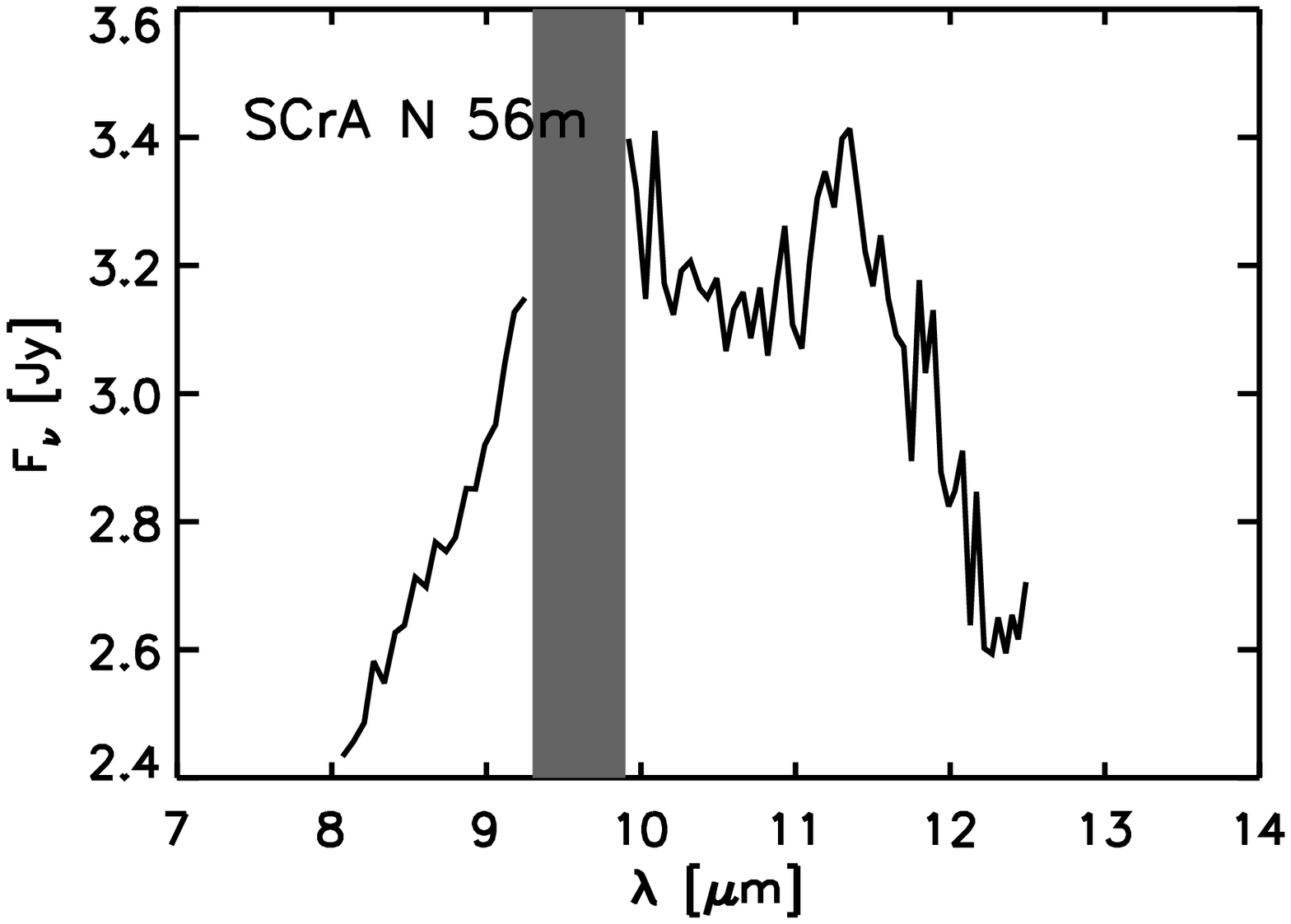}
    \includegraphics[scale=0.29]{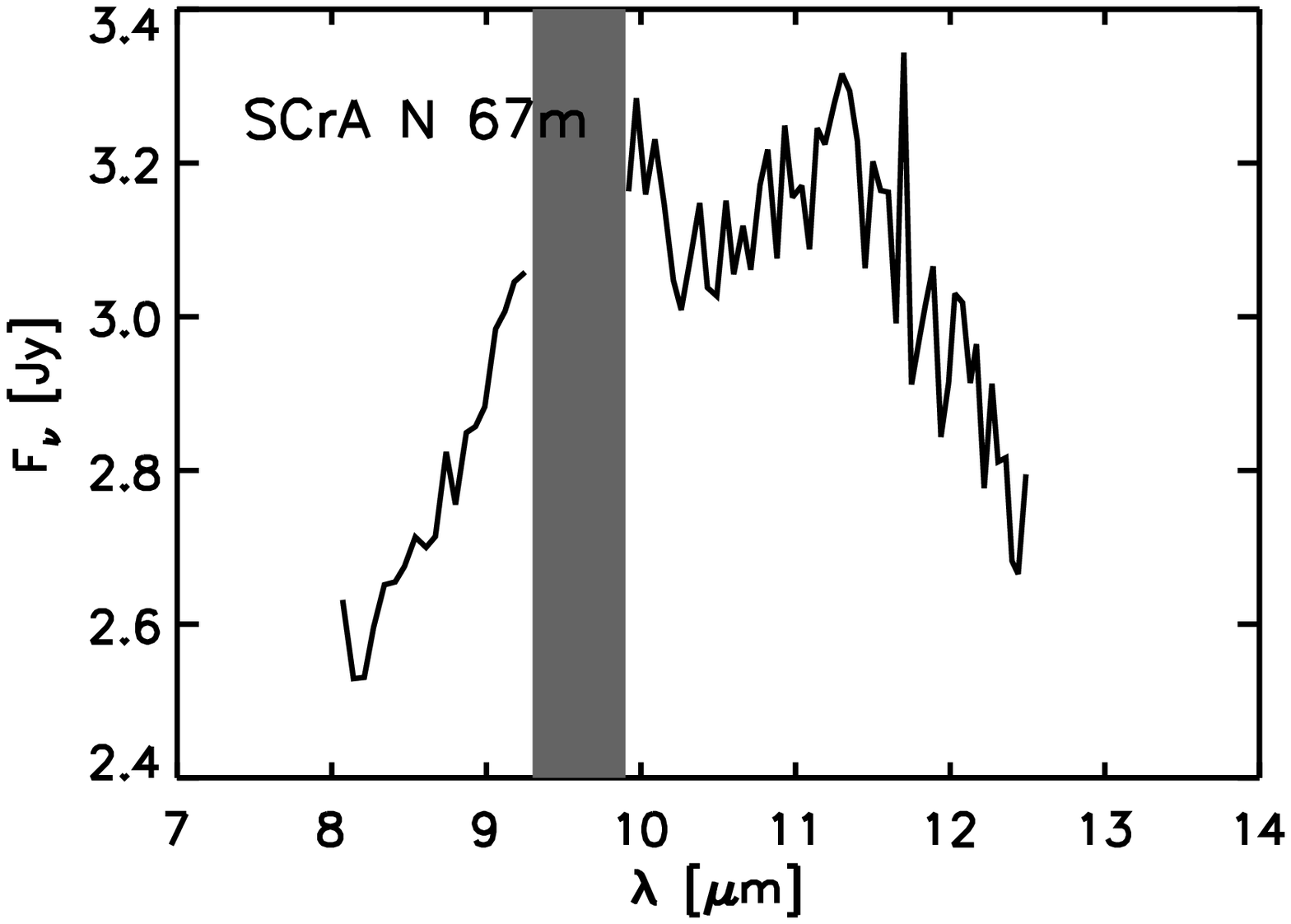}\newline
    \includegraphics[scale=0.29]{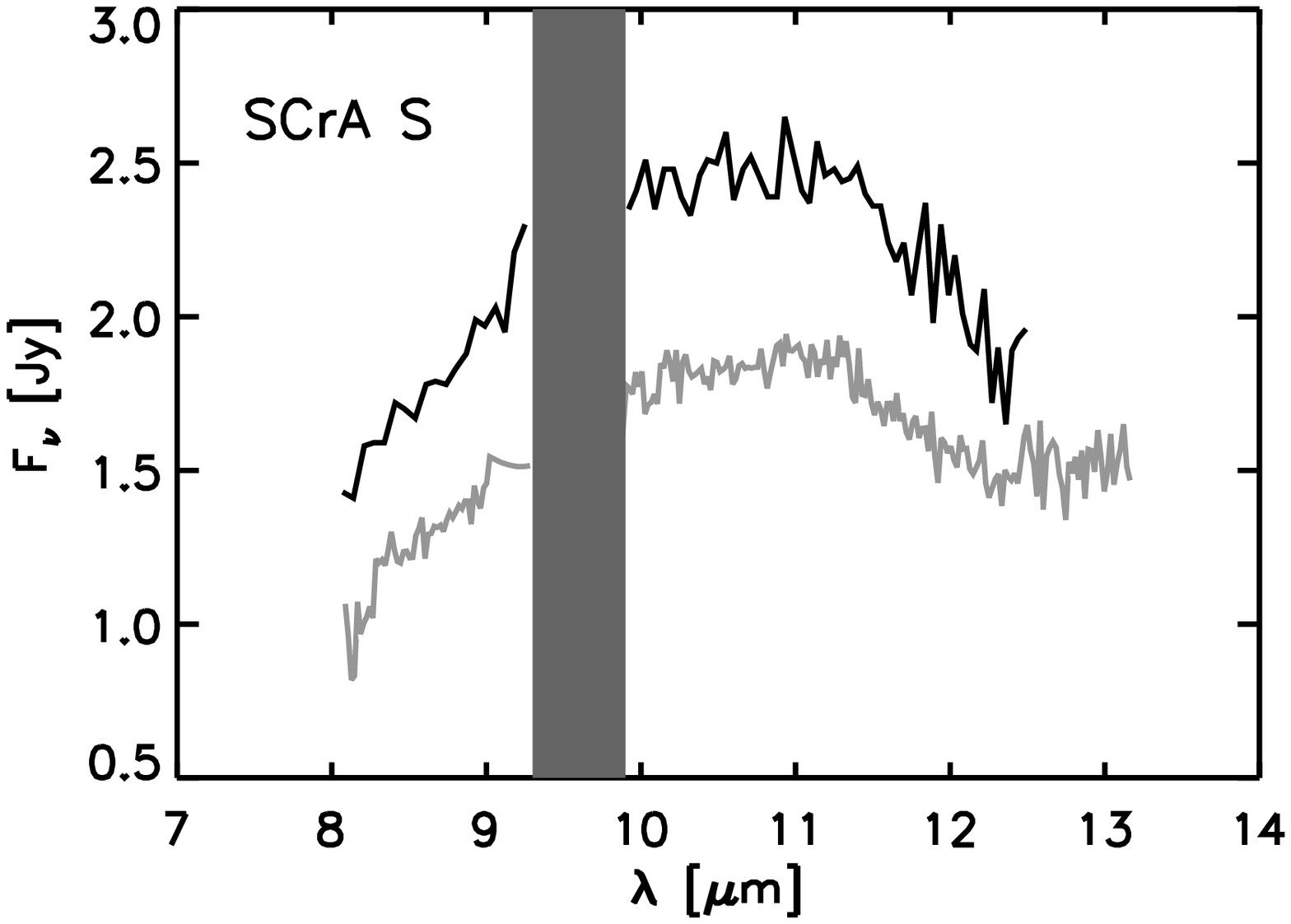}
    \includegraphics[scale=0.29]{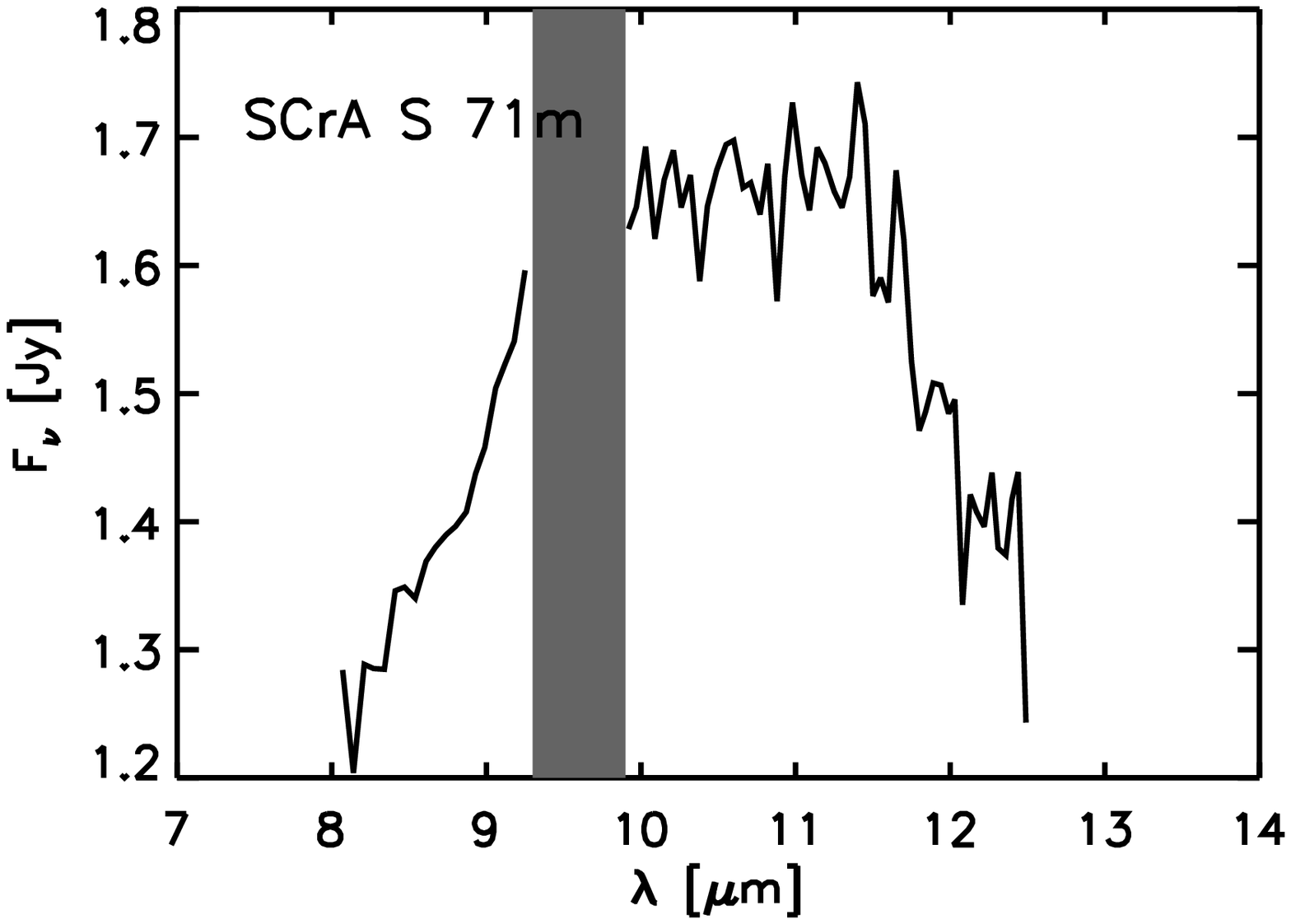}
    \caption{Continuation of Fig.~\ref{figure:midi-10umfit}.}
    \label{figure:midi-10umfit2}
  \end{figure*} 

  \begin{figure*}[!tb]
    \centering
    \includegraphics[scale=0.29]{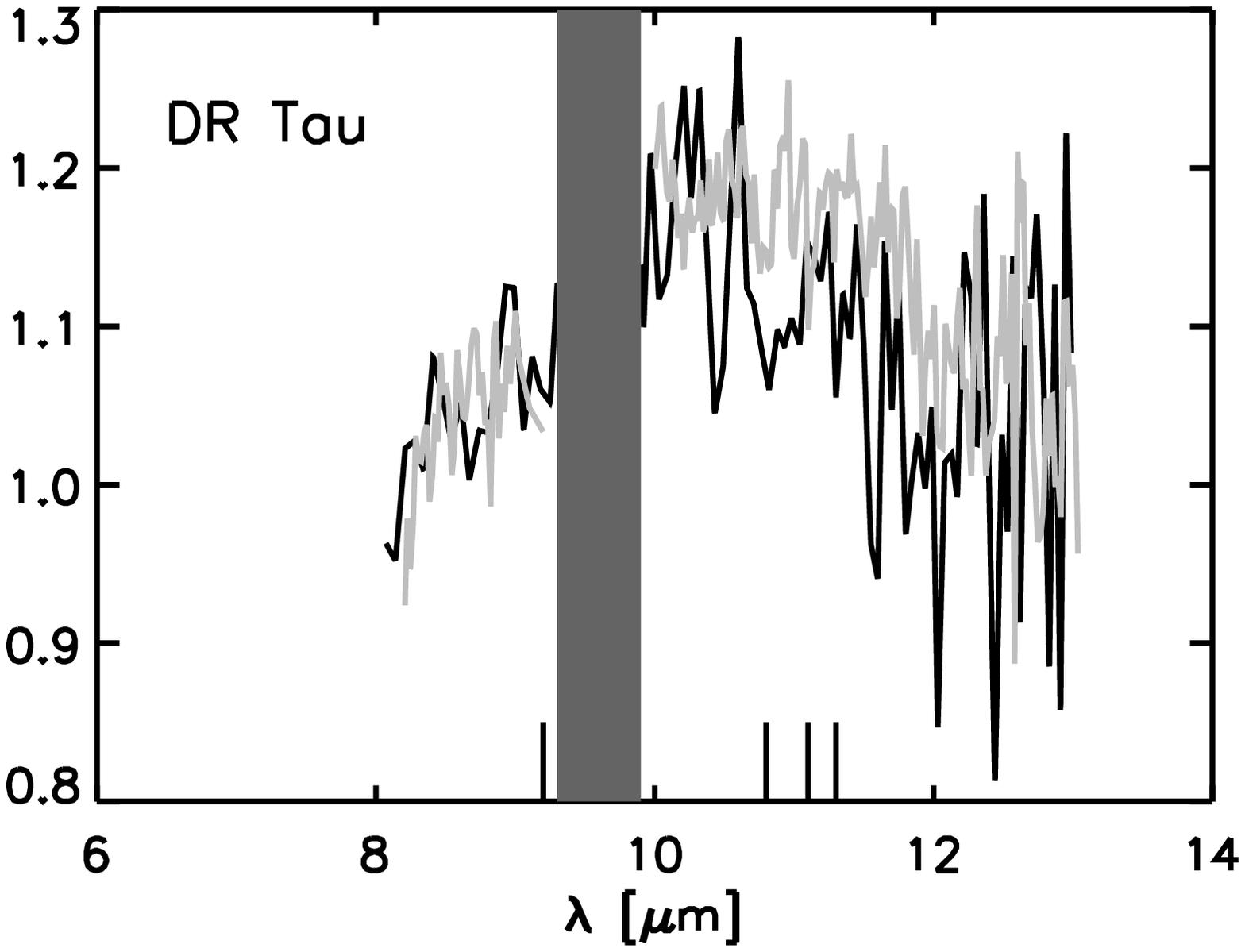}
    \includegraphics[scale=0.29]{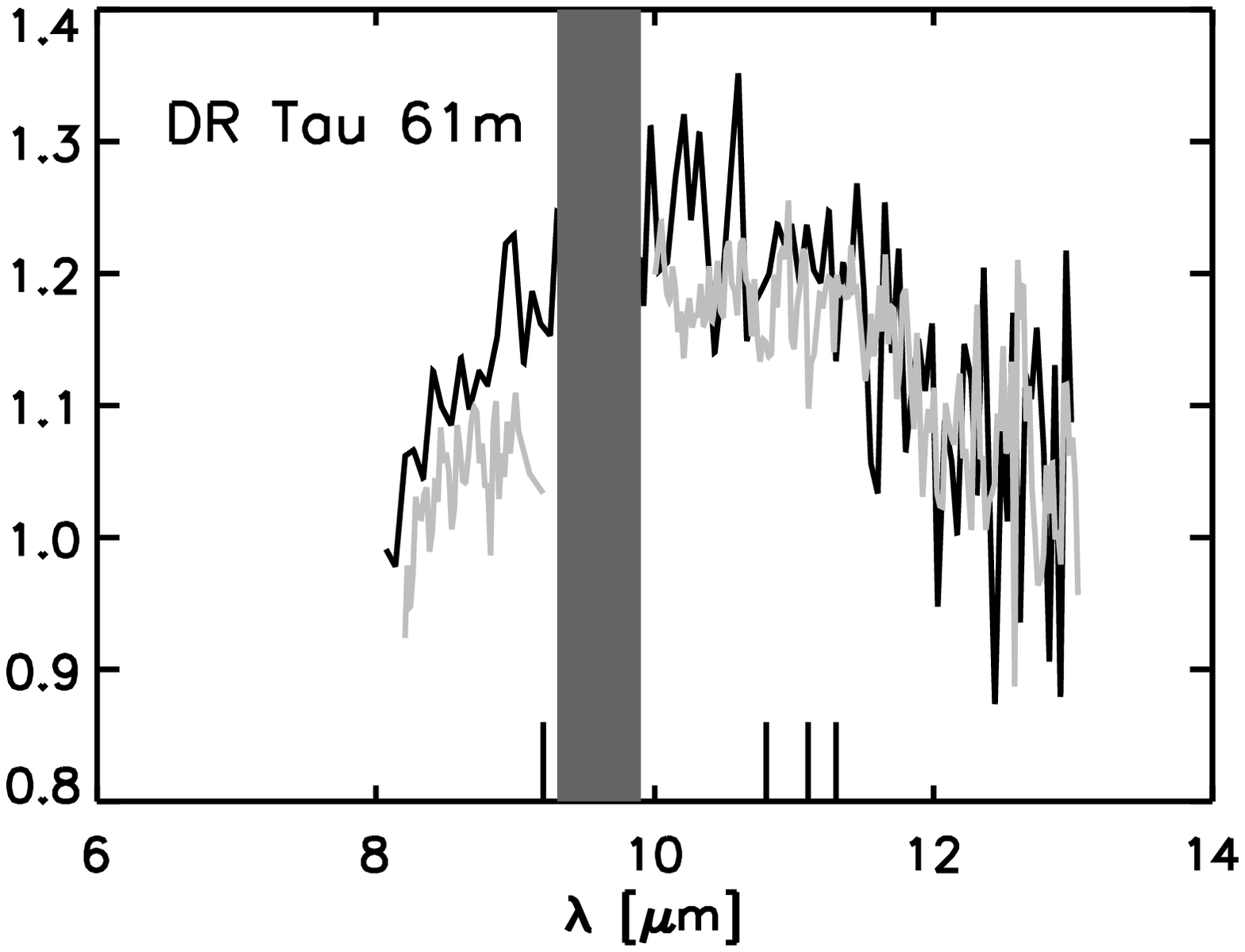}\newline
    \includegraphics[scale=0.29]{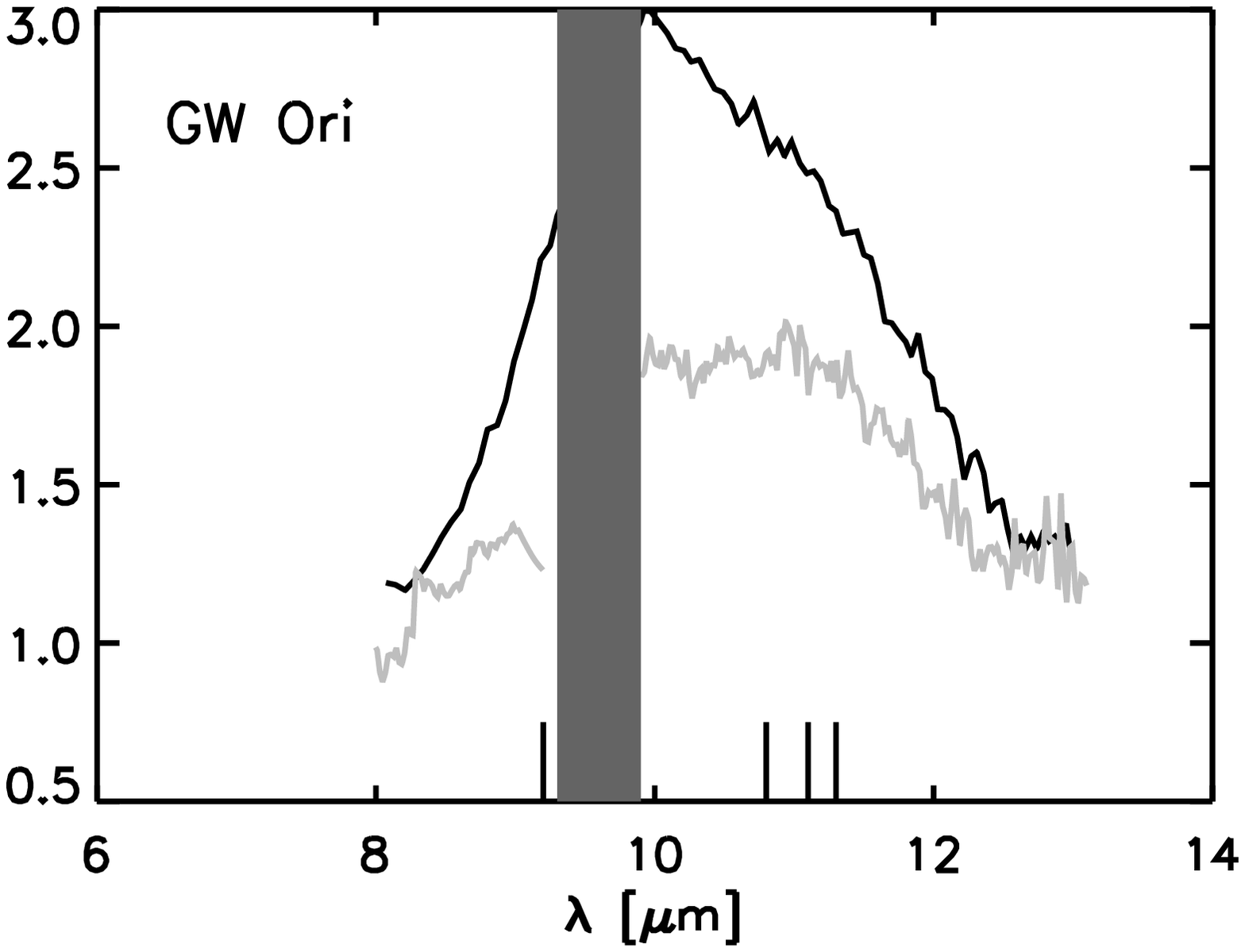}
    \includegraphics[scale=0.29]{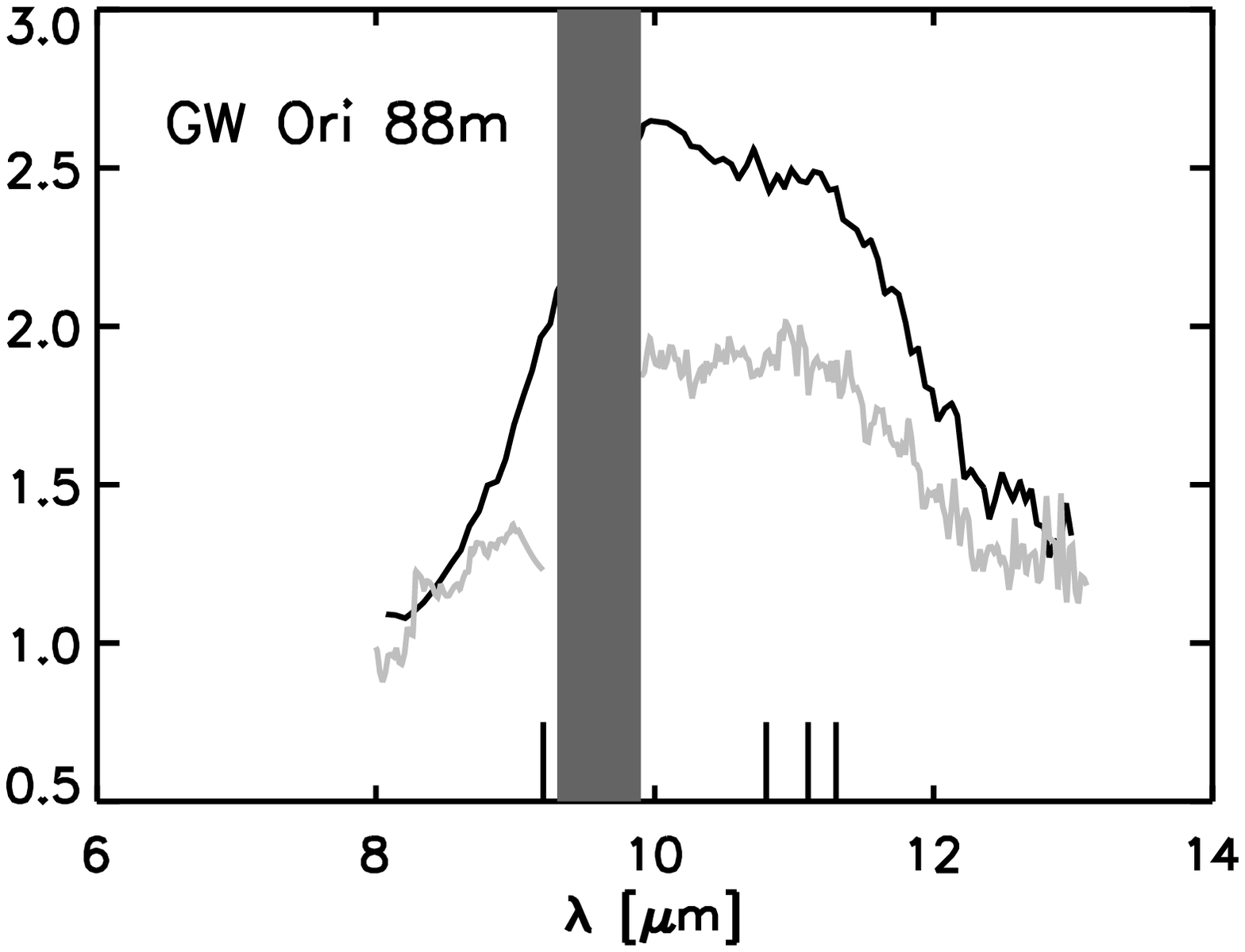}\newline
    \includegraphics[scale=0.29]{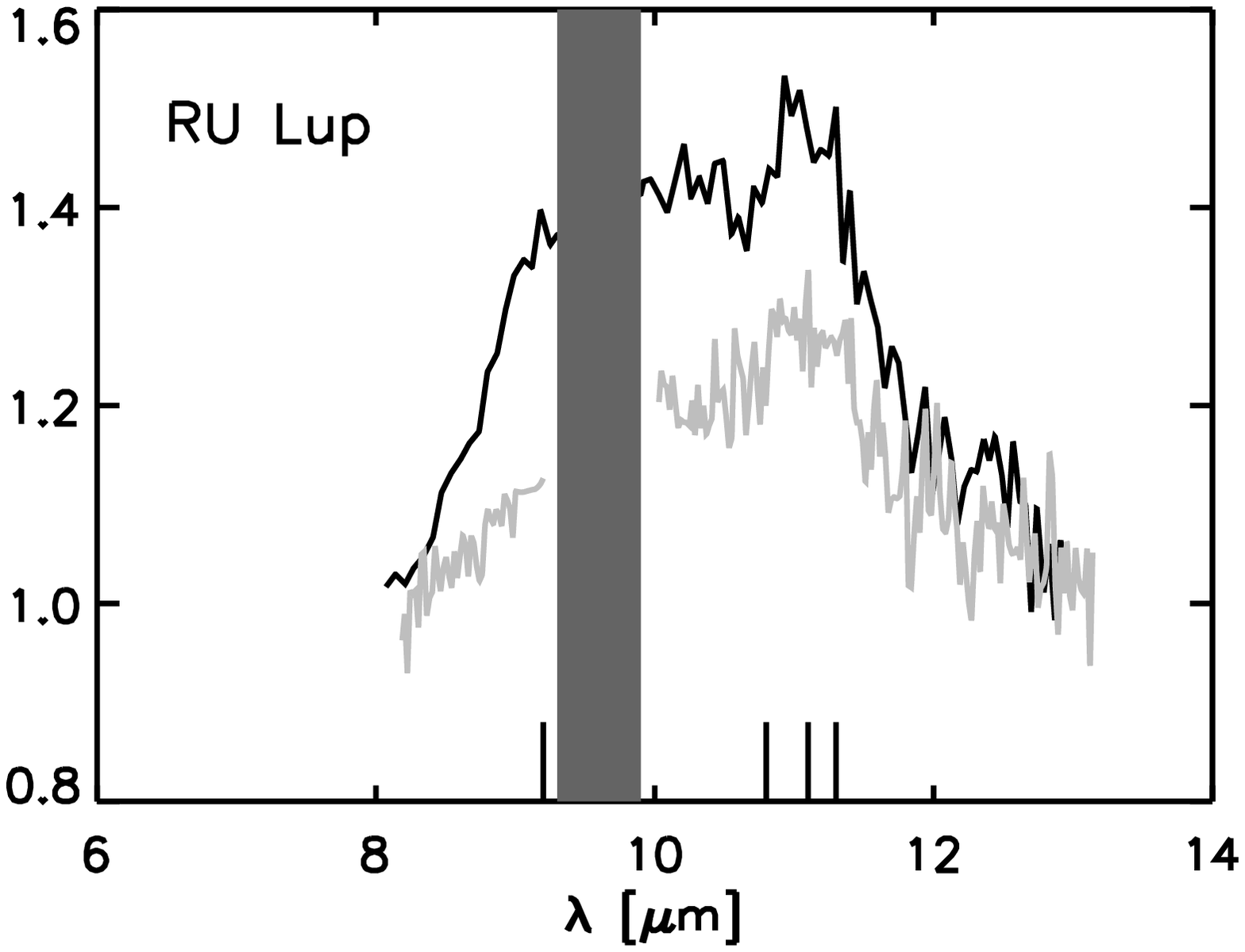}
    \includegraphics[scale=0.29]{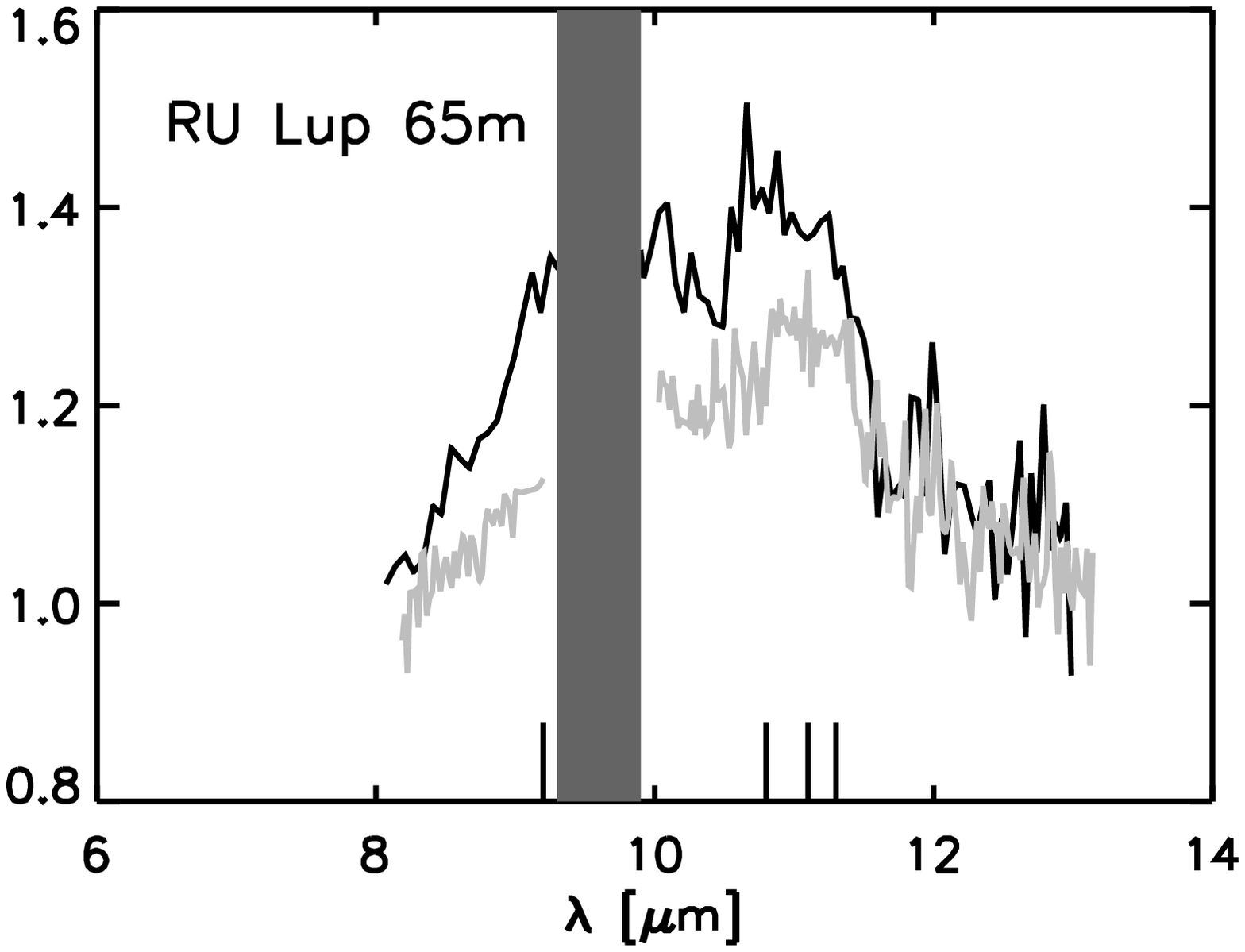}\newline
    \caption{Uncorrelated and correlated MIDI spectra after subtraction of a
      black body that represents the underlying continuum and is fitted to
      the extended wings of the emission features (black lines). The gray
      curves represent TIMMI\,2 data (Przygodda et al.~\cite{przygodda}) after
      subtraction of a black body {\it and} the spectral contribution of small
      amorphous dust grains found in the preceeding study of
      Schegerer et al.~(\cite{schegererI}). For clarity, we only
      show the results obtained for the single-dish spectra and the
      correlated spectra that were obtained for the longest effective baseline
      length. The vertical lines designate features of enstatite (at
      $9.2\,\mathrm{\mu m}$, $10.8\,\mathrm{\mu m}$, and $11.1\,\mathrm{\mu m}$) and
      forsterite (at $11.3\,\mathrm{\mu m}$).}
    \label{figure:10um_abzug}
  \end{figure*} 

  \begin{figure*}[!tb]
    \centering
    \includegraphics[scale=0.29]{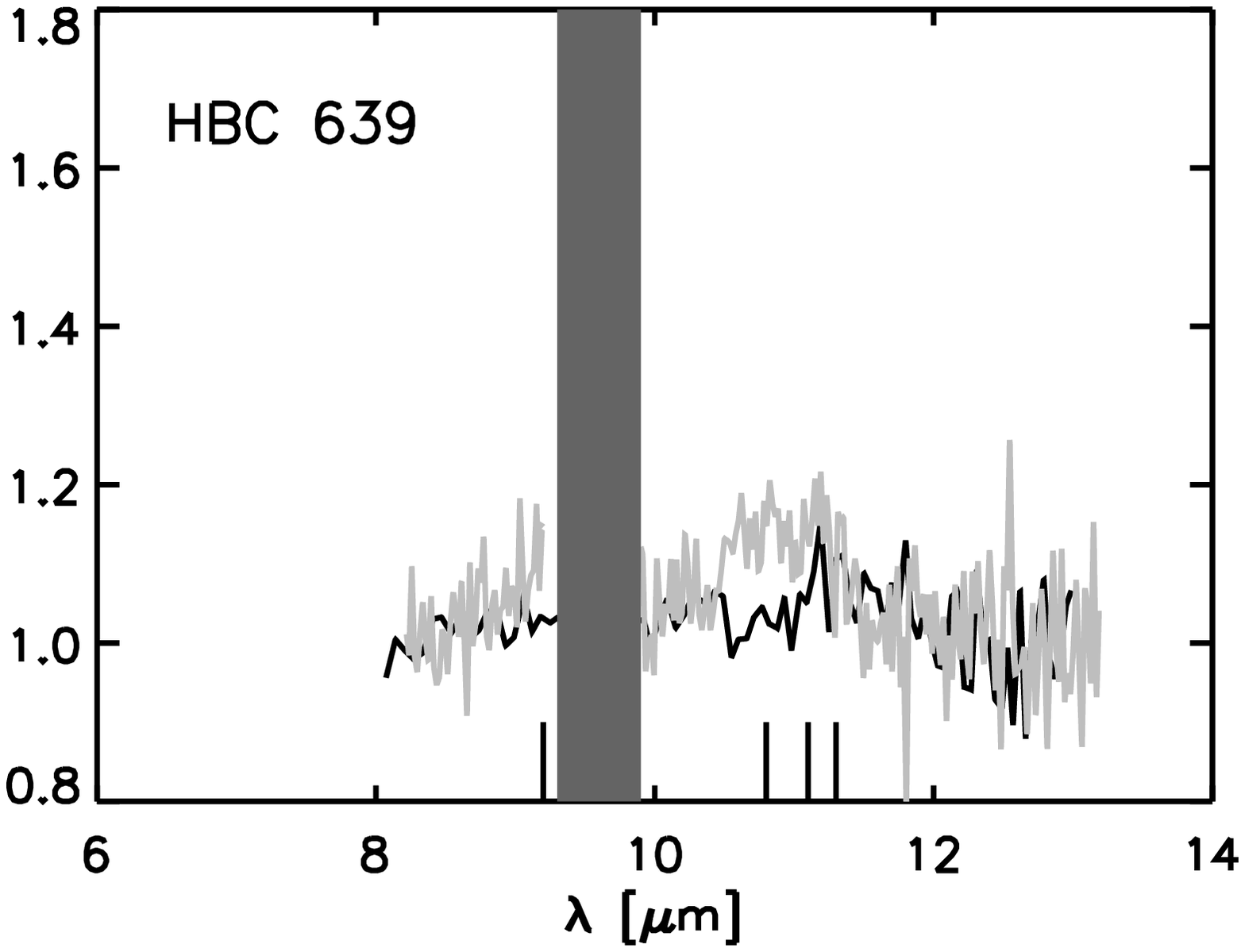}
    \includegraphics[scale=0.29]{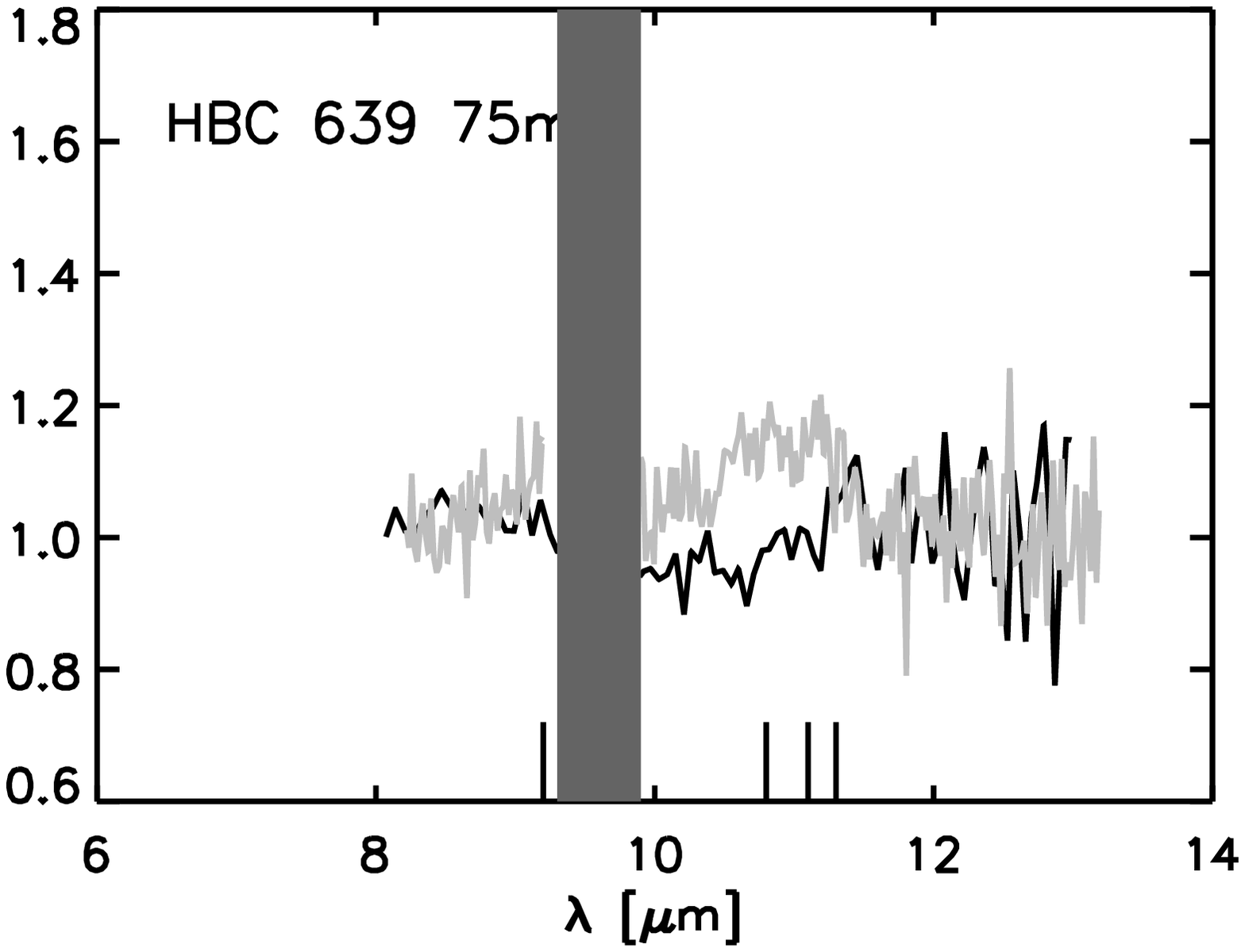}\newline
    \includegraphics[scale=0.29]{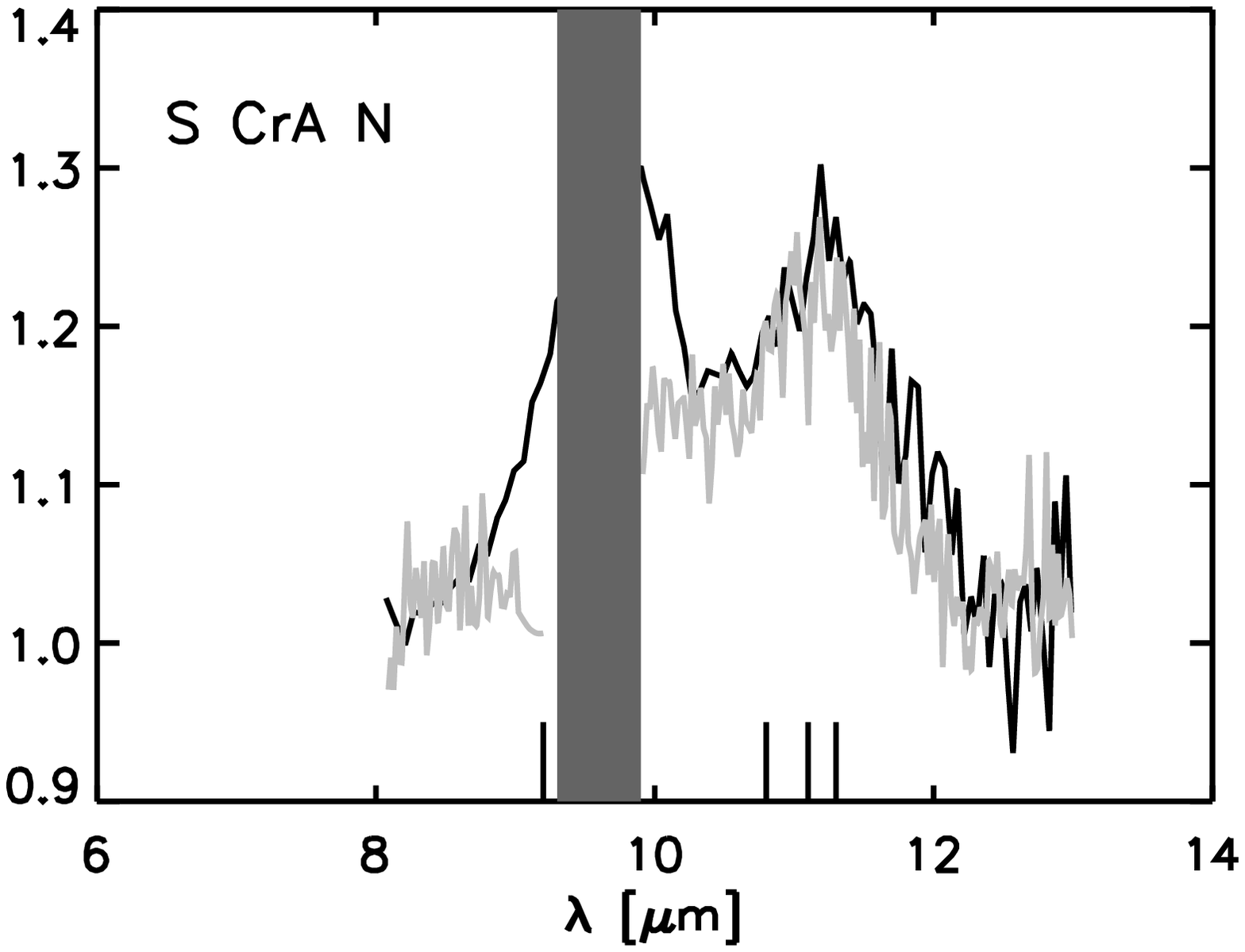}
    \includegraphics[scale=0.29]{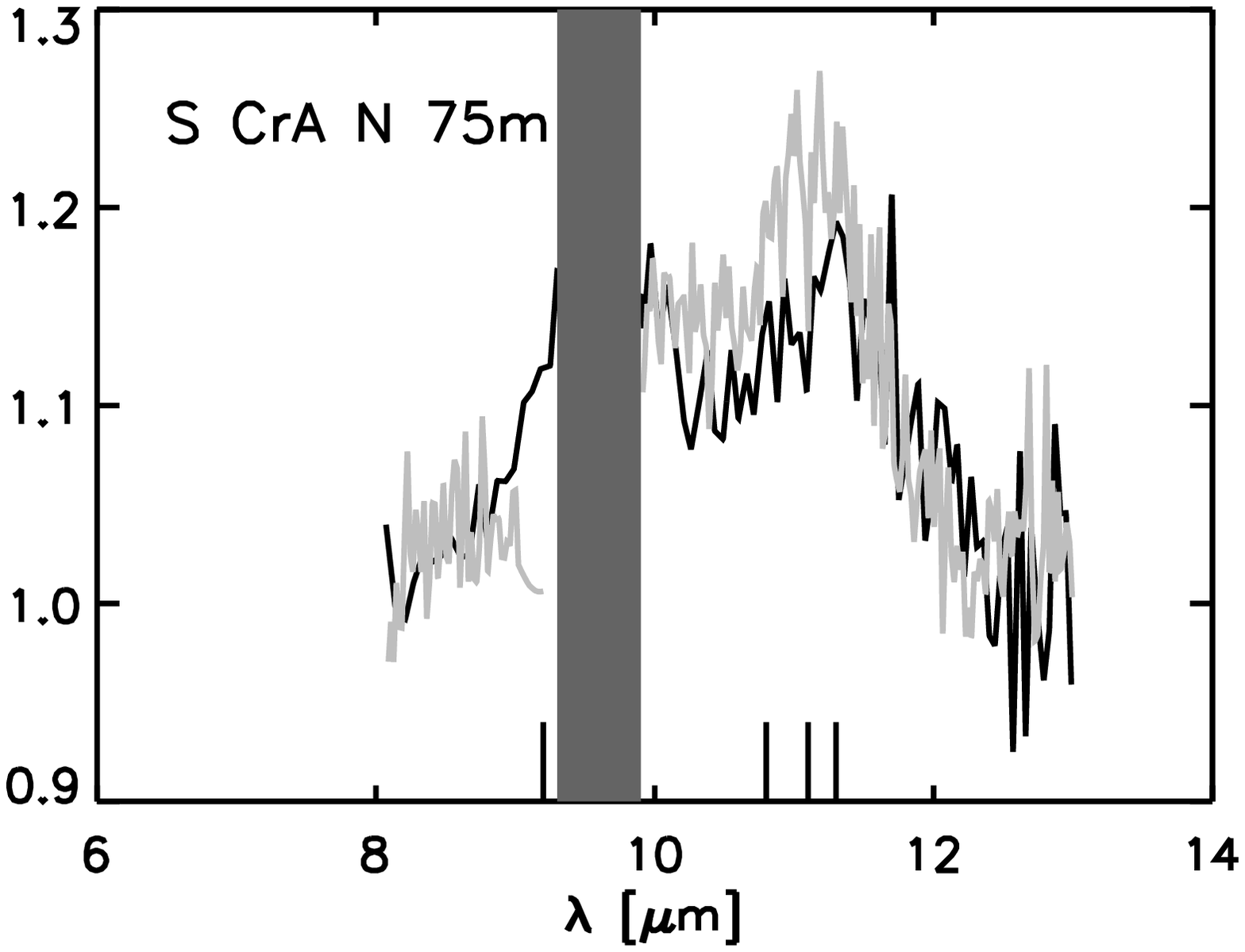}\newline
    \includegraphics[scale=0.29]{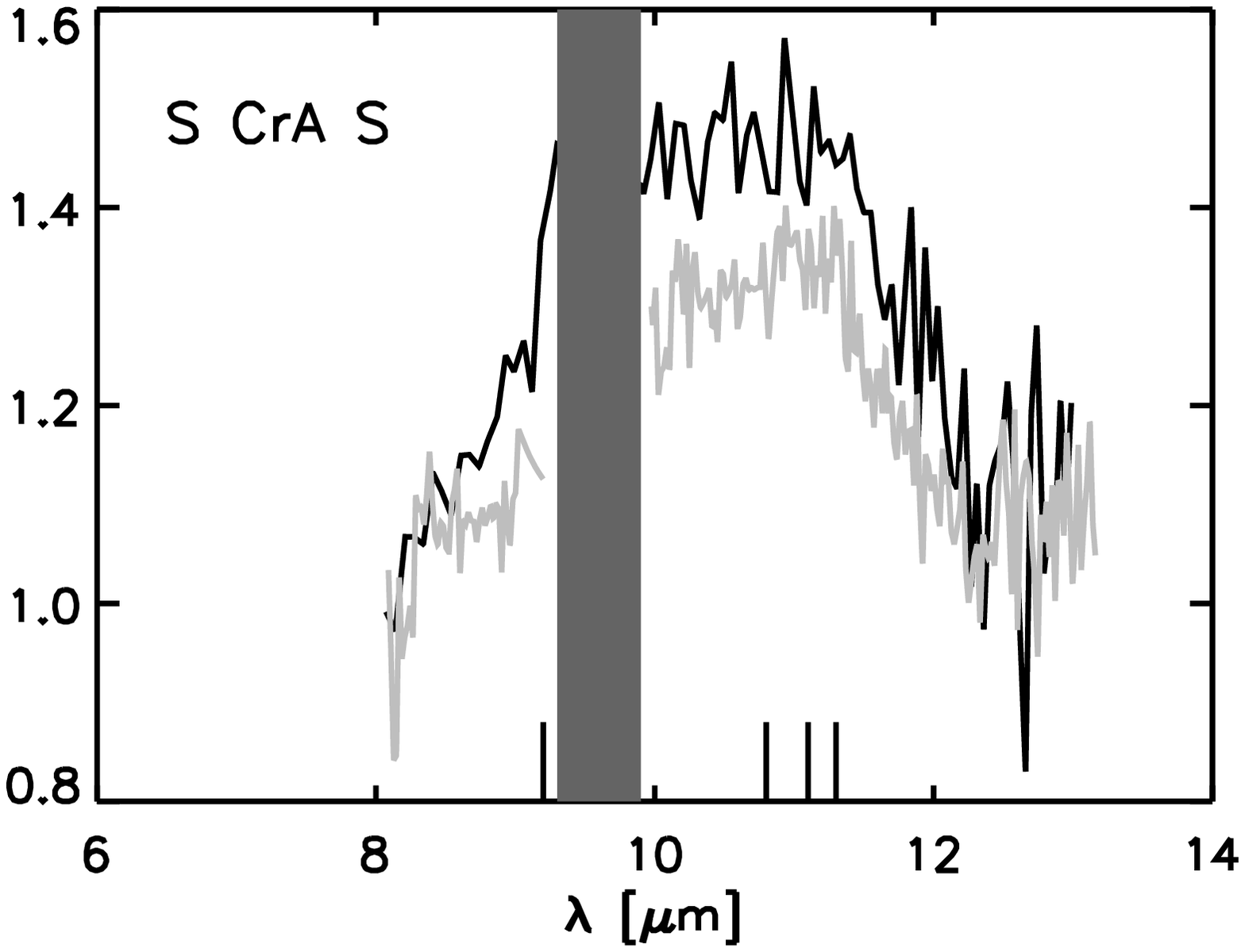}
    \includegraphics[scale=0.29]{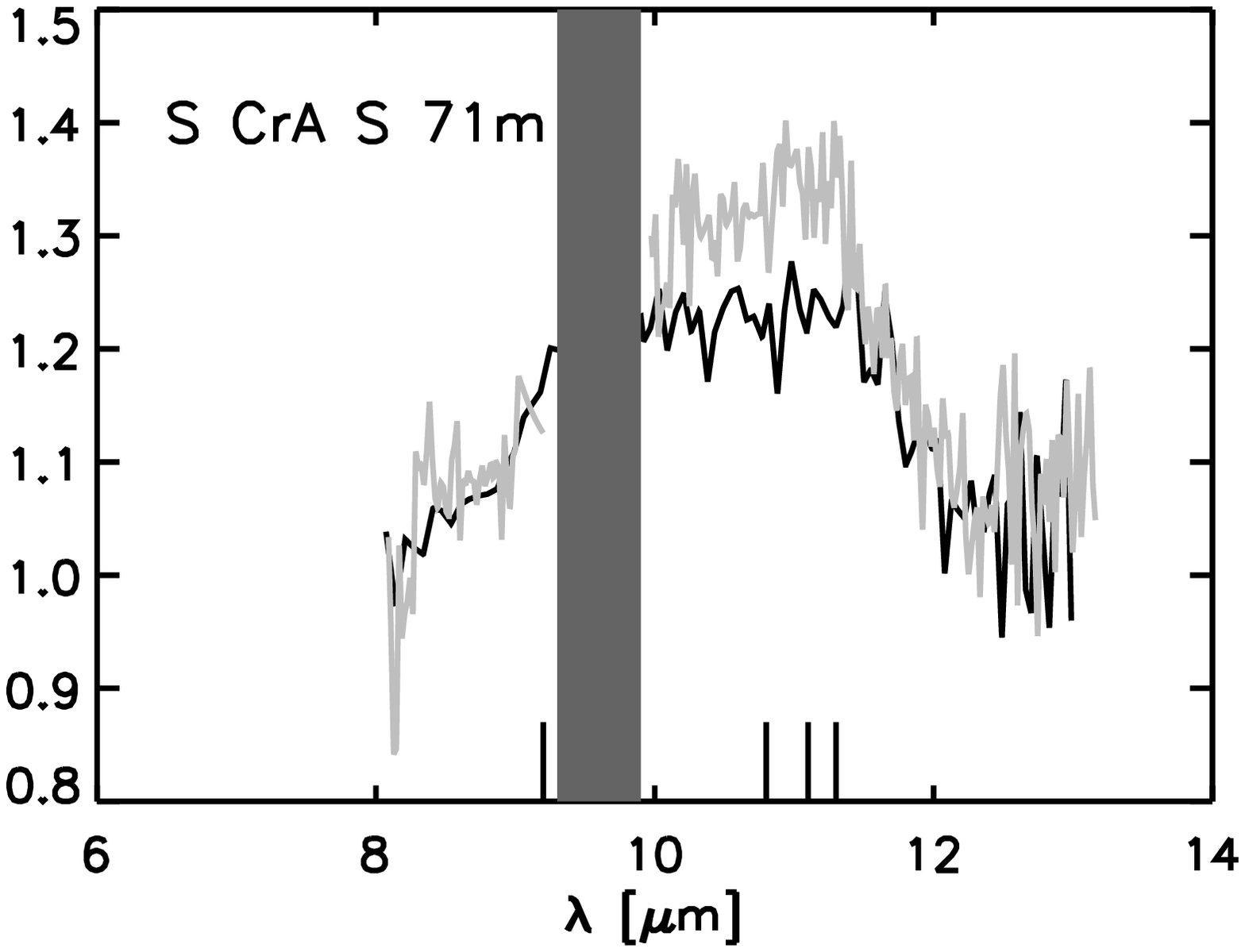}\newline
    \caption{Continuation of Fig.~\ref{figure:10um_abzug}. }
    \label{figure:10um_abzug2}
  \end{figure*}

  \section{Summary}\label{section:conclusion}
  We have presented MIR visibilities of seven pre-main-sequence stars
  observed with MIDI at the VLTI. 
  We modeled the SEDs and the spectrally resolved MIR visibi\-lities of these 
  YSOs, and in particular T\,Tauri objects. The  
  density distributions of the circumstellar disks were derived using the
  parametrized approach  of Shakura \& Sunyaev~(\cite{shakura}, Eq.~\ref{eq:shakura}). 
  The results of this study are:
  \renewcommand{\labelenumi}{\roman{enumi}.}
  \begin{enumerate}    
  \item All objects could be spatially resolved with MIDI. 
  \item We showed that the results of highly spatially resolved observations
    in the MIR range probing small-scale structures in the range of several
    AUs, and the SED from the visual to the millimeter wavelength
    range can simultaneously be reproduced by a single model except for
      one source (see vi.). We note that the SED does not generally
    provide any spatial information. The number of models that simulate
    the SED, solely, can be strongly reduced by the measurements with
    MIDI. Furthermore, the results of previous investigations based
    on large-scale observations in the range of several $10\,\mathrm{AUs}$,
    were confirmed by our modeling results in most cases. Any differences
    between the results of previous measurements and our modeling approach
    (e.\/g., the accretion rate of RU\,Lup) could be caused by a time-dependent
    varia\-bi\-lity.  
  \item For five of seven objects, the modeling approach of a purely passive
    disk is insufficient to reproduce the SED and the MIR visibilities of
    the sources. The implementation of accretion effects also significantly
    improves the simulation of the measurements. 
  \item For two objects in this study, i.\/e., HD\,72106\,B and HBC\,639, the
    approach of a passive disk without accretion is sufficient. 
    HD\,72106\,B is $10\,$million years old and its low accretion rate
    and high mass contributions from crystalline and large dust grains, have
    already been mentioned in previous
    investigations (Vieira  
    et al.~\cite{vieira}; Sch\"utz et al.~\cite{schuetz}). 
    In contrast, HBC\,639 is still a relatively young object (Gras-Vel\'azquez
    \& Ray al.~\cite{gras-velazquez}) but belongs to the class of 
    weak-line T\,Tauri stars. 
  \item For some targets (e.\/g., DR\,Tau, and S\,CrA\,N), the MIR data do
      not constrain all model parameters. However, further modeling constraints
      can be obtained from additional measurements, e.\/g., NIR
    visibilities as shown for our modeling approach for DR\,Tau, where
    PTI data were additionally taken into account.     
  \item The modeling result for the source GW\,Ori illustrates the need for an
    individual extension to our 
    approach. According to Mathieu et al.~(\cite{mathieu}), GW\,Ori has a
    stellar companion at a (projected)  
    distance of $\sim$$1\,\mathrm{AU}$. Therefore, in a second model we
    considered a combination of a  
    circumstellar and circumbinary disk. The circumstellar and circumbinary
    disk are determined by the same disk parameters $\beta$  
    and $h_\mathrm{100}$ and are seperated by a dust-free gap. However, a model 
    that reproduces all the measurements could not be found. We recommend
    acquiring 
    additional high spatial resolution observations including phase
    measurements to clarify the complex inner-disk structure
    of this object.
  \item Our measurements of RU\,Lup can be simulated equally with an active disk model
    with or without a dust-free gap. A disk model with a gap was formerly
    proposed by Takami et al.~(\cite{takamiII}) based on spectro-astrometric
    observations.    
  \item Each single component of the binary system S\,CrA could be seperately
    observed with MIDI. Considering inner disk regions on 
    small scales that could be resolved with MIDI, basic differences between
    this binary system and other objects in our sample   
    without a companion could not be 
    found. Only the SEDs of the sources HD\,72106\,B, HBC\,639, and S\,CrA\,N
    that decline strongly at $\lambda >  
    10\,\mathrm{\mu m}$ support the idea that a stellar companion truncates the
    outer   
    disk regions of this sources. 
  \item The relative mass contribution of evolved dust in the systems of fainter 
    T\,Tauri objects is higher in the inner disk regions close to the central
    star than in the outer regions. However, enstatite could already be enriched
    and stimulated in outer disk regions and the spectral contribution from
    non-evolved dust grains could still be high in the inner regions. 
  \end{enumerate}

  \begin{acknowledgements}
    A.~A.~Schegerer and S.~Wolf were supported by the German Research
    Foundation (DFG) through the Emmy-Noether grant WO 857/2 ({\it ``The
      evolution of circumstellar dust disks to planetary systems''}). We
    gratefully acknowledge O.~Chesneau, Ch.~Leinert, 
    S.~A.~Lamzin, G.~Meeus, F.~Przygodda, Th.~Ratzka, and O.~Sch\"utz who have
    generously given their time to provide valuable assistance during the
    proposal and observation preparation phase as well as during data
    analysis and discussions. We thank R.~Akeson who send us the reprocessed
    PTI data. We also thank the anonymous referee for her/his suggestions for
    improvement. 
  \end{acknowledgements}

  \onecolumn
  \appendix
  \section{Previous measurements:}\label{appendix}
  {\bf DR\,Tau:}
  The $2.5\,$million-year-old object DR\,Tau belongs to the Taurus-Auriga star-forming region at a distance 
  of $\sim$$140\,\mathrm{pc}$ (Siess et al.~\cite{siess}). Numerous publications illustrate that DR\,Tau is one of the most 
  well-studied T\,Tauri objects. Strong veiling has been found in the visual and NIR wavelength range: the flux 
  in V band, for instance, exceeds the intrinsic stellar flux by a factor of five (Edwards et 
  al.~\cite{edwards}). Apart from the excessive flux in the visual, the profiles of several emission 
  lines, such as the Pf$\gamma$-line, points to material accreting onto the central star while the profiles of 
  further emission lines such as the He\,I line (Kwan et al.~\cite{kwan}) and the H$\alpha$ lines (Vink et 
  al.~\cite{vink}) are evidence of outflowing stellar/disk winds. Based on a modeling study, Edwards et al.~(\cite{edwards}) 
  determined an accretion rate of $\dot{M}=7.9 \times 10^{-6}\,\mathrm{M_{\odot}yr^{-1}}$, while a mass of 
  $\dot{M}_\mathrm{wind} = 2.5 \times 10^{-9}\,\mathrm{M_{\odot}yr^{-1}}$ is lost by stellar/disk winds. A strongly collimated 
  outflow was also found by Kwan \& Tademaru~(\cite{kwanII}). DR\,Tau is photometrically and spectrally 
  variable on short terms, i.\/e., in the range of weeks ($\Delta V = 1.3\,\mathrm{mag}$; Grankin et 
  al.~\cite{grankin}; Eiroa et al.~\cite{eiroa}; Smith et al.~\cite{smithIII}). This variability is also ascribed to 
  the formation and movement of stellar spots on this star (Ultchin et al.~\cite{ultchin}). \smallskip\\
  {\bf GW\,Ori:}
  GW\,Ori, also known as HD\,244138, belongs to the star-forming region B\,$30$ in a ring-shaped molecular cloud 
  close to $\lambda$~Ori. According to Dolan \& Mathieu~(\cite{dolan}), the ring shape has its origin in a  
  central supernova explosion $1\,$million years, ago. As GW\,Ori is $\sim$$1\,$million years old 
  (Mathieu et al.~\cite{mathieuIII}), the supernova explosion could cause the formation of this object. 
  Considering a stellar luminosity of several tens of solar luminosities (Mathieu et al.~\cite{mathieu}; Calvet 
  et al.~\cite{calvet}), GW\,Ori is one of the most luminous YSOs with a spectral type of G$0$. Using theoretical 
  evolutionary tracks, a stellar mass of $2.5\,\mathrm{M_\mathrm{\odot}}$ (Mathieu et al.~\cite{mathieuIII}) and 
  $3.7\,\mathrm{M_\mathrm{\odot}}$ (Calvet et al.~\cite{calvet}) could be derived. GW\,Ori is a 
  spectroscopic binary (Mathieu et al.~\cite{mathieuIII}). The companion with a mass of $0.5\,\mathrm{M_{\odot}}$ up to 
  $1\,\mathrm{M_{\odot}}$ orbits the primary at a projected distance of $\sim$$1\,\mathrm{AU}$ in $242\,$days. Mathieu et 
  al.~(\cite{mathieu}) used two different modeling approaches to reproduce the SED of this system. In 
  their first approach, the secondary creates a (gas and dust free) gap between $0.17\,\mathrm{AU}$ and 
  $3.3\,\mathrm{AU}$. Their second modeling approach, where the circumbinary
  disk was replaced by a spherical  
  envelope, disagreed with subsequent millimeter measurements at the James
  Clerk Maxwell Telescope (Mathieu et  
  al.~\cite{mathieuII}). A disk mass of $0.3\,\mathrm{M_{\odot}}$ could be
  derived using the latter set of millimeter 
  measurements, where the object could be spatially resolved. An outer radius
  of $500\,\mathrm{AU}$ and an inclination  
  angle of $\vartheta \approx 27\degr$ were determined. Artymowicz \&
  Lubow~(\cite{artymowicz}) showed that the  
  disk gap of GW\,Ori proposed by Mathieu et al.~(\cite{mathieuIII})
  cannot be explained by a  
  theoretical modeling study of tidal forces. Calvet et
  al.~(\cite{calvet}) measured an accretion rate  
  of $\dot{M} > 2.5 \times 10^{-7}\,\mathrm{M_{\odot}a^{-1}}$, but the source
  does not reveal any veiling in the  
  visual. To stabilize a high accretion rate for several $100,000\,$years, Gullbring et 
  al.~(\cite{gullbring}) proposed the existence of an additional massive
  envelope where an inner cavity enables  
  the observation of the inner disk edge. GW\,Ori is only weakly variable
  ($\Delta V = 0.2\,\mathrm{mag}$;  
  Grankin et al.~\cite{grankin}). Mathieu et al.~(\cite{mathieuIII}) pointed
  to a second companion with a period  
  of $1,000\,$days that was found by the movement of the center of gravity in
  the system.\smallskip\\ 
  {\bf HD\,72106\,B:}
  The object HD\,72106 is a visual binary (angular distance $0.78\arcsec$,
  $PA=199.8\degr$) in the Gum nebula at a distance of
  $288^{+490}_{-204}\,\mathrm{pc}$ (Torres et  
  al.~\cite{torres}; Hartkopf et al.~\cite{hartkopf}; Fabricius \& Makarov~\cite{fabricius}). The 
  H$\alpha$-emission as well as the infrared excess are ascribed to the
  visually fainter B-component ($\Delta V  
  = 0.8\,\mathrm{mag}$), while the A component already belongs to the main
  sequence (Vieira et al.~\cite{vieira}; Wade et  
  al.~\cite{wade}) and shows a strong magnetic field that was found with
  spectropolarimetry (Wade et al.~\cite{wade}; Wade et al.~\cite{wadeII}).  
  The faint H$\alpha$ emission line as well as the broad $10\,\mathrm{\mu m}$
  silicate emission band  
  that can be compared with the silicate band of the comets Hale-Bopp and
  Halley point to the advanced  
  evolutionary status of the B component as a YSO (Vieira et al.~\cite{vieira}; Sch\"utz et 
  al.~\cite{schuetz}). In particular, Sch\"utz et al.~(\cite{schuetz}) found
  larger amounts of enstatite  
  ($\sim$$50\%$) that effectively contributes to the silicate band. This
  large spectral contribution of enstatite has only  
  been found for the evolved Herbig Ae/Be objects, HD\,100546 and HD\,179218, so
  far. Folsom~(\cite{folsom}) intensively studied this binary system. \smallskip\\ 
  {\bf RU\,Lup:} 
  RU\,Lup is a classical T\,Tauri star in the star-forming region
  Lupus. Visual and millimeter measurements  
  showed that the object does not have a (remaining) circumstellar envelope (Giovannelli et 
  al.~\cite{giovannelli}; Lommen et al.~\cite{lommen}). A secondary could not be found with speckle 
  interferometry down to a minimal distance of $0.1\arcsec$ in the NIR range
  (Ghez et al.~\cite{ghez}) and with the Hubble Space Telescope (Bernacca et
  al.~\cite{bernacca}). Broad 
  visual emission lines as well as flux variations in the U and V band are
  evidence of accreting material. Lamzin et al.~(\cite{lamzin}) determined a
  mass accretion rate of $\dot{M} = 3 \times  
  10^{-7}\,\mathrm{M_\mathrm{\odot}yr^{-1}}$. The 
  stellar magnetic field of the object has a strength of $3\,\mathrm{kG}$ (Stempels et al.~\cite{stempels}). 
  Absorption lines shifted to longer wavelengths results from an outflowing stellar wind (Herczeg et 
  al.~\cite{herczeg}). \smallskip\\
  {\bf HBC\,639:}
  HBC\,639, also known as DoAr\,24\,E, belongs to the star-forming region $\rho$ Ophiuchi. It is a class II 
  object (McCabe et al.~\cite{cabe}) with an age of between $1$ and $3$ million years (Gras-Vel\'azquez \& 
  Ray~\cite{gras-velazquez}). Because the H$\alpha$ line reveals an equivalent
  width of $\sim$$5\,$\AA, HBC\,639 is a  
  weak-line T\,Tauri object, where the circumstellar material has evolved
  more rapidly than in classical T\,Tauri stars  
  of similar age. However, the infrared excess points to a remaining circumstellar disk in the system. The accretion rate 
  inferred from the width of the Pa$\beta$ and Br$\gamma$ lines is low: $\dot{M} = 6 \times 
  10^{-9}\,\mathrm{M_\mathrm{\odot}yr^{-1}}$ (Natta et al.~\cite{natta}). HBC\,639 has an infrared companion at an angular 
  distance of $2\arcsec$ ($320\,\mathrm{AU}$ for $160\,\mathrm{pc}$) and a position angle of $PA=150\degr$ 
  measured from North to East (Reipurth \& Zinnecker~\cite{reipurth}). The brightness of the secondary 
  increases in the infrared and exceeds the brightness of the primary already in L band (Prato et 
  al.~\cite{prato}). Chelli et al.~(\cite{chelli}) assumed that the primary
  effectively contributes only to 
  the NIR and MIR flux of the system. Polarimetric measurements in the K band showed that both components have a 
  circumstellar disk with almost identical position angles ($PA \approx 12,5\degr$; Jensen et al.~\cite{jensen}). The 
  companion is a class I object and active (Prato et al.~\cite{prato}). Using the 
  speckle-interferometric technique in K band, Koresko~(\cite{koresko}) found
  a second companion close to the 
  secondary. Both latter companions have a similar brightness in K band. \smallskip\\
  {\bf S\,CrA:}
  The source S\,CrA belongs to the Southern Region of the Corona Australis Complex (e.\/g., Chini et 
  al.~\cite{chini}). The source was already defined as a T\,Tauri object by Joy et al.~(\cite{joy})
  who pointed to an infrared companion at a projected distance of $169\,\mathrm{AU}$ and at a position angle of 
  $PA=149\degr$ (Joy \& Biesbrock~\cite{joyII}; Reipurth \& Zinnecker~\cite{reipurth}). Highly spatially 
  resolved observations in the NIR wavelength range showed that both objects have an active circumstellar disk 
  (Prato et al.~\cite{prato}). The spectral lines of both objects indicate
  that they have similar shapes and depths. 
  Both components are probably coeval (Takami et al.~\cite{takami}). S\,CrA is a 
  YY\,Ori object, i.\/e., the emission lines are asymmetric and shifted to longer wavelengths. These lines arise 
  from material that accretes onto the central star. The measured spectral variability is another hint of 
  non-continuous accretion process. Prato \& Simon~(\cite{pratoII}) showed that only an infalling circumbinary 
  envelope provides enough material for accretion in the long-term. 
  
  \begin{table*}[!tb]
    \centering
    \caption{Photometric flux measurements of DR\,Tau.}
    \label{table:photo-drtau}
    \begin{tabular}{llr} \hline\hline
      wavelength [$\mathrm{\mu m}$] & flux [Jy] & Ref. \\ \hline
      $0.36	$ & $0.038	\pm 0.021	     $ & { 1} \\
      $0.44	$ & $0.050	\pm 0.049	     $ & { 1} \\
      $0.55	$ & $0.083	\pm 0.060	     $ & { 1} \\
      $0.64	$ & $0.13	\pm 0.06	     $ & { 1} \\
      $0.79	$ & $0.21	\pm 0.01	     $ & { 1} \\
      $1.25	$ & $0.45	\pm 0.01	     $ & { 2} \\
      $1.65	$ & $0.77	\pm 0.04            $ & { 2} \\
      $2.20	$ & $1.13	\pm 0.02            $ & { 2} \\
      $3.50	$ & $2.07	\pm 0.82             $ & { 1} \\
      $3.60 $ & $1.86       \pm 0.20             $ & { 7} \\
      $4.50 $ & $1.89       \pm 0.15             $ & { 7} \\
      $4.80	$ & $2.41	\pm 1.10	     $ & { 1} \\
      $5.80	$ & $1.27	\pm 0.01	     $ & { 3} \\
      $8.00	$ & $1.77	\pm 0.20	     $ & { 7} \\
      $12.0	$ & $3.16	\pm 0.25	     $ & { 4} \\
      $25	$ & $4.30	\pm 0.29	     $ & { 4} \\
      $60	$ & $5.51	\pm 0.54	     $ & { 4} \\
      $100	$ & $6.98	\pm 0.63	     $ & { 4} \\
      $200     $ & $3.98      \pm      0.84	     $ & { 5} \\
      $450	$ & $2.38	\pm 	0.17	     $ & { 6} \\
      $729  $ & $0.40       \pm     0.08        $ & { 8} \\
      $850	$ & $0.53	\pm 	0.01	     $ & { 6} \\
      $1300	$ & $0.16	\pm 	0.01	     $ & { 6} \\
      \hline
    \end{tabular}
        {\newline \scriptsize {References}~-- { 1}: Kenyon \&
          Hartmann~(\cite{kenyon}); { 2}:  
          2\,MASS catalogue~(Cutri et al.~\cite{cutri}); { 3}: Hartmann et
          al.~(\cite{hartmannIV}); { 4}:  
          Gezari catalogue~(\cite{gezari}); { 5}: ISO data archive; { 6}: Andrews \& 
          Williams~(\cite{andrewsII}); { 7}: Robitaille et
          al.~(\cite{robitaille}); { 8}: Beckwith et  
          al.~(\cite{beckwith})}
  \end{table*} 
  
  \begin{table*}[!bt]
    \centering
    \caption{Photometric flux measurements of GW\,Ori. The value that is
      marked with the symbol ``$\downarrow$''  
      is an upper flux limit.}
    \label{table:photo-gwori}
    \begin{tabular}{llr} \hline\hline
      wavelength [$\mathrm{\mu m}$] & flux [Jy] & Ref. \\ \hline
      $0.36   $ & $0.068\pm0.014$ & { 1} \\ 	
      $0.44    $ & $ 0.21\pm0.02$ & { 1} \\ 
      $0.55    $ & $ 0.43\pm0.02$ & { 1} \\
      $0.70$ & $0.61\pm0.03$ & { 1}\\
      $0.90    $ & $ 0.82\pm0.05$ & { 1} \\
      $1.25    $ & $ 1.31     \pm 0.04   $ & { 2} \\
      $1.65    $ & $ 1.47     \pm 0.04   $ & { 2} \\
      $2.20    $ & $ 1.47     \pm 0.04   $ & { 2} \\
      $3.50$ & $2.45\pm0.09$ & { 3} \\
      $4.80     $ & $ 2.7\pm1.2$ & { 4} \\
      $18$ & $17\pm3$ & { 5} \\
      $25     $ & $ 20.5     \pm   1.2   $ & { 6} \\
      $60     $ & $ 31.5     \pm   4.1   $ & { 6} \\ 
      $100     $ & $ 35^{\downarrow}$ & { 6} \\   	
      $350     $ & $ 5.0     \pm  0.6   $ & { 7} \\ 
      $450     $ & $ 3.5    \pm  0.4   $ & { 7} \\  
      $800     $ & $0.9     \pm 0.1   $ & { 7} \\  
      $850     $ & $ 1.0     \pm  0.1   $ & { 7} \\ 
      $1100     $ & $0.29     \pm 0.03   $ & { 7} \\
      $1360     $ & $0.26     \pm 0.06   $ & { 7} \\   
      \hline
    \end{tabular}
        {\newline \scriptsize {References}~-- { 1}: Calvet et 
          al.~(\cite{calvetII}); { 2}: 2\,MASS catalogue~(Cutri et al.~\cite{cutri}); { 3}: Rydgren \& 
          Vrba~(\cite{rydgrenII}); { 4}: Cohen \& Schwartz~(\cite{cohen}); 
          { 5}: Cohen~(\cite{cohenIV}); { 6}: IRAS catalogue~(\cite{iras}); 
          { 7}:~Mathieu et al.~(\cite{mathieuII})}
  \end{table*} 
  
  \begin{table*}[!bt]
    \centering
    \caption{Photometric flux measurements of HD\,72106\,B. The value that is marked with the symbol 
      ``$\downarrow$'' is an upper limit.}
    \label{table:photo-hd72106}
    \begin{tabular}{llr} \hline\hline
      wavelength [$\mathrm{\mu m}$] & flux [Jy] & Ref. \\ \hline
      $0.36$ & $0.80$ & { 1} \\
      $0.44$ & $1.67$ & { 1} \\
      $0.55$ & $1.45$ & { 1} \\
      $0.79$ & $1.32$ & { 1} \\
      $0.90$ & $0.89$ & { 1} \\
      $1.23	$ & $0.61	 $ & { 3} \\
      $1.25$ & $0.67 \pm 0.01$ & { 2} \\
      $1.63	$ & $0.45	 $ & { 3} \\
      $1.65$ & $0.50 \pm 0.02$ & { 2} \\ 
      $2.19	$ & $0.41	 $ & { 3} \\
      $2.20$ & $0.43 \pm 0.01$ & { 2} \\ 
      $3.78	$ & $0.39	 $ & { 3} \\
      $4.66	$ & $0.44	 $ & { 3} \\
      $8.36	$ & $1.74	 $ & { 3} \\
      $9.67	$ & $1.45	 $ & { 3} \\
      $12.0	$ & $2.23\pm0.13 $ & { 3, 4} \\
      $25	$ & $3.62\pm0.21 $ & { 3} \\
      $60	$ & $1.88\pm0.11 $ & { 4} \\
      $100	$ & $16.8^{\downarrow}$ & { 4} \\
      \hline
    \end{tabular}
        {\scriptsize \newline {References}~-- { 1}: Torres et al.~(\cite{torres}); { 2}: 
          2\,MASS catalogue~(Cutri et al.~\cite{cutri}); { 3}: Gezari et al.~(\cite{gezari}); 
          { 4}:~IRAS catalogue~(\cite{iras})} 
  \end{table*}

  \begin{table*}[!t]
    \centering
    \caption{Photometric flux measurements of RU\,Lup.}
    \label{table:photo-rulup}
    \begin{tabular}{llr} \hline\hline
      wavelength [$\mathrm{\mu m}$] & flux [Jy] & Ref. \\ \hline
      $0.36  $ & $0.016  $ & { 1} \\ 
      $0.44  $ & $0.073  $ & { 1} \\ 
      $0.55  $ & $0.115  $ & { 1} \\ 
      $0.64  $ & $0.184  $ & { 1} \\ 
      $0.79  $ & $0.290  $ & { 1} \\ 
      $1.25$ & $0.51 \pm 0.01$ & { 2} \\
      $1.25$ & $0.72 \pm 0.02$ & { 3} \\
      $1.25$ & $0.41 \pm 0.01$ & { 3} \\
      $1.65$ & $0.76 \pm 0.03$ & { 2} \\
      $1.65$ & $1.02 \pm 0.03$ & { 3} \\
      $1.65$ & $0.62 \pm 0.02$ & { 3} \\
      $2.20$ & $0.89 \pm 0.02$ & { 2} \\
      $2.20$ & $1.29 \pm 0.02$ & { 3} \\
      $2.20$ & $0.75 \pm 0.03$ & { 3} \\
      $3.50  $ & $1.16  $ & { 4} \\ 
      $3.50  $ & $1.77 \pm 0.07  $ & { 3} \\
      $3.50  $ & $1.0 \pm 0.10 $ & { 3} \\
      $4.80  $ & $1.5  $ & { 4} \\
      $4.80  $ & $1.5 \pm 0.2  $ & { 3} \\
      $4.80  $ & $0.6 \pm 0.3 $ & { 3} \\
      $15$ & $2.22$ & { 5} \\
      $20$ & $2.80$ & { 5} \\
      $25  $ & $  4.64 $ & { 4} \\
      $60  $ & $  4.68 $ & { 4} \\ 
      $100 $ & $  5.70 $ & { 4} \\ 
      $1300$ & $  0.197\pm0.007 $ & { 6} \\
      $1400$ & $0.159\pm0.010$ & { 7}\\
      \hline
      \end{tabular}
      {\newline \scriptsize {References}~-- { 1}: Gahm et al.~(\cite{gahm}); 
        { 2}: 2\,MASS catalogue~(Cutri et al.~\cite{cutri}); { 3}:
        Giovannelli et al.~(\cite{giovannelli}); { 4}: 
        Gezari catalogue~(\cite{gezari}); { 5}: Gras-Vel\'azquez \& 
        Ray~(\cite{gras-velazquez}); { 6}: N\"urnberger et al.~(\cite{nuernberger}); { 7}: 
        Lommen et al.~(\cite{lommen})}
  \end{table*} 
  
  \begin{table*}[!t]
    \centering
    \caption{Photometric flux measurement of HBC\,639. The flux can be ascribed to the main component up to 
      wavelengths in L band.}
    \label{table:photo-hbc639}
    \begin{tabular}{llr} \hline\hline
      wavelength [$\mathrm{\mu m}$] & flux [Jy] & Ref. \\ \hline
      $0.36	$& $0.0001 $ & { 1} \\
      $0.44	$& $0.0008	  $ & { 1} \\
      $0.55	$& $0.0053		  $ & { 1} \\
      $0.64	$& $0.018		  $ & { 2} \\
      $0.79	$& $0.065		  $ & { 2} \\
      $1.25	$& $0.33	\pm  0.10  $ & { 3}\\
      $1.65	$& $0.77	\pm 0.16  $ & { 3} \\
      $2.20	$& $0.92	\pm 0.09  $ & { 3} \\
      $3.50	$& $0.75	\pm 0.07  $ & { 4} \\
      $4.80	$& $0.98		  $ & { 5} \\
      $7.70	$& $0.60		  $ & { 5} \\
      $10.0	$& $1.06		  $ & { 5} \\
      $12.0	$& $1.87		  $ & { 5} \\
      $12.8	$& $2.57		  $ & { 5} \\
      $15.0	$& $2.41		  $ & { 5} \\
      $20.0	$& $2.30		  $ & { 5} \\
      $25	$& $5.91		  $ & { 6} \\
      $60	$& $35	        	  $ & { 6} \\
      $100	$& $40		  $ & { 6} \\
      $800	$& $0.037	\pm 0.008 $ & { 7} \\
      $1300	$& $0.065	\pm 0.019 $ & { 8} \\           
      \hline
    \end{tabular}
        {\newline \scriptsize {References}~-- { 1}: Bouvier \& 
          Appenzeller~(\cite{bouvier}); { 2}: Rydgren~(\cite{rydgren}); { 3}: 
          Prato et al.~(\cite{prato}); { 4}: Mc\,Cabe et al.~(\cite{cabe}); 
          { 5}:~Gras-Velazquez \& Ray~(\cite{gras-velazquez}); { 6}: Gezari Catalogue~(\cite{gezari}); 
          { 7}:~Jensen et al.~(\cite{jensenII}); { 8}: 
          N\"urnberger et al.~(\cite{nuernberger})}
  \end{table*} 
  
  \begin{table*}[!t]
    \centering
    \caption{Photometric flux measurement of S\,CrA\,N and S\,CrA\,S. For the photometric observations of 
      wavelengths $\lambda<0.64\,\mathrm{\mu m}$ and $\lambda>12\,\mathrm{\mu m}$, both components could not be 
      spatially resolved.}
    \begin{tabular}{lllr} \hline\hline
      & S\,CrA\,N & S\,CrA\,S & \\
      wavelength [$\mathrm{\mu m}$] & flux [Jy] & flux [Jy] & Ref.\\ \hline
      $0.44$ &   \multicolumn{2}{c}{$0.0089$} & { 1} \\
      $0.55$ &   \multicolumn{2}{c}{$0.13$} & { 2} \\
      $0.64$ &   \multicolumn{2}{c}{$0.09$} & { 1} \\
      $1.25	$ & $0.57 \pm 0.02$ & $0.28 \pm 0.02$ & { 3} \\
      $1.65	$ & $1.02 \pm 0.04$ & $0.50 \pm 0.03 $ & { 3} \\
      $2.22	$ & $2.44 \pm 0.21$ & $0.74 \pm 0.09 $ & { 4} \\
      $3.50	$ & $3.99 \pm 0.23$ & $1.15 \pm 0.03 $ & { 4} \\
      $8.80	$ & $2.90 \pm 0.20$ & $1.59 \pm 0.17 $ & { 4}\\
      $10.6	$ & $3.19 \pm 0.20$ & $1.27 \pm 0.06$ & { 4} \\
      $12.0	$ & \multicolumn{2}{c}{$3.97$} & { 5} \\
      $12.0	$ & \multicolumn{2}{c}{$4.88 \pm 0.59$} & { 5} \\
      $12.5	$ & \multicolumn{2}{c}{$1.50 \pm 0.10$} & { 5} \\
      $25	$ & \multicolumn{2}{c}{$9.21 \pm 0.64$} & { 5} \\
      $25	$ & \multicolumn{2}{c}{$8.52 $} & { 5} \\
      $60	$ & \multicolumn{2}{c}{$1.81 \pm 0.16$} & { 5} \\          
      $60 $ & \multicolumn{2}{c}{$2.7 \pm 4.5 $} & { 6}\\   
      $60 $ & \multicolumn{2}{c}{$3.32 \pm 3.1$} & { 6} \\
      $90 $ & \multicolumn{2}{c}{$2.87 \pm 2.85$} & { 6} \\      
      $100$ & \multicolumn{2}{c}{$2.79 \pm 0.75$} & { 5} \\
      $170$ & \multicolumn{2}{c}{$8.29 \pm 22.24$} & { 6} \\      
      $450$ & \multicolumn{2}{c}{$2.75 \pm 0.38$} & { 7} \\
      $800$ & \multicolumn{2}{c}{$0.77 \pm 0.05$} & { 7} \\
      $1100$ & \multicolumn{2}{c}{$0.36 \pm 0.03$} & { 7} \\
      \hline
    \end{tabular}
        {\newline \scriptsize {References}~-- { 1}: Denis data base~(\cite{denis}); 
          { 2}: Takami et al.~(\cite{takami}); { 3}: Prato et 
          al.~(\cite{prato}); { 4}: Mc\,Cabe et al.~(\cite{cabe}); { 5}: 
          IRAS catalogue~(\cite{iras});  { 6}:~ISO data archive; 
          { 7}: Jensen et al.~(\cite{jensenII}) } 
        \label{table:photo-scra}
  \end{table*} 

\end{document}


\maketitle
\section*{Preface}

\texttt{lineno.sty} is a macro package made by 
Stephan~I.~B\"ottcher for attaching line numbers to 
\LaTeX\ documents. Some people have used it for revising 
submittings in collaboration with referees or co-authors. 
Documentations are nowadays preferred to be in 
Adobe's \texttt{PDF}---so \texttt{lineno.sty}'s 
documentation is \lcurl{lineno/lineno.pdf\,.}

\texttt{ednotes.sty} uses \texttt{lineno.sty} for critical 
editions, combining it with Alexander~I.~Rozhenko's 
\texttt{manyfoot.sty}---this was Christian Tapp's idea, 
who then hired me for adding the \TeX nical details. 
In doing this, I had to change some internals of 
\texttt{lineno.sty}, so Stephan transferred maintenance 
to me; then some of my macro files that I originally had 
made for \texttt{ednotes.sty} wandered into the 
\texttt{lineno} directory of CTAN---because they turned 
out not to need \texttt{ednotes.sty}, 
just to work as extensions of \texttt{lineno.sty}\,. 

Now, I haven't had the time for making \texttt{.dtx} versions 
of the \texttt{.sty} files for \texttt{ednotes}. 
Therefore, ordinary \texttt{.pdf} documentation for 
the remaining \texttt{.sty} files of \lcurl{lineno/}
is missing. 
What you see here is nothing but a somewhat structured listing 
of the additional \texttt{.txt} and \texttt{.sty} files in 
\texttt{PDF}, deriving from the \texttt{verbatim} package and 
its \cs{verbatiminput} command. I hope the high quality 
(scalable) output is worth it. 

\leavevmode\hfill \textit{U.\,L.}

\newpage 
\tableofcontents
 
\section{The \texttt{.txt} files}
\subsection{\texttt{README.txt}}
\begin{verbatim}
README
                      lineno.sty  v4.31

                         2005-10-03 

The LaTeX package lineno.sty provides line numbers on paragraphs.
After TeX has broken a paragraph into lines there will be line numbers
attached to them, with the possibility to make references through the
LaTeX \ref, \pageref cross reference mechanism.

% Copyright 1995--2003 Stephan I. B"ottcher <boettcher@physik.uni-kiel.de>; 
% Copyright 2002--2005 Uwe L"uck, http://www.contact-ednotes.sty.de.vu, 
%                      for versions 4.x and code from former Ednotes 
%                      bundle--author-maintained. 
% 
% The files listed below can be redistributed and/or modified under 
% the terms of the LaTeX Project Public License; either 
% version 1.3a of the License, or any later version.
% The latest version of this license is in
%     http://www.latex-project.org/lppl.txt
% We did our best to help you, but there is NO WARRANTY. 

** [ UL: A few festive words on history and responsibility   **
**       are next preceding hard facts as to (see lower)     ** 
**       o `Files'  and                                      ** 
**       o `Installation and usage'.                       ] ** 

2004-09-13 Uwe L"uck  [(UL)]  is new maintainer for lineno.sty. 

lineno.sty served the purpose for which I wrote it years ago.  Uwe
L"uck uses lineno.sty with his Ednotes package, which required quite a
few changes and fixes.  His package depends on lineno, therefore
Uwe agreed to take over the maintenance of lineno.sty.

lineno.sty v4.0 includes most of the well tested changes that Uwe
needs for Ednotes.  These changes blend well into the concepts of the
package, so I am happy to let it go.  

From here on it is Uwe's, and he may proceed to mangle it as he likes.
Expect some radical changes.  You may find him to be quite a bit
friendlier towards the poor souls who still use Windows :-).
Currently, you need some kind of Unix environment to extract the
source documentation from the sty file.

If some version 3 users run into difficulties with Uwe's newer
versions, but need a minor bug-fix in version 3, please do not
hesitate to ask me for help.  But all requests for new features or
major changes shall go to Uwe.

Cheers
Stephan


[ Thanks!  And please let me know as well should 
  compatibility problems arise!  The announced radical 
  changes are postponed again this time (v4.1). 
  --Ednotes is in CTAN:macros/latex/contrib/ednotes.
                                                   Uwe ] 
 


Changes:

   2004-10-19  UL: package options for tabular and math 
   2004-09-03  UL: merge Ednotes changes, taking over lineno.sty

   2002 .. 2003 FMi, UL, SiB: bug fixes
   2001-08-04 SiB: linenomath wrapping for \[ \]
   2001-07-30 SiB: [hyperref] option obsolete.
   2001-01-17 SiB: LaTeX class option [twocolumn] support
   2001-01-04 SiB: LaTeX class option [fleqn] support
   2000-12-18 SiB: longtable compatibility
   2000-07-01 SiB: extra \newlabel items, [hyperref] option
   2000-03-10 SiB: indirect call of \output, to work with multicol.
   1999-08-28 SiB: fixed the footnote problem using \holdinginserts
   1999-06-11 SiB: included the extensions into lineno.sty
   1999-03-02 SIB: Added LPPL License


Files:

   README        This file.                              (v4.1: UL)
   CHANGES       Differences to previous versions.       (v4.1: UL)
   COPYING       The LPPL header.
   lineno.sty    The package itself, ready to use.
   edtable.sty   Module for tabular environments.        (UL, v4.1) 
   ednmath0.sty  Module for \linelabel in math mode.     (UL, v4.1) 
   lineno.tex    The source for the documentation.
   lineno.pdf    PDF deriving from the  former.          (v4.1: UL)
   lnosuppl.pdf  PDF listing of present non-PDF files 
   ulineno.tex   The pathetic attempt of a users' manual. 
                 [Describes v3.1 currently.]             (UL, v4.1) 
   ulineno.pdf   PDF of former.
   vplref.sty    Conditionally include page number in 
                 line number references                  (UL, v4.2) 

     As of version v3.00, the extension packages for lineno have all been
     incorporated into lineno.sty itself.  Except for itemrule.sty,
     which was removed.

Home:

   CTAN:macros/latex/contrib/lineno

Authors:

   Uwe L"uck <http://contact-ednotes.sty@web.de>
   Stephan I. B"ottcher <boettcher@physik.uni-kiel.de>

Compatibility with other packages:

   wrapfig.sty   works since v2.05

   multicol.sty  works partly since 3.02.  
                 Do NOT use \linelabel.  Do NOT put a multicol in
                 internal vertical mode {table}, {figure}, etc.

   hyperref.sty  \ref to a \linelabel works since v3.03.

   longtable.sty broke with lineno.sty loaded, but not enabled.  
                 This is fixed in v3.04.

Installation and usage (UL, v4.1): 

   For being able to use ALL the new lineno.sty options, the 
   following files must be "visible" to (La)TeX ("visible" 
   explained below for beginners): 

     lineno.sty, edtable.sty, ednmath0.sty, ltabptch.sty 
 
   as above ("Home", note that clicking on "entire directory" 
   suffices); 

     longtable.sty -- from the standard LaTeX Tools bundle. 

   Usage always starts with loading lineno.sty by \usepackage. 
   The remaining .sty files are loaded automatically on the 
   lineno.sty options (and we recommend not to load them through 
   [the mandatory argument of] \usepackage). 
 
   For details, see lineno.tex/pdf and the .sty files mentioned 
   above--search especially for tabular and math mode. 
   (+ `print' below here). 
 
   "Visible to (LaTeX)": Some users don't understand this 
   "visibility" for a while, and indeed it may be somewhat 
   non-trivial. These users may find help in 

      CTAN:macros/latex/ednotes/visible.txt .

   E.g., former users of lineno.sty may just put all the .sty 
   files into the folder (at their workplace) where they had 
   placed lineno.sty before.                     (/UL, /v4.1)

To print the documented source:

   Take the style-file `lineno.sty', and feed it to a Un*x shell. 
   (Or download the extracted source documentation `lineno.tex'.)

      csh> source ./lineno.sty
      sh>  . ./lineno.sty

   [ I.e., type `sh lineno.sty' (e.g.) as a UNIX command line. 
     Problems with awk may arise. I therefore switched to nawk 
     in lineno.sty v4.00, but this may trouble you as well.    (UL) ] 

   (Please ignore the error message at the beginning about the iffalse.)


Please have a look at a similar work of Michal Jaegermann and James
Fortune:
         CTAN:macros/latex/contrib/numline/





==============================================================================
<- End of official README.txt (UL) 

Changes for v4.1 (Oct. 2004): minor fixes (removed `supported/', e.g.); 
bracketed text; lines with `(UL)'; redistribution with README only. 
Formerly: 

                      lineno.sty  v4.0

     $Id: README,v 3.6.2.1 2004/09/13 20:15:47 stephan Exp $

   CTAN:macros/latex/contrib/supported/lineno
         CTAN:macros/latex/contrib/supported/numline/


\end{verbatim}
\subsection{\texttt{COPYING.txt}}
\begin{verbatim}
% lineno.sty
%
% The files in this directory are
%
% Copyright 1995--2003 Stephan I. B"ottcher <boettcher@physik.uni-kiel.de>; 
% Copyright 2002--2005 Uwe L"uck, http://contact-ednotes.sty.de.vu, 
%                      for versions 4.x and code from former Ednotes 
%                      bundle--author-maintained. 
% 
% The files can be redistributed and/or modified under 
% the terms of the LaTeX Project Public License; either 
% version 1.3a of the License, or any later version.
% The latest version of this license is in
%     http://www.latex-project.org/lppl.txt
% We did our best to help you, but there is NO WARRANTY. 
%
% ============================================================================
<- (UL) End of official declaration. 

History: 
This file -> The files 2004/10/12 ul 
$Id: COPYING,v 3.0.2.1 2004/09/13 20:15:47 stephan Exp $
LPPL v1.3a 2004/10/26 

\end{verbatim}
\subsection{\texttt{CHANGES.txt}}
\begin{verbatim}
CHANGES  for lineno.sty v4.41  2005/11/02: 

1. Loadable after amsmath. 

2. Removed some nonsense from documentation. 


CHANGES  for lineno.sty v4.4  2005/10/27 
[failed to be uploaded to CTAN, but distributed by mail]: 

1. Proper effective line depth at end of paragraphs. 
   The spacing bug was quite obvious in two-column mode 
   when a paragraph end was at a column bottom. 

2. Another bug concerning two-column mode that had been 
   introduced in v4.22 has been removed again. 

3. Support for \addvspace introduced more and more bugs 
   in versions of v4.32 and v4.33. The reasons seem to 
   be clear now, and v4.4 should be stable. 


CHANGES  for lineno.sty v4.32  2005/10/17: 

1. Support for \addvspace 
   (a math display or a list meets a heading -- or a sub-heading 
   follows a heading -- or the like) 

2. Clearly explained former option `displaymath' and its change 
   to a default. 


CHANGES  for lineno.sty v4.31  2005/10/01: 

1. \modulolinenumbers* and a package option `modulo*' for 
   printing first line number after interrupting editor's 
   text, regardless of the modulo. 

2. Improved explanation of \firstlinenumber and package 
   options. 


CHANGES  for lineno.sty v4.3  2005/05/16: 

1. Option `displaymath' (proper numbering at paragraphs 
   containing math displays) becomes default. 

2. Compatibility with hyperref now indeed (at least much more). 

3. Tidied up documentation: terrible confusion of \newcounter 
   vs. \stepcounter; sections on the same matter written at 
   different times; ... 

4. Additional internal improvements that perhaps hardly are 
   observable (no more spurious linenumbers in math displays 
   from vertical mode; some compatibity with packages that use 
   \holdinginserts; \linelabel in headings etc.). 


CHANGES  for lineno.sty v4.22 2005/05/09: 

1. Restored "global" version of numbering lines of a \parbox or 
   minipage or ... (I had missed and disabled this in taking 
   over lineno.sty), explained in documentation (lineno.pdf/dvi 
   subsec. 7.2). 

2. Enabled \flushbottom in two-column pagewise or switching 
   mode. 

3. Re-implemented modulo mode -- disabling certain users' tricks 
   see lineno.pdf sec. 5.5, also for a still supported 
   substituting trick. 

4. Tidied up setting the next line number globally vs. locally 
   (TeXbook p. 301). 

5. Tidied up discussions (in documentation) of possible changes 
   in implementation. 


CHANGES  for lineno.sty v4.21 2005/04/28: 

Removed serious flaws with math display, + something ... 


CHANGES  for lineno.sty v4.2 2005/04/26: 

1. Re-enabled package option `displaymath' (needed rearrangement 
   after sec. 5). 

2. New package option `addpageno' for adding page numbers to 
   line number references -- see sec. 6.1 of lineno.pdf. 

3. Improved support for \includeonly (and improved sec. 5.3, 
   p. 27). 

4. Improved compatibility with other packages that change \output 
   (tameflts.sty, e.g., for saving footnotes against \marginpar 
   and floats), added advice on this matter -- sec. 2.3, 
   pp. 7, 14f.


CHANGES  for lineno.sty v4.11 2005/03/08: 

1. The `edtable' option now supports math environments like 
`array'. This requires updating edtable.sty to v1.3. (The claim 
in previous versions of edtable on this support simply was 
wrong, sorry.) For how to make use of this support, we urge you 
to read the usage instructions in edtable.sty (v1.3). These have 
been extended very much, structured more clearly, and supplied 
with examples. 

2. \linelabel now complains when appearing outside line 
numbering mode. This counters Stephan Boettcher's original 
intention, but I generalize from my own experience that it is 
helpful to be told when you have forgotten to switch into line 
numbering mode and (e.g.) wonder why all the line number 
references are 1 ... ednotes.sty users profit as well (at least 
I expect and hope that anybody profits, which, to be sure, does 
not mean that I hope that anybody forgets to switch ...). 

3. The subsection on `edtable' in lineno.sty/tex/pdf was not 
quite correct or complete -- corrected, improved. 

4. The final list of user commands in lineno.sty/tex/pdf has not 
been complete. This has not changed, but it is briefly explained 
what is missing and where it can be found. 


CHANGE  2005/01/20: 

We have devised macros for indexing with line numbers, 
yet we don't take the time to release them officially. 
If you are interested, please ask via 

    http://contact-ednotes.sty.de.vu 


CHANGES  for lineno.sty v4.1 2004/10/19: 

Extension packages from the Ednotes bundle for enabling 
\linelabel in math mode and tabular environments are now 
handled through new package options `mathrefs', `edtable', 
`longtable', and `nolongtablepatch'. Two of these extension 
packages moved from the ednotes folder to the lineno folder. 
lineno.tex/pdf has been updated accordingly. 


CHANGES  for lineno.sty v4.00 2004/09/03: 

o incorporated earlier extension packages linenox0.sty, 
  linenox1.sty, lnopatch.sty (which belonged to the 
  Ednotes bundle before); 

o adopted LaTeX Project Public License v1.3. 

\end{verbatim}

\section{Tabular and array environments}
\texttt{lineno.sty}'s package options \texttt{edtable}, 
\texttt{longtable}, and \texttt{nolongtablepatch} 
redefine \LaTeX\ tabular and array environments 
such that \texttt{lineno} and \texttt{ednotes} commands 
can be used inside. The code for these options resides 
in separate files at present. We are listing them here. 
\subsection{\texttt{edtable.sty}}
\begin{verbatim}
%% `edtable.sty'---Uwe L"uck, direction Christian Tapp. 
%% LaTeX package for tables with line numbers and 
%% editorial notes. 
%% 
%% Copyright (C) 2003-2005 Uwe L"uck--author-maintained. 
%% 
\def\fileversion{1.3c} \def\filedate{2005/10/03} 
%% 
%% This file can be redistributed and/or modified under 
%% the terms of the LaTeX Project Public License; either 
%% version 1.3 of the License, or any later version.
%% The latest version of this license is in
%%   http://www.latex-project.org/lppl.txt
%% We did our best to help you, but there is NO WARRANTY. 
%% 
%% Please send your comments via 
%% 
%%   http://www.contact-ednotes.sty.de.vu 
%% 
%% * USAGE: * 
% 
% *Requirements/overview:* 
% 
% The package is made for the background of the LaTeX2e macro 
% package and enhances functionality of the following packages: 
% 1.) Stephan I. B"ottcher's `lineno.sty' for printing line 
%     numbers in the margin and a \label version \linelabel 
%     relating labels to line numbers; 
% 2.) our `ednotes.sty' for indicating variant readings and 
%     other editorial remarks in separate footnote apparatuses; 
% 3.) David Carlisle's `longtable.sty' for multi-page tables. 
%                                                 %% TODO: supertabular etc.!? 
% 4.) our `ltabptch.sty' for a patch of `longtable.sty' 
%     (i.e., optionally, see below). 
% 
% Actually, only the first package is necessary for using the 
% present one. The present one needs your line numbering commands 
% according to the first one and its documentation to which we 
% refer here. We likewise refer to the remaining packages for the 
% details of their functionality. 
% 
% `lineno.sty' version 4.1 and (in case it is used) `ednotes.sty' 
% version 0.8 (onwards) are needed. For obtaining recent versions 
% of required packages, see the CTAN folder 
%     /macros/latex/contrib/ednotes. 
% 
% *Install/load:* 
% 1.) To be used, the present file must be put into a folder 
%     that (La)TeX searches. You should have obtained a file 
%     containing more detailed hints about this. 
% 2.) We recommend loading this file *not* by 
%     \usepackage[<options>]{edtable} but by loading `lineno.sty' 
%     or `ednotes.sty' with package option `edtable'. 
% 3.) To use the package options `longtable' and 
%     `nolongtablepatch' that are described below, enter them 
%     as options for `lineno.sty' or `ednotes.sty'. 
% 
% *User Commands:* 
% 
% 1. The package defines an environment `edtable' -- for its 
% syntax we consider two cases: 
% (a) Let <stdtable> be a tabular environment "like `tabular'". 
% "Like" here means: (i) we have tested it with `tabular', but 
% it should work with many more (e.g., from the `array' and 
% `tabularx' packages) -- which share certain properties of 
% implementation and requirements. Sorry, you must try, or we 
% hope you will (and tell us). (For wizards: <stdtable> may 
% probably be anything using a single \halign and \tabskip=0pt.) 
% Definitively: 
% (ii) LaTeX's standard `array' and other environments are *not* 
% "like" `tabular'. Namely, environments that work only in math 
% mode are not meant here. They are considered below in `b.'. 
% (iii) `longtable' from the `longtable' package of the `tools' 
% bundle neither is meant. An option of the present package 
% deals with it as described below. Neither `supertabular' is 
% meant. 
% (iv) In general, usage with tabular environments that can 
% break across pages as in (iii) is not recommended, at least 
% when working with `ednotes.sty'. Even if some worked here, a 
% shortcoming with `ednotes.sty' would be that the footnotes can 
% appear on bad pages. 
% -- Now the syntax is: 
%    \begin{edtable}{<stdtable>}<args><entries>\end{edtable}
% -- where <args> and <entries> are usual arguments and entries 
% for <stdtable>. So an example is: 
%    \begin{edtable}{tabular}{cc} 
%      left upper & right upper \\ left lower & right lower 
%    \end{edtable} 
% This produces just what <stdtable> does in an extra line, 
% only adding line numbers in the margin and processing `ednotes' 
% commands in the entries appropriately. 
% (b) Let <stdarray> be an "array environment" like -- standard 
% LaTeX's `array'! "Like" here means (i) ... analogously to 
% <stdtable> above. (ii) "Text" ... oh, sorry, this is not clear 
% to me at present, and I am in a hurry.                 %% TODO 
% (iii) Nothing from the `amsmath' package seems to work here, 
% sorry, I have tried a lot ... (iv) For standard LaTeX's 
% `eqnarray', it is better to try the `linenomath' environment 
% of `lineno.sty' [with package option `mathlines']. Its 
% modifications in `amsmath' produce (at best) a spurious line 
% number with the `linenomath' environment (up to now). 
% -- Now the syntax is: 
%    \begin{edtable}[<$$$>]{<stdarray>}
%      <args><entries>\end{edtable}
% -- where <args> and <entries> are usual arguments and entries 
% for <stdarray>. <$$$> can be one of `$' or `$$'. Indeed, the 
% <stdtable> syntax above (without [<$$$>]) *does not work* 
% inside $...$ or $$...$$. -- An example: 
%    \begin{edtable}[$$]{array}{cc}
%      a_{11} & a_{12} \\ a_{21} & a_{22}
%    \end{edtable}
% Now, this produces just what <stdarray> does in an extra line 
% etc. as with <stdtable>. However, \linelabel and `ednotes.sty' 
% footnotes may need the `mathrefs' option of `lineno.sty'. For 
% the difference between `[$]' and `[$$]', see below. 
% 
% 2. When working with `ednotes.sty', in 
%   \Anote{L1\<L2\>L3}{NOTE}
% L2 may contain &'s and \\'s, but L1, L3 must not! Analogously 
% for \Bnote, etc., \Anotelabel...\donote..., etc. -- Well this 
% holds for <stdtable>, hardly for <stdarray> ...        %% TODO 
% 
% 3. On positioning: You may have wondered about `extra line' 
% above. This means that `edtable' starts a new line at 
% \begin{edtable} and at \end{edtable}.          %% TODO: sure!? 
% (a) It should be placed within a `center', `flushleft', or 
% `flushright' environment for proper (vertical) spacing, but 
% also works without. However, with <stdarray> and optional 
% parameter `$$' of `edtable' (i.e., \begin{edtable}[$$]...), 
% you can obtain the usual math display spacing *within a 
% paragraph*; especially, the vertical space is less when the 
% previous line is short ... and so on. 
% (b) Horizontal positioning of line numbers (usually) needs a 
% second run (after changing a table)! %% TODO: automatic warning. 
% (For wizards: It may fail if some of \leftskip, \rightskip, 
% \linewidth, and \@totalleftmargin are used in an unusual way.) 
% A <dimen> register \ETextraoffset is provided whose value is 
% 0pt by default and which additionally moves line numbers 
% to the left (right) if given a positive (negative) value. 
%                                               %% TODO: any use? 
% 
% *Package options:* 
% 
% 1. Option `longtable' adjusts the `longtable' environment 
% defined by David Carlisle's longtable.sty for use with 
% `lineno.sty' and `ednotes.sty'. The option makes `longtable' 
% environments appear with line numbers in the margin, according 
% to `lineno.sty', and process `ednotes.sty' commands if line 
% numbering is active according to `lineno.sty'. 
% [We maintain an alternative package with just this function 
% in a slightly different implementation.] %% -> ulnltab.sty. 
% Lemmas may go across table entries as with `edtable' (see above). 
% ---There might be options like `supertabular' for adjusting 
% other tabular environments that cannot be handled by the 
% `edtable' environment provided here---they have not been 
% implemented yet!                                             %% TODO 
% 
% 2. Option `nolongtablepatch' avoids loading/asking for 
% `ltabptch.sty'. I.e.: according to LaTeX bugs database, 
% tools/3180 and tools/3485, there have been problems with 
% vertical spacing around `longtable' environments. By default 
% the present package loads our `ltabptch.sty' or asks for it. 
% Option `nolongtablepatch' overrides this. This is useful when 
% you don't want to have the patch, especially when you use an 
% "emergency stop" installation of TeX. 
% 
% *Wizard interface:* 
% Macros \@ET@makeLineNumber, \@ET@use@outerhook, 
% \@ET@execute@outerhook, \@ET@ampnotes \@ET@startlinewith, 
% \@ET@trivialize@linelabel are provided as tools for adjusting 
% tabular environments for use with `lineno.sty' and 
% `ednotes.sty'. \@ET@step@linenumber is \let \stepLineNumber 
% (from `lineno.sty'), but could be used as a hook for 
% something different. See Environment `edtable' and Option 
% `longtable' below for examples of application. 
%                                        %% TODO: specification here. 
% 
%
%% * Acknowledgements: * 
% 
% Stephan I. B"ottcher told us how to do it in extensive discussion 
% and by providing some first code lines. We changed these essentially 
% in some respects, but kept his general ideas and some parts of his 
% macros, the latter even without knowing what we are doing. 
% 
% v1.3 is due to a request by Martin Brandenburg. 
% 
%  * Now for internals: * 
% 
\NeedsTeXFormat{LaTeX2e}[1994/12/01] 
% 1994/12/01: \newenvironment* etc. %% TODO: more recent needed? 
\ProvidesPackage{edtable}[\filedate\space v\fileversion\space 
  arrays with lineno + ednotes (ul)] 
% 
% Alternative ideas for implementation: 
% 1. There is a German package `TABMAC' enhancing `EDMAC' (cf. 
% documentation of `ednotes.sty'). We could have rewritten its 
% macros so it would work with `lineno.sty' and `ednotes.sty' 
% in place of `EDMAC'. [TODO: "tabmac.sty"] 
% 2. We redefine `longtable' by Option `longtable' because 
% there seems to be only one reasonable use of the `longtable' 
% environment in the course of a critical edition. This may be 
% different for environments like `array' and `tabular' which 
% can be used within text lines. Therefore, these environments 
% are not redefined. 
% 3. One might use \everycr, however: (a) it is executed 
% before the first row, (b) it is executed in fake lines that 
% `longtable' uses. 
% 4. As an alternative to \everycr, there is the approach of 
% redefining \cr and \crcr for setting \noalign. This needs 
% redefining \@tabularcr etc.---not nice. We use this approach 
% here for `longtable' which uses a stretching \tabskip. 
% For \tabskip=0pt, we attach line numbers by a template. 
% 
% \RequirePackage{lineno}[2004/10/06] 
% Wanted to check lineno version--causes unknown option error. 
% See LaTeX bug latex/3730. 

% Options: 

%% TODO: underful page with `longtable.sty'!? 
\let\if@ET@longtable\iffalse % \newif\if@ET@longtable@ 
\DeclareOption{longtable}{\let\if@ET@longtable\iftrue} 
\let\if@ET@ltabptch\iftrue 
\DeclareOption{nolongtablepatch}{\let\if@ET@ltabptch\iffalse}
\ProcessOptions

% Redefine \longtable if option: 
\if@ET@longtable
  % Stephan's direction for attaching a line number to each row 
  % is using \noalign in the course of \\. 
  % However, (1) the user should not worry about closing the table 
  % with \\ or without, and (2) the line number thing should not 
  % happen twice if \\ is the last token before \end{longtable}. 
  % Our solution: attach something to the beginning of \endlongtable 
  % which attaches line number unless an unsucceeded \\ has done it. 
  % This thing is activated by each row start and turned off by \\. 
  % Since the switch at row start may happen to be in a pbox, 
  % let it act globally on \endlongtable. (The \crcr at the beginning 
  % of \endlongtable must not be changed directly, since original 
  % \crcr may be required in table entries using \oalign etc.) 
  % 
  \RequirePackage{longtable}%
%   \IfFileExists{longtable.sty}{\RequirePackage{longtable}}%
%                               {\RequirePackage{longtabl}}%
% For Atari problem, it suffices to rename `longtable.sty' 
% into `longtabl.sty'. 
% 
% Patch for tools/3180 and tools/3485 of LaTeX bug database:
 \if@ET@ltabptch
  \IfFileExists{ltabptch.sty}{\RequirePackage{ltabptch}}{% 
    \PackageError{edtable}{% 
      ltabptch.sty (for improving spacing around\MessageBreak
      longtable) missing! Be sure to use it always\MessageBreak 
      or never!%
    }{% 
      To omit ltabptch.sty *and* escape this error,\MessageBreak 
      use package option `nolongtablepatch'.%
    }%
 }
 \fi 
  \let\@ET@@longtable\longtable 
  \def\longtable{% 
    \ifLineNumbers 
      \expandafter\@ET@longtable 
    \else 
      \let\@ET@sw@cr\@ET@crcr % ... in \endlongtable. 
      \expandafter\@ET@@longtable 
    \fi 
  } 
  \def\@ET@longtable{% 
  % Since we have made it anyway, we use the method of redefining 
  % \halign for inserting the activating row start as well. 
  % However, redefinition must be repeated before every \LT@bchunk. 
  % \longtable and \LT@get@withs are good places for this. 
    \@ET@startlinewith\@ET@sw@cr@on 
    \@ET@trivialize@linelabel 
    \@ET@use@outerhook 
    \@ET@ampnotes 
    \let\LT@tabularcr\@ET@LT@tabularcr 
    \expandafter\def\expandafter\LT@get@widths\expandafter 
      {\LT@get@widths\@ET@startlinewith\@ET@sw@cr@on}
    % Admittedly, we could make it less of a hack by using @{...} 
    % ---we could then leave \LT@get@widths untouched. 
    \@ET@@longtable 
  } 
  \def\@ET@sw@cr@on{\global\let\@ET@sw@cr\@ET@cr@attach} 
  % We are hacking a longtable version offering 
  %   \def\LT@tabularcr{%
  %     \relax\iffalse{\fi\ifnum0=`}\fi
  %     \@ifstar
  %       {\def\crcr{\LT@crcr\noalign{\nobreak}}\let\cr\crcr
  %        \LT@t@bularcr}%
  %       {\LT@t@bularcr}}
  % We need a redefinition of \cr which is not overridden by \@ifstar: 
  \def\@ET@LT@tabularcr{%
    \relax\iffalse{\fi\ifnum0=`}\fi
    \def\cr{\@ET@sw@cr}% 
    % So \crcr is affected by change of \@ET@sw@cr, cf. below. 
    \let\crcr\cr 
    \@ifstar
      {\gdef\@ET@sw@nobreak{\nobreak 
       \global\let\@ET@sw@nobreak\relax}% 
  %     {\def\cr{\@ET@sw@cr\noalign{\nobreak}}\let\cr\crcr
  % Why didn't this work? %% TODO: try shorter again. 
       \LT@t@bularcr}%
      \LT@t@bularcr 
  %     {\let\cr\@ET@sw@cr \let\crcr\cr
  %      \LT@t@bularcr}% This accompanied \def\cr... above. 
  }
  % Attach \@ET@sw@cr to beginning of \endlongtable: 
  \expandafter\def\expandafter\endlongtable\expandafter 
    {\expandafter\@ET@sw@cr\endlongtable}
  \def\@ET@cr@attach{% Actually attaching line numbers. 
    \@ET@crcr\noalign{% Basically Stephan's approach. 
  % If we were not careful, following box containing line number 
  % could fool interline glue after longtable, even if tools/3485 
  % is repaired in some way. This box should have same depth as 
  % the line composed previously. 
      \nobreak %% \@ET@ex... might cause page break. 2003/10/30. 
      \setbox\z@\@ET@makeLineNumber 
      \ht\z@-\prevdepth \dp\z@\prevdepth \box\z@ 
      \@ET@step@linenumber 
      \global\let\@ET@sw@cr\@ET@crcr % ... if called by \\. 
  % This also resets the \crcr starting \LT@echunk following 
  % \@xargarraycr or \@yargarraycr in \LT@argtabularcr. 
  % \let...\relax seems to suffice at well, but in case ... 
      \@ET@execute@outerhook %% 2003/10/30. 
      \@ET@sw@nobreak 
%       \@ET@execute@outerhook 
    }% 
  } 
  \let\@ET@crcr\crcr 
  \global\let\@ET@sw@nobreak\relax % Just to remind ... 
\fi 
 
\let\@ET@step@linenumber\stepLineNumber 
% \def\@ET@step@linenumber{\global\advance\c@linenumber\@ne}
\def\@ET@makeLineNumber{\hb@xt@\z@{\makeLineNumber}} 
%% TODO: export to lineno.sty!? 
% Special \halign: 
% For `array' etc., we insert line numbers by a leftmost template. 
% @{...} in the last argument of `array' etc. is difficult for this 
% since `tabular*' has an additional argument. So we redefine 
% \halign to put something to the right of the next \bgroup. 
% Now, as a macro, \halign would break in an \edef or \xdef. 
% Horribly, this danger has become quite real in longtable.sty's 
% definition of \LT@bchunk. Only its latest version 4.10 has 
% preceded \halign by a \noexpand (for mathtext.sty which 
% redefines \halign as well). We need not rely on such a 
% provision if we let \halign expand to \the<token register> 
% for some token register. However, using \toks@ or \@temptokena 
% is neither very safe---might be filled by macro ahead with 
% something new. So reserve an own token register. 
\newtoks\@ET@toks 
\@ET@toks{\@ET@specialhalign} 
% v1.3: In 'AMSmath', there are \halign's followed by explicit 
% left braces. Thus '\def\@ET@specialhalign#1\bgroup{%' broke. 
%% TODO ... 
\def\@ET@specialhalign{%
  \ifmeasuring@ \expandafter\@firstoftwo 
% v1.3 2005/03/04, for `amsmath'. 
  \else \expandafter\@secondoftwo 
  \fi
  {\let\halign\@ET@@halign \halign}% 
  {\@ifnextchar\bgroup\@ET@replace@arg\@ET@sphalign@to}}
%% TODO: change back, commenting out code above, report 
%%       error with `alignat'. 
%   \@ifnextchar\bgroup\@ET@replace@arg\@ET@sphalign@to}
% If `amsmath' not loaded (v1.3): 
\AtBeginDocument{\@ifundefined{measuring@true}{% 
  \let\ifmeasuring@\iffalse}\relax}
\def\@ET@replace@arg#1{%
  \def\@EN@tempa{#1}\def\@EN@tempb{\bgroup}% Corr. after v1.3b. 
  \ifx\@EN@tempa\@EN@tempb 
    \def\@EN@tempa{\@ET@sphalign@to\bgroup}%
  \else 
    \def\@EN@tempa{\@ET@sphalign@to{#1}}%
% This may be wrong, #1 might be `t' from `to' ... 
  \fi 
  \@EN@tempa}
\def\@ET@sphalign@to#1\bgroup{% 
  \let\halign\@ET@@halign % Reset for nested arrays. 
  \halign#1\bgroup 
%  \ifmeasuring@\else % For `amsmath', moved backwards. 
    \@ET@startline % [Wizard interface, via \@ET@starlinewith.] 
%  \fi 
}
% Wizard interface: 
\def\@ET@startlinewith{% Next token precedes preamble. 
%   \let\@ET@@halign\halign % Or \AtBeginDocument? 
% ... indeed: bad loop with long longtable (\LT@get@widths!?) 
  \def\halign{\the\@ET@toks}% 
  \let\@ET@startline 
} 
\AtBeginDocument{\let\@ET@@halign\halign} 
% Or save it at start of environment only 
% (don't repeat inside longtable). 
% 
% Outer hook for inserts: (wizard interface) 
\def\@ET@execute@outerhook{% To be placed outside inner mode. 
  \@ET@outerhook \global\let\@ET@outerhook\@empty 
} 
\global\let\@ET@outerhook\@empty % Just to remind: global! 
% \@EN@hookfn (ednotes) sends to outer hook: 
\def\@ET@use@outerhook{% Wizard interface. 
  \def\@EN@hookfn{\g@addto@macro\@ET@outerhook}% 
}%% TODO: move last line to `ednotes', \let\@ET@use...\@empty. 
%%        But this requires re-installing both packages 
%%        at the same time. See the `ampnotes' thing as well. 
% 
% Trivialize \linelabel: (wizard interface) 
\def\@ET@trivialize@linelabel{\let\linelabel\@ET@linelabel}
\def\@ET@linelabel{% 
  \protected@edef\@currentlabel{\theLineNumber}% 
  \label
}
% Change ampersand at \Anote etc.: (wizard interface) 
%% 2003/10/31. 
% & is changed by \Anote etc. No \begin...\endgroup, 
% next \\ or & after % note/donote command switches back. 
% ---Global change seems to break table setup. 
\begingroup 
\gdef\@ET@hideamps{\catcode`\&\active}
\@ET@hideamps 
\gdef\@ET@ampnotes{% 
  \let&\@ET@hideamp 
  %% & would be undefined after \\ in lemma. 
  \let\@ET@EN@rnote\@EN@note 
  \def\@EN@note{\@ET@hideamps\@ET@EN@rnote}%
  \let\@ET@EN@rnotelb\@EN@notelabel 
  \def\@EN@notelabel{\@ET@hideamps\@ET@EN@rnotelb}%
} 
%% TODO: & and \\ in \@EN@lemmaexpands!? 
\endgroup 
\def\@ET@hideamp{&}

\newenvironment*{edtable}[2][]{% 
% #1 $ or $$, #2 standard environment name. 
%% TODO: star version!? cf. lineno.sty's `numquote'. 
  \def\@ET@mode{#1}% 
  \def\@ET@currenvir{#2}% 
  \ifhmode \ifnum\the\lastpenalty=-\@M\else 
% \par must be executed for printing/stopping line numbers: 
    \@centercr\relax \noindent 
  \fi \fi 
  \ifvmode \noindent \fi % Otherwise linenumbers are indented 
% -- however, not recommended. 
  \@ET@use@outerhook 
  \@ET@trivialize@linelabel 
  \@ET@ampnotes 
  \@ET@startlinewith\@ET@normal@startline 
  \global\advance\c@linenumber\m@ne % Stepping correction. 
% (Must come after closing previous paragraph.)
% Calculate offset: 
  \global\advance\c@ET@array\@ne 
% Saving/calling width of table (`longtable's algorithm). 
% Test if there is \@flushglue on the left (center or flushleft): 
  \edef\@EN@tempa{\the\@flushglue}% 
% ... v1.3: fool this for "$$". 
  \in@{$$}{#1\@nil}% \in@{#1}{#2} searches #1 in #2#1 ... 
  \edef\@EN@tempb{\the\ifin@\@flushglue\else\leftskip\fi}% 
  \ifx\@EN@tempa\@EN@tempb 
% Assume \@ET@offset is 0pt else here. 
    \@ET@offset\linewidth 
    \advance\@ET@offset-% 
      \@ifundefined\@ET@arraywidth@csn % Shortened for v1.3. 
        {\linewidth\G@refundefinedtrue 
         \PackageWarningNoLine{edtab}{^^JLine numbers at 
           \@ET@currenvir\space need \jobname.aux}}% 
        {\csname\@ET@arraywidth@csn\endcsname}% 
% Test if \@flushglue on the right (center): 
% ... v1.3: fool it again. Repeat \in@ in case ... 
    \in@{$$}{#1\@nil}%
    \edef\@EN@tempb{\the\ifin@\@flushglue\else\rightskip\fi}% 
    \ifx\@EN@tempa\@EN@tempb \divide\@ET@offset\tw@ \fi 
  \fi 
  \advance\@ET@offset\@totalleftmargin 
  \advance\@ET@offset\ETextraoffset % Offset ready. 
  \setbox\@tempboxa\hbox\bgroup 
% v1.3: `alignat' complains later when this is \vbox instead. 
%    \vbox\bgroup % Tried in vain. 
    \ifx\@ET@mode\@empty\else $\fi 
    \csname#2\endcsname 
}{% 
    \csname end\@ET@currenvir \endcsname 
    \ifx\@ET@mode\@empty\else $\fi 
  \egroup 
  \@ET@step@linenumber % Stepping correction. 
  \if@filesw
    \immediate\write\@auxout{%
      \gdef\expandafter\noexpand
        \csname\@ET@arraywidth@csn\endcsname
          {\the\wd\@tempboxa}}%
%% TODO: warning if changed!? 
  \fi
% No automatic line number at array line: 
  \nolinenumbers 
  \@ET@mode\ensuremath{\vcenter{\box\@tempboxa}}\@ET@mode % v1.3. 
  \@ET@execute@outerhook
  \ifhmode \@centercr\relax \fi 
% \par must be executed for proper restart of printing line numbers 
% (think so). 
  \@endpetrue % Outside \@flushglue environment, avoid indent ahead. 
%% TODO: slightly change for math mode around? Or give an error 
%%       in math mode. 
} 
%% TODO: make \edtable \outer!? 
\def\@ET@normal@startline{%
  \@ET@step@linenumber 
  \llap{\@ET@makeLineNumber \hskip\@ET@offset}% 
} 
\newdimen\@ET@offset 
\newdimen\ETextraoffset 
%% TODO: export to lineno.sty!? 
\newcount\c@ET@array 
\def\@ET@arraywidth@csn{ET@a@\romannumeral\c@ET@array}
% 
%% TODO: .dtx 
\endinput 

%% VERSION HISTORY: 
v0.22 2003/01/13  First version, named `edtab02.sty', sent around, 
                  together with ednotes.sty. 
v0.23 2003/01/22  Added version history. 
v0.24 2003/01/27  Copyright notice. 
%%% v0.3  2003/03/03  Requires linenox0.sty. %% RETREATED 
%% TODO: reconsider,  see `lineno[x].sty'. 
v0.25 2003/03/05  Added <ednotes.sty@web.de>. 
v0.26 2003/03/23  Added \@ET@startlinewith to wizard command list; 
                  moved redefinition \LT@get@widths from option code 
                  into \@ET@longtable so this really happens in 
                  line number mode only; moved \let\@ET@@halign\halign 
                  from \@ET@startlinewith to \AtBeginDocument. 
v0.27 2003/06/02  Added \ifvmode \noindent \fi to `edtable' definition; 
      2003/06/10  added `lemma cannot run' to instructions. 
v0.28 2003/10/31  \@ET@cr@attach: moved \@ET@execu..., added \nobreak. 
                  Added \@ET@ampnotes and documentation: now working 
                  across entries. 
v1.0  2004/05/13  `ednotes.sty' etc., updated copyright. Renamed as 
                  `edtable.sty' for TUGboat article. 
                  \RequirePackage{lineno}, changed doc. accordingly. 
                  Removed `may be released under different name'. 
v1.10 2004/07/27  Looking for `longtabl.sty' for my Atari! 
      2004/08/22  Added warning (screen/documentation) concerning 
                  ltabptch.sty. 
      2004/08/23  Added suggestion of option `nolongtablepatch'; 
                  LPPL v1.3. 
v1.2  2004/08/31  Rearranged preamble concerning maintenance, 
                  removed `preliminary release'. 
      2004/10/06  \newif -> \let...; option `nolongtablepatch'; 
                  changed documentation accordingly and for 
                  loading edtable.sty by lineno/ednotes option; 
                  \let\@ET@step@linenumber\stepLineNumber; use 
                  \ifLineNumbers. 
      2004/10/07  Needed version numbers. 
      2004/10/08  ltabptch warning -> error! 
      2004/10/11  Disabled \RequirePackage{lineno}[...]; undid 
                  the `longtabl.sty' thing from 2004/07/27. 
      2004/10/12  latex/3730. 
v1.2a 2004/11/07  LPPL v1.3a. 
v1.2b 2005/01/10  Contact via http. 
v1.3b 2005/03/04  Support for environments like `array' 
                  environment. The description of `edtable' was 
                  wrong in this respect! 
v1.3  2005/03/07  Adaptations in documentation, and extended usage 
                  instructions very much, correcting some 
                  errors as well. Acknowledgement to M. B. 
v1.3a 2005/03/09  Corrected numbering in `usage'. 
v1.3c 2005/03/15  \@centercr\relax, TODO on math mode. 
      2005/04/09  `editory' -> `editorial'. 
\end{verbatim}
\subsection{\texttt{ltabptch.sty}}
\verbatiminput{ltabptch.sty}

\section{\cs{linelabel} and notes from \textit{math} mode: 
         \notinaux{\\} \texttt{ednmath0.sty}}
\begin{verbatim}
%% Macro package `ednmath0.sty' for LaTeX2e, 
%% copyright (C) 2004 Uwe L\"uck, 
%% http://www.contact-ednotes.sty.de.vu 
%% --author-maintained; 
%% math support for `lineno.sty' and `ednotes.sty'. 
%% 
\def\fileversion{v0.2b} \def\filedate{2005/01/10}
%% This program can be redistributed and/or modified under the 
%% terms of the LaTeX Project Public License distributed from 
%% CTAN archives in directory macros/latex/base/lppl.txt; either
%% version 1.3a of the License, or any later version.
%% The latest version of this license is in
%%   http://www.latex-project.org/lppl.txt
%% There is NO WARRANTY. 
%% This code is very EXPERIMENTAL! 
%%
%% Please report bugs, problems, and suggestions via 
%% 
%%   http://www.contact-ednotes.sty@web.de 
% 
%% * MAIN FEATURE * 
% 
% lineno.sty's \linelabel and ednotes.sty's commands are enabled 
% to work in math mode if it's "entered in outer mode" 
% (including `displaymath' and `equation' environments). 
% (lineno.sty is the package by Stephan Boettcher.) 
% They will even work in tabular environments that are adjusted 
% to notes by package `edtable.sty'. 
% 
% CAVEATS: 
% -- Does not work yet in environments like LaTeX's 
% `eqnarray'. (This could probably repaired along the lines 
% of Edtable.sty--we're short of time and will try later.) 
% -- Useful error messages when (i) math mode is entered from 
% inner mode or when (ii) a math display gets not line number 
% are missing at present. 
% 
%% * USAGE: *
% 
% * Most simple: * 
% --If you are working with ednotes and want to use its 
% commands in math mode, load ednotes.sty--version 0.8 
% onwards--with its package option `mathnotes'. 
% --If you don't work with ednotes, only with lineno, you 
% get the main feature of making \linelabel work in math mode 
% by loading lineno.sty--version 4.1 onwards--with its 
% package option `mathrefs'. 
% 
% * Switch off and on: * 
% To reduce danger resulting from missing error messages 
% ("caveat" above), you may switch these new math facilities 
% off by \NoNotesToMath where you don't expect to need them. 
% You may switch them on again by \NotesToMath where you want 
% to use them, being aware of the danger. Both commands work 
% locally, so you can replace one of them by enclosing it in 
% a group. E.g., even, after \NoNotesToMath you can use an 
% environment as follows: 
%   \begin{NotesToMath}
%     <text> 
%   \end{NotesToMath} 
% (I am not quite sure that this is useful.)
% 
% * Customize ellipsis: * 
% ednotes' \lemmaellipsis is changed to expand to 
% \mathlemmaellipsis when entering math, and this is preset 
% to be LaTeX's \mathellipsis. (This is three dots as 
% \mathinner.) You can change this by redefining 
% \mathlemmaellipsis, e.g.: 
%   \renewcommand{\mathlemmaellipsis}{\cdots}
% If you need \cdots as the ellipsis at a single place only, 
% you may, of course, use the `<...>' option of \<, e.g.: 
%   $ x = \Anote{a\<<\cdots>bcd\>e}{Indeed?} - y $
% 
% * Customize note mode: *
% For variant readings, you may want that the note is 
% usually set in math mode--so you may want that you 
% needn't type the dollar signs in the note text. 
% Note that you can do this by customizing \notefmt, 
% and you can do this by customizing \Anotefmt (e.g.) 
% to have this feature for \Anote only. 
% 

\NeedsTeXFormat{LaTeX2e}
\ProvidesPackage{ednmath0}[\filedate\space\fileversion\space 
  math support for lineno/ednotes (ul)] 
% 
%% User commands: 
\def\NotesToMath{\let\@LN@mathhook\@LN@labelinmath
  \@bsphack \@esphack 
% For \begin{NotesToMath}
} 
\def\NoNotesToMath{\@bsphack
  \def\@LN@mathhook{\@parmoderr\@gobble}% 
  \@esphack 
} 
\def\endNotesToMath{\@bsphack\@Esphack}
\let\endNoNotesToMath\endNotesToMath 
% 
%% Core code for lineno.sty: 
\@ifundefined{@LN@postlabel}{% 
  \PackageError{ednmath0}{% 
    Bad lineno.sty version% 
  }{% 
    lineno.sty from 2004/08/16 or later 
    must be loaded earlier.% 
  }%
}{% 
  \def\@LN@labelinmath#1{% 
    \ifmmode 
      \@LN@postlabel{#1}% 
    \else 
      \@parmoderr 
    \fi 
  }
} 
% 
%% Core code for ednotes.sty: 
\@ifundefined{@EN@note}{% 
% v0.01 sent a warning in this case. Considered superfluous now. 
}{% 
  \def\@EN@themathlemmatag{%
    \ifmmode 
      \toks@\expandafter{\@EN@lemmatag}% 
      \edef\@EN@lemmatag{%
        $% 
          \def\noexpand\lemmaellipsis{% 
            \noexpand\mathlemmaellipsis}% 
          \the\toks@ 
        $% 
      }% 
%       \expandafter \def \expandafter \@EN@lemmatag 
%         \expandafter {\expandafter $\expandafter 
%         \def \expandafter \lemmaellipsis \expandafter {% 
%           \expandafter \mathlemmaellipsis \expandafter }% 
%         \@EN@lemmatag $}% 
    \fi 
  } 
% To be sure, \lemmaellipsis doesn't need to be changed when 
% ednotes `\<...\>' feature is not used. Though I prefer to 
% use one hook only in ednotes for both situations, with and 
% without `\<...\>'. 
% 
% The final \unskip in ednotes' \@EN@lemmatag would undo a final 
% \quad. That's OK: outside math the same happens. 
% In v0.01, \NoNotesToMath undid ednotes changes for math mode. 
% However, re-appearence of \linelabel error messages suffices. 
% 
% Now add lemma switch to the left of \[No]NotesToMath: 
  \toks@\expandafter{\NotesToMath}
  \edef\NotesToMath{% 
    \let \noexpand\@EN@mathlemmatag \noexpand\@EN@themathlemmatag 
    \the\toks@
  }
%  \typeout{\string\NotesToMath: \meaning\NotesToMath}
  \toks@\expandafter{\NoNotesToMath}
  \edef\NoNotesToMath{% 
    \let \noexpand\@EN@mathlemmatag \relax 
    \the\toks@
  }
%  \typeout{\string\NoNotesToMath: \meaning\NoNotesToMath}
} 
% We need no extra device for a choice for users whether the *note* 
% should be set in math mode or in horizontal mode by default 
% (which might depend on the kind ["layer"] of notes). 
% This can be done already by customization of ednotes' \notefmt. 
% However, we might change ednotes' default \notefmt to default 
%   \renewcommand*{\notefmt}[1]{$#1$}
% 
\let\mathlemmaellipsis\mathellipsis 
%% TODO: Since when has LaTeX provided \mathellipsis? 
%% -> \Needs... 
% 
% Default: 
\NotesToMath
% 
\endinput 

%% TODO: Without \linenumberdisplaymath, in displaymath, 
%% an error should be shown. Use, e.g., that in a displaymath 
%% \ifinner is false. 
%% TODO: E.g., by changing \everymath, perhaps can be warned 
%% that the math group is in a box already, so the vertical 
%% items will get lost. 
%% TODO: Adjust `eqnarray' (in Edtable?) as well. 

%% VERSION HISTORY: 
v0.01 2004/08/16  First version, sent to Christian. 
v0.02 2004/08/16  Considerably simplified for ednotes. 
      2004/08/19  Added ellipsis stuff, documentation, and 
                  instructions. Uncapitalized package names. 
                  Added \end[No]NotesToMath. 
      2004/08/20  Added \@bsphack and \@esphack; corrected 
                  ednotes extension (too much deleted, completely 
                  wrong), introducing \@EN@themathlemmatag. 
v0.02b .../08/31  Rearranged preamble concerning maintenance. 
v0.1  2004/09/20  Removed mentions of `linenox0.sty'. 
v0.2  2004/10/07  Removed another mention of `linenox0.sty'; 
                  Instructions: `lineno' or `ednotes.sty' option. 
v0.2a 2004/11/07  LPPL v1.3a. 
v0.2b 2005/01/10  Contact via http. 


\end{verbatim}

\section{Extended line number references: \texttt{vplref.sty}} 
\texttt{vplref.sty} is input through the \texttt{lineno} 
package option \texttt{addpageno}. This adds page numbers 
to line number references to distant sides---using the 
\texttt{varioref} package from the \LaTeX\ distribution. 
\begin{verbatim}
%% `vplref.sty' 
%% -- extended line number referencing with lineno.sty. 

\def\filedate{2005/04/25} \def\fileversion{0.2}

%% Copyright (C) 2004, 2005 Uwe Lueck, 
%% http://contact-ednotes.sty.de.vu --author-maintained 
%% -- support of lineno.sty for varioref.sty. 

%% This file can be redistributed and/or modified under 
%% the terms of the LaTeX Project Public License; either 
%% version 1.3 of the License, or any later version.
%% The latest version of this license is in
%%   http://www.latex-project.org/lppl.txt
%% We did our best to help you, but there is NO WARRANTY. 

%% USAGE: 
%
% \vpagelineref{<label>} expands to 
%
% a) \ref{<label>} 
%    -- if on same page as \linelabel{<label>}
% 
% b) \LineWithPage{<label>} -- otherwise. 
% 
% \LineWithPage{<label>} expands -- by default -- to 
% 
%   \pageref{<label>}.\ref{<label>}
% 
% This can be customized by editing 
% 
%   \renewcommand*{\LineWithPage}[1]{\pageref{#1}.\ref{#1}} 
% 
% in your document preamble, after vplref.sty has been loaded 
% (which may have happened through lineno.sty with option 
% `addpageno'). 

%% IMPLEMENTATION: 

\NeedsTeXFormat{LaTeX2e}[1994/12/01] %% \Declare...* 
\ProvidesPackage{vplref}[\filedate\space v\fileversion] 

\AtBeginDocument{\RequirePackage{lineno,varioref}} 

%% Anderer Ansatz: GPNo (\FirstOnPage) 

\DeclareRobustCommand*\vpagelineref[1]{{% 
%   \def\reftextcurrent{\lineref{#1}}%% First vpageref arg. 
  \let\reftextfaraway\LineWithPage 
  \def\reftextafter{\reftextfaraway{#1}}% 
  \let\reftextbefore\reftextafter 
  \let\reftextfaceafter\reftextafter 
  \let\reftextfacebefore\reftextafter 
%% <- Looks somewhat stupid, but varioref.sty has its merits 
%%    as compared with the mechanism in ednotes.sty. 
  \vpageref[\ref{#1}][]{#1}%% The robust alternative. 
%% Here and with \LineWithPage, \lineref seems more appropriate 
%% than \ref, but it produces errors when labels have not been 
%% defined. This seems to be an incompatibility with lineno.sty. 
}} 

%% Customizable format for different page: 
\newcommand*\LineWithPage[1]{\pageref{#1}.\ref{#1}}

\endinput 

VERSION HISTORY: 
v0.1   2004/10/19  First, sent to Sergei Mariev. 
v0.11  2004/10/19  Fit to recent varioref version; 
                   sent to Sergei. 
v0.2   2005/04/25  \Require... \AtBeginDocument. 

\end{verbatim}